\documentclass[useAMS,usenatbib]{mn2e}
\usepackage{url}
\usepackage{graphicx}

\def\dim#1{\mbox{\,#1}}

\font\japit = cmti10 at 10truept

\title
     [Baryon-rich structures]
{\vglue-3.0truecm
\centerline{\japit Submitted to Monthly Notices}
\vglue 2.5truecm
\noindent
Intergalactic baryon-rich regions at high redshift
\author[A. G. Harford et al.]
	{A. Gayler Harford$^1$, Andrew J. S. Hamilton$^2$, Nickolay Y. Gnedin$^3$ \\
	$^1$JILA and National Institute of Standards \& Technology,
	  Boulder,CO 80309, USA \\
	$^2$JILA, Dept.\ Astrophysical \& Planetary Sciences,
	Box 440, U. Colorado, and \\
	National Institute of Standards \& Technology, Boulder CO 80309, USA; \\ 
	$^3$Particle Astrophysics Center, Fermilab, Batavia, IL
            60510, USA; \\
	    Kavli Institute for Cosmological Physics,
	    Chicago, IL 60637, USA; \\
	    Dept. \ Astronomy \& Astrophysics,
            U. of Chicago, Chicago, IL 60637 USA}
}

\newcommand{\unit}[1]{\, {\rm #1}}

\newcommand{\cmdcartoon}{
    \begin{figure}
    \begin{center}
    \leavevmode
    \includegraphics[scale=0.5]{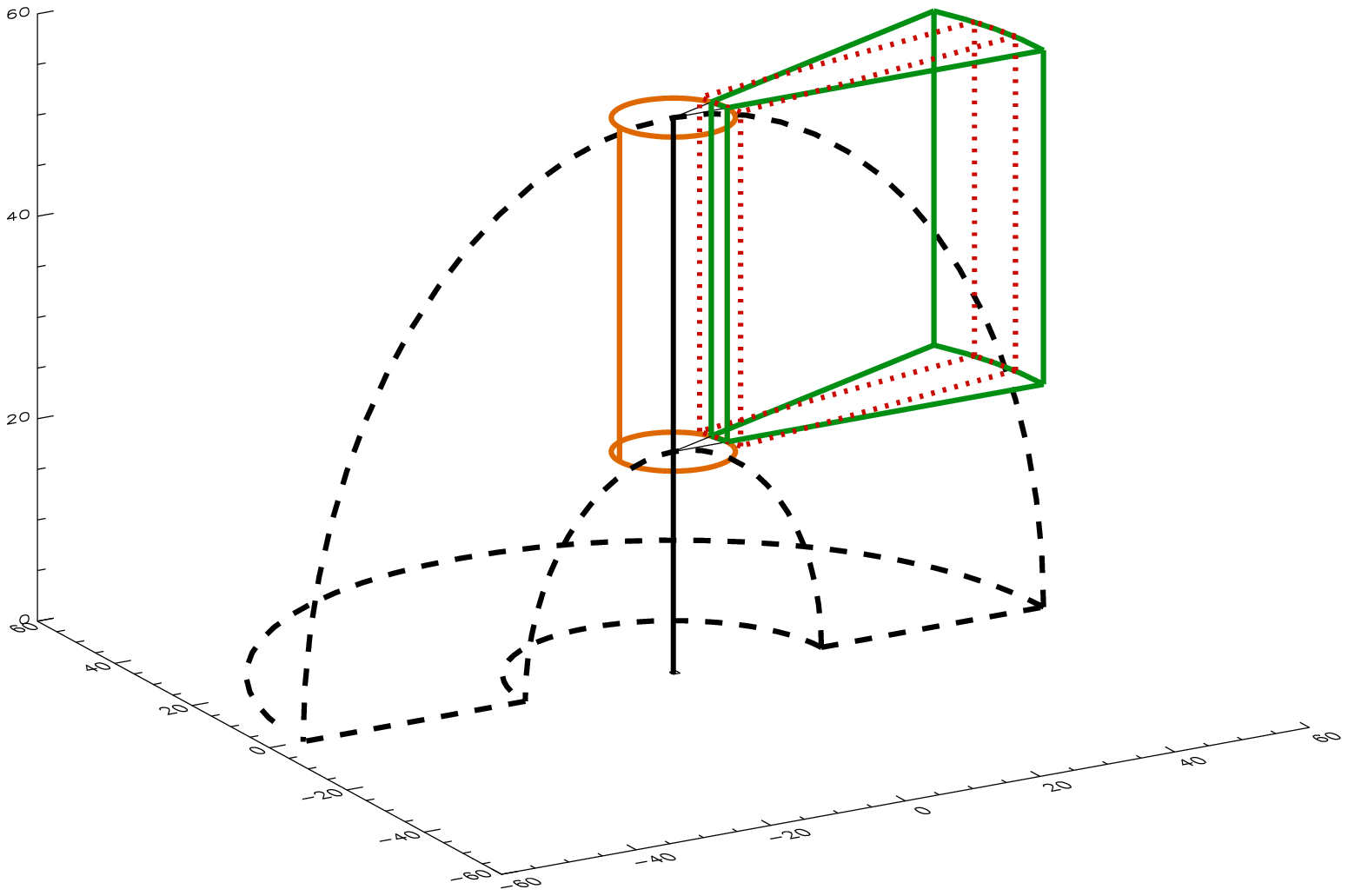}
    \end{center}
    \caption[1]{
    \label{cartoon}
    Cutaway of analyzed regions surrounding a galaxy.
    The solid vertical line represents a filament 
    extending $55.4 \unit{kpc}$ from the center of a galaxy
    at its lower end.  
    The dashed black lines show the shell about the galaxy
    that contributes to the HealPix contour map.
    Linear density along the filament is computed for the
    $8 \unit{kpc}$ orange cylinder.
    The green wedge portrays one of the 16 radial sectors
    used to select the sheets.  The red dotted box depicts
    a slab centered on a selected sheet.
    }
    \end{figure}
}

\newcommand{\cmdsimbox}{
    \begin{figure}
    \begin{center}
    \leavevmode
    \includegraphics[scale=0.39]{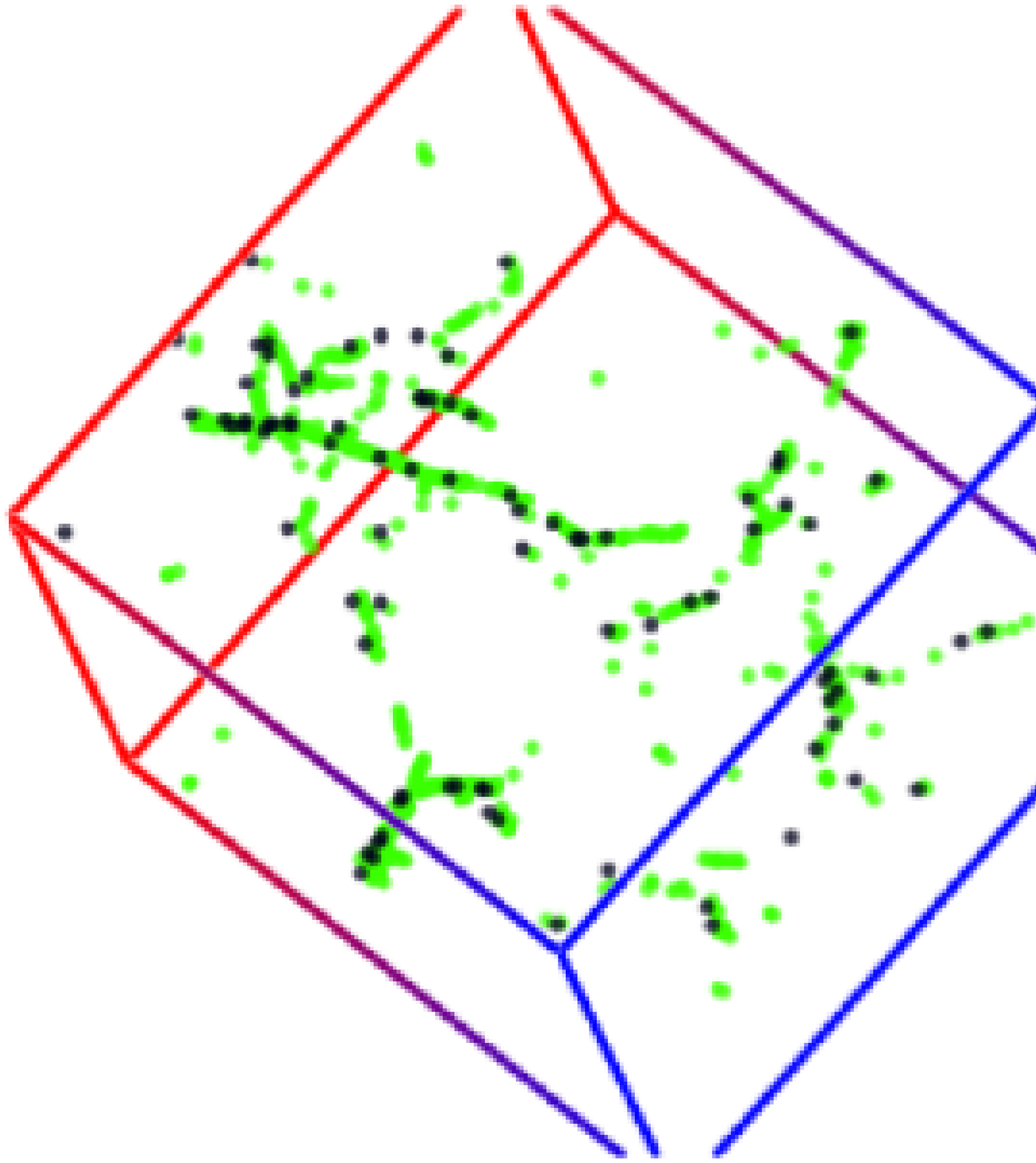}
    \includegraphics[scale=0.39]{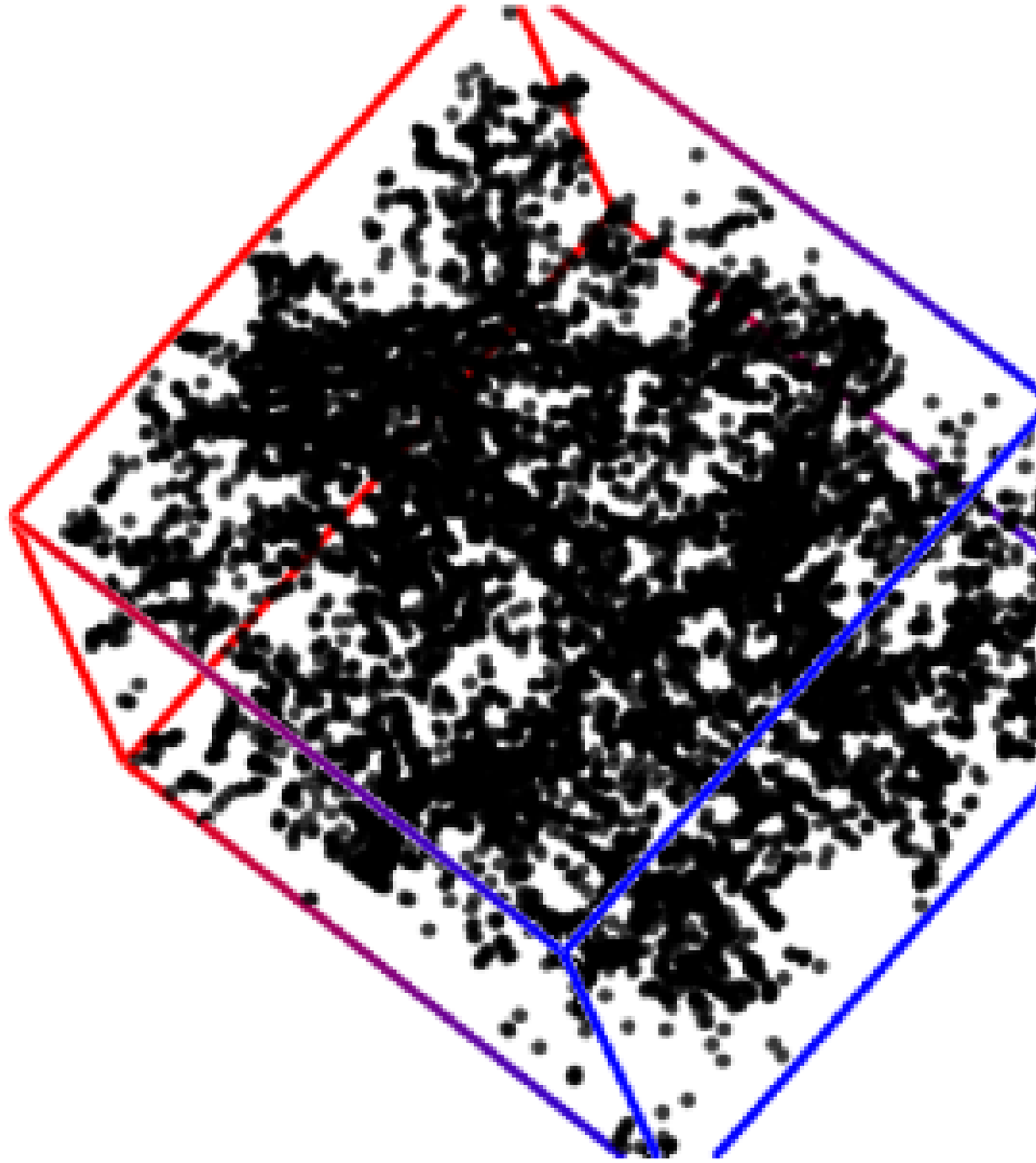}
    \end{center}
    \caption[1]{
    \label{simbox}
    The dense regions in the simulation box at redshift 5.1.
    Upper:  All regions of 
    overdensity 56 or higher that are enriched in baryons to 
    at least twice the cosmic mean are shown in translucent
    green.  The centers of the 88 largest galaxies, those
    having a minimum total mass of $10^{10} \unit{M}_{\sun}$, are shown
    as black dots.  Densities and baryonic fractions were computed
    on grid cells $6.5 \unit{kpc}$ proper on a side.  The simulation box is
    1837 proper kpc on a side at this redshift.
    Lower: Same as upper but showing instead all regions whose baryonic
    fraction are less than twice the cosmic mean.
    }
    \end{figure}
}

\newcommand{\cmdgalacct}{
    \begin{figure}
    \begin{center}
    \leavevmode
    \includegraphics[scale=0.5]{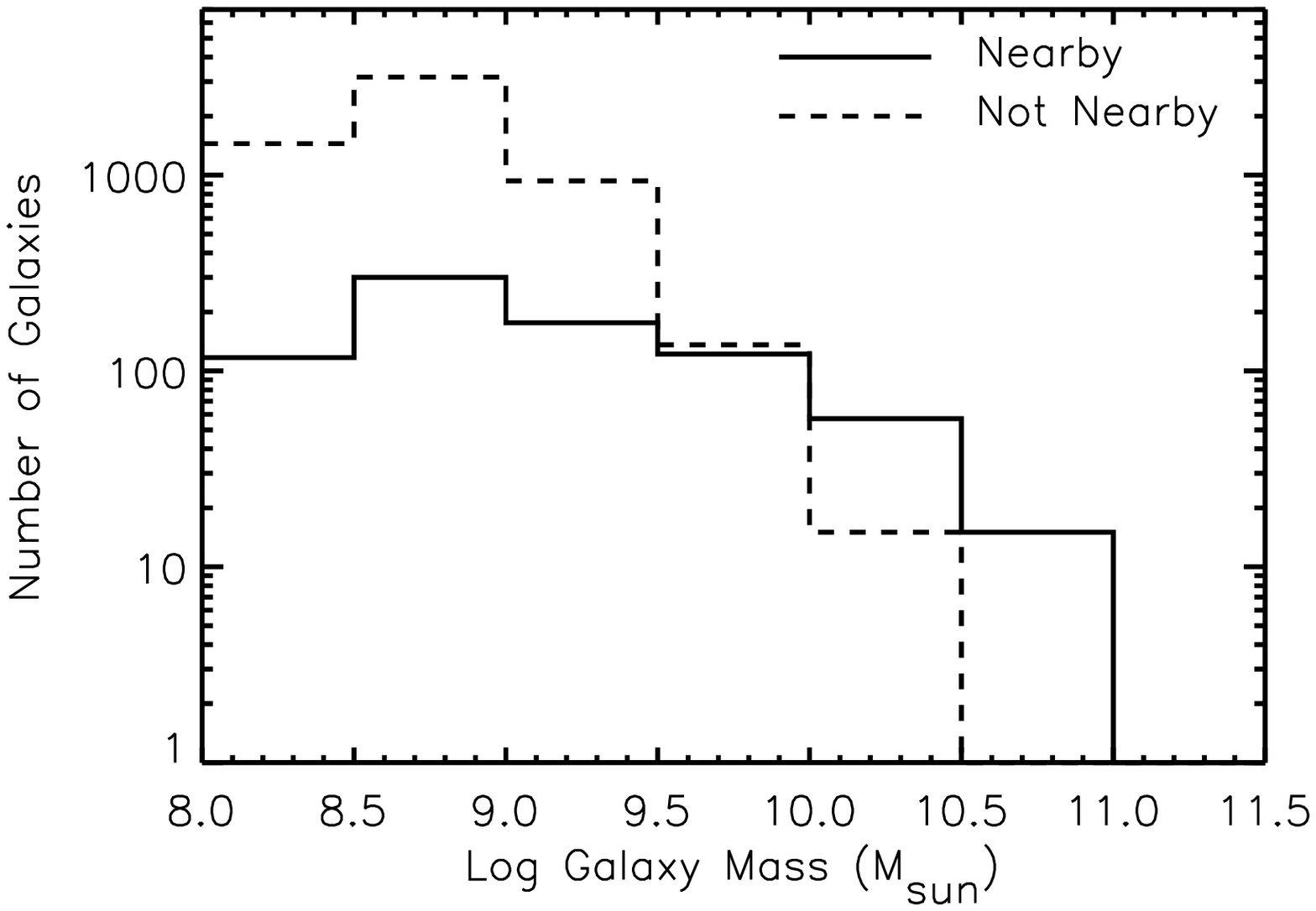}
    \end{center}
    \caption[1]{
    \label{galacct}
    Number of galaxies with centers that are (solid line) or are not
    (dashed line) within two grid cells of a
    baryon-rich cell shown in Figure \ref{simbox}.  Numbers are shown
    as a function of total galaxy mass.  All galaxies above
    $10^{10.5} \unit{M}_{\sun}$ are nearby including one above
    $10^{11} \unit{M}_{\sun}$.
    }
    \end{figure}
}

\newcommand{\cmdrichfract}{
    \begin{figure}
    \begin{center}
    \leavevmode
    \includegraphics[scale=0.45]{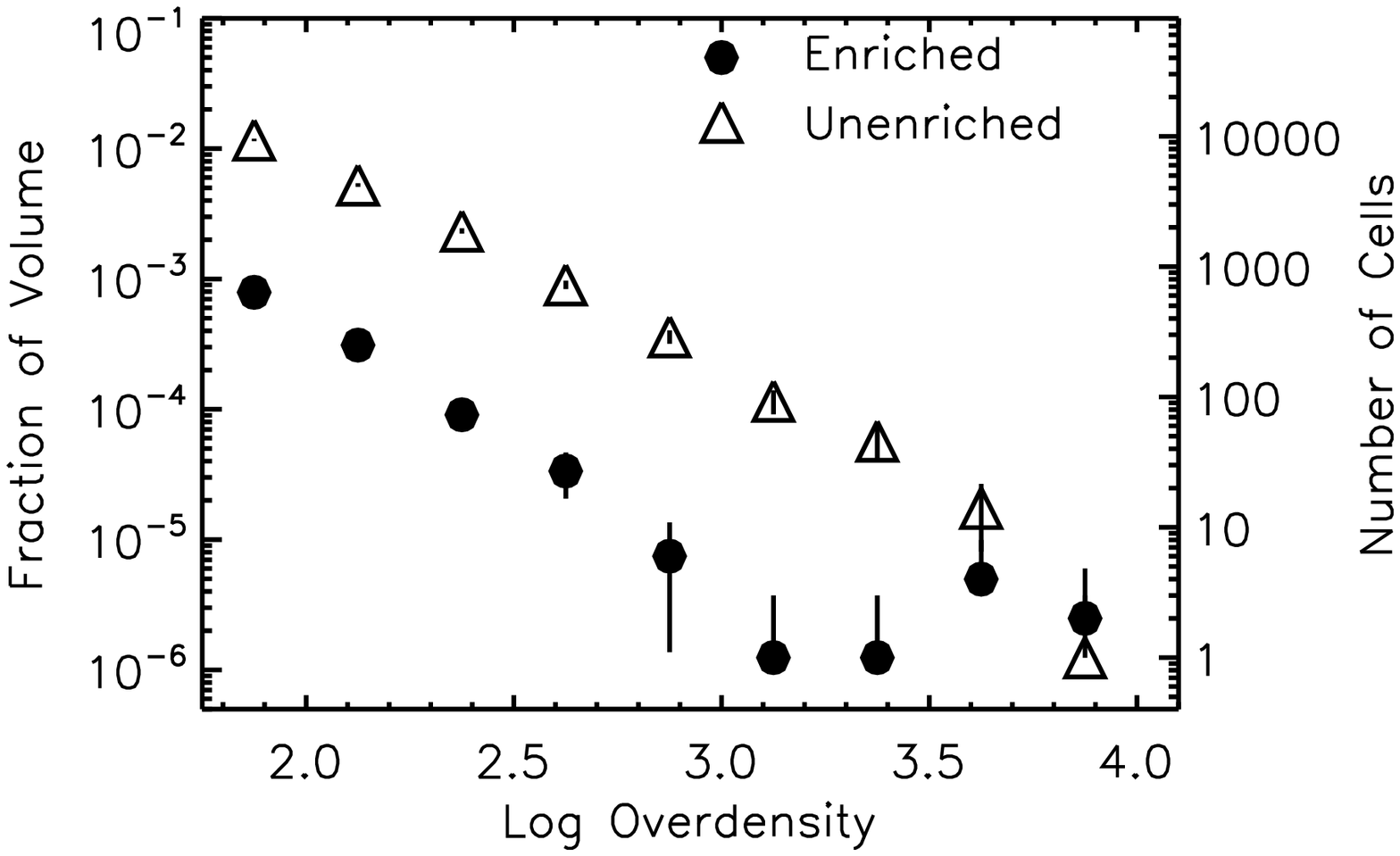}
    \end{center}
    \caption[1]{
    \label{rich_fract}
    Fraction of the volume of the simulation box
    occupied by gas that is enriched or
    unenriched in baryons, as a function of overdensity in
    $6.5 \unit{kpc}$ grid cells having an overdensity of 56 or higher.
    Threshold for baryon enrichment is twice the cosmic mean. 
    Vertical lines show two sigma errors.
    Symbols are placed at the centers of the overdensity bins.
    }
    \end{figure}
}

\newcommand{\cmdfiga}{
    \begin{figure}
    \begin{center}
    \leavevmode
    \includegraphics[scale=0.44]{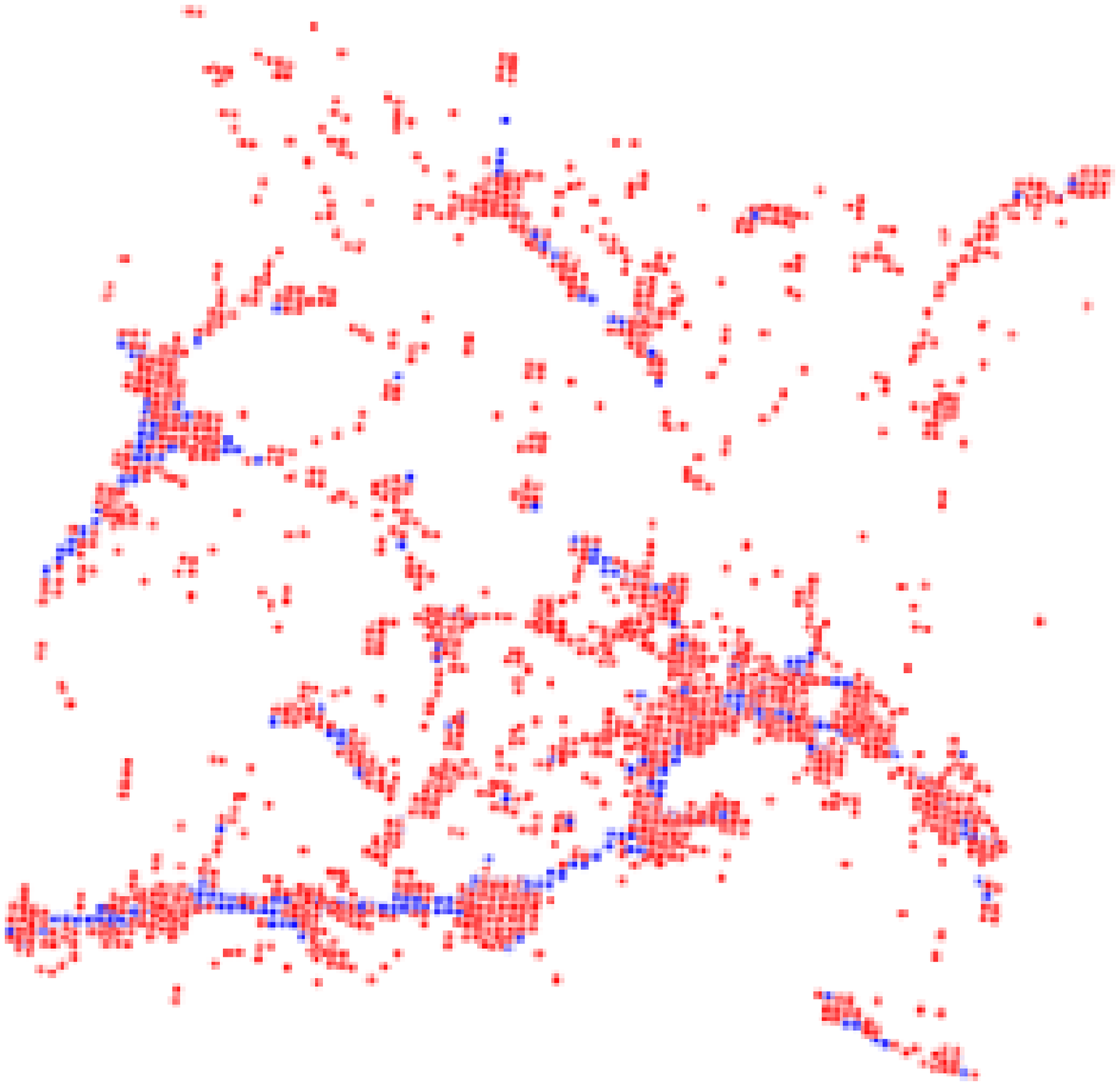}
    \includegraphics[scale=0.44]{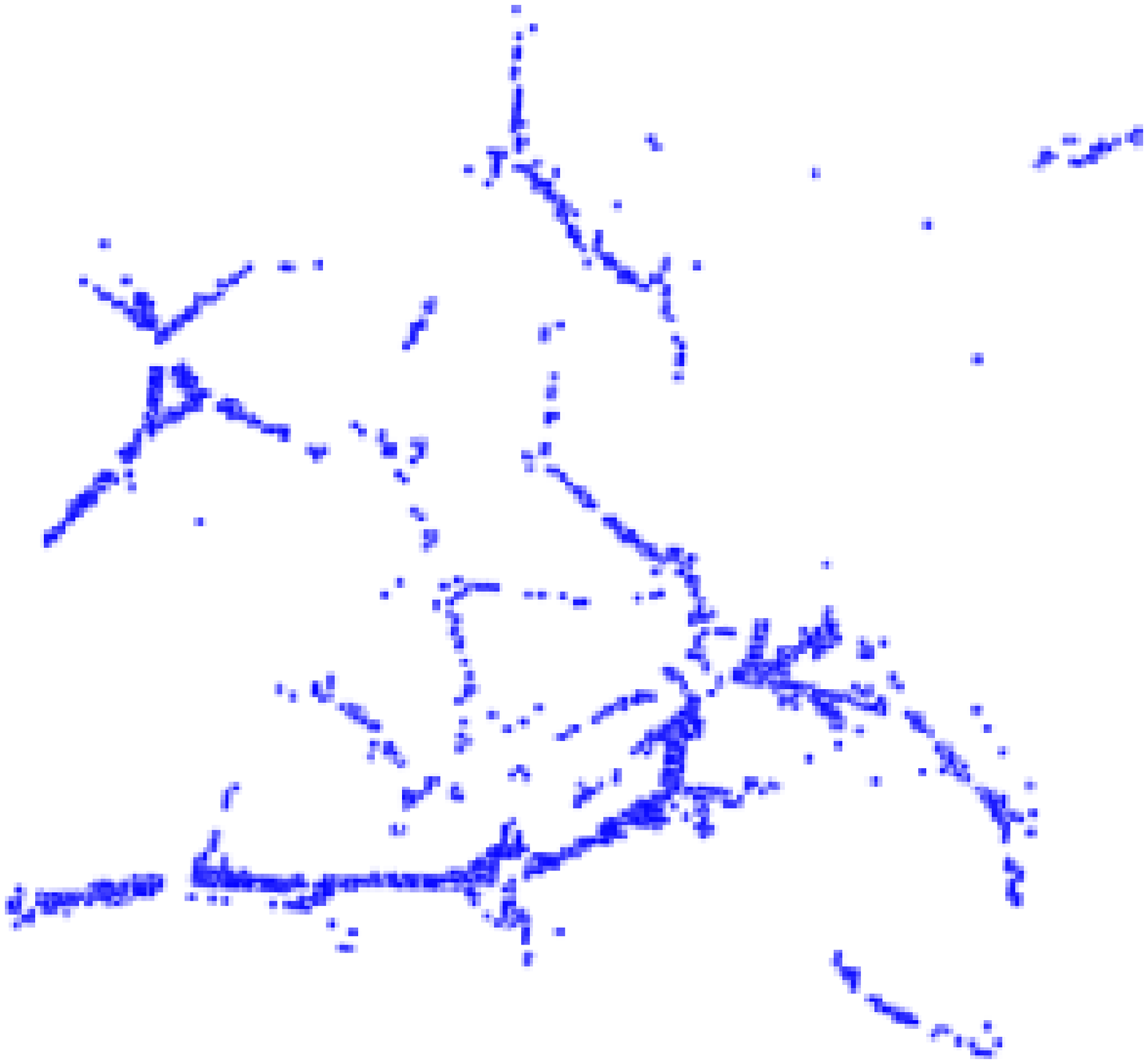}
    \end{center}
    \caption[1]{
    \label{fig_1_2_2}
Zoom-in of the longest filament of Figure~\ref{simbox}, again showing
only $6.5 \unit{kpc}$ grid cells with an overdensity of 56 or higher.
Upper: Cells enriched in baryons to at least twice the
cosmic mean are shown in dark blue; those not so enriched are in 
lighter red.
Lower:  Same, but only the dark blue enriched cells are shown for
clarity.  The width of the images is about $600 \unit{kpc}$.
    }
    \end{figure}
}

\newcommand{\cmdsuballden}{
    \begin{figure}
    \begin{center}
    \leavevmode
    \includegraphics[scale=0.35]{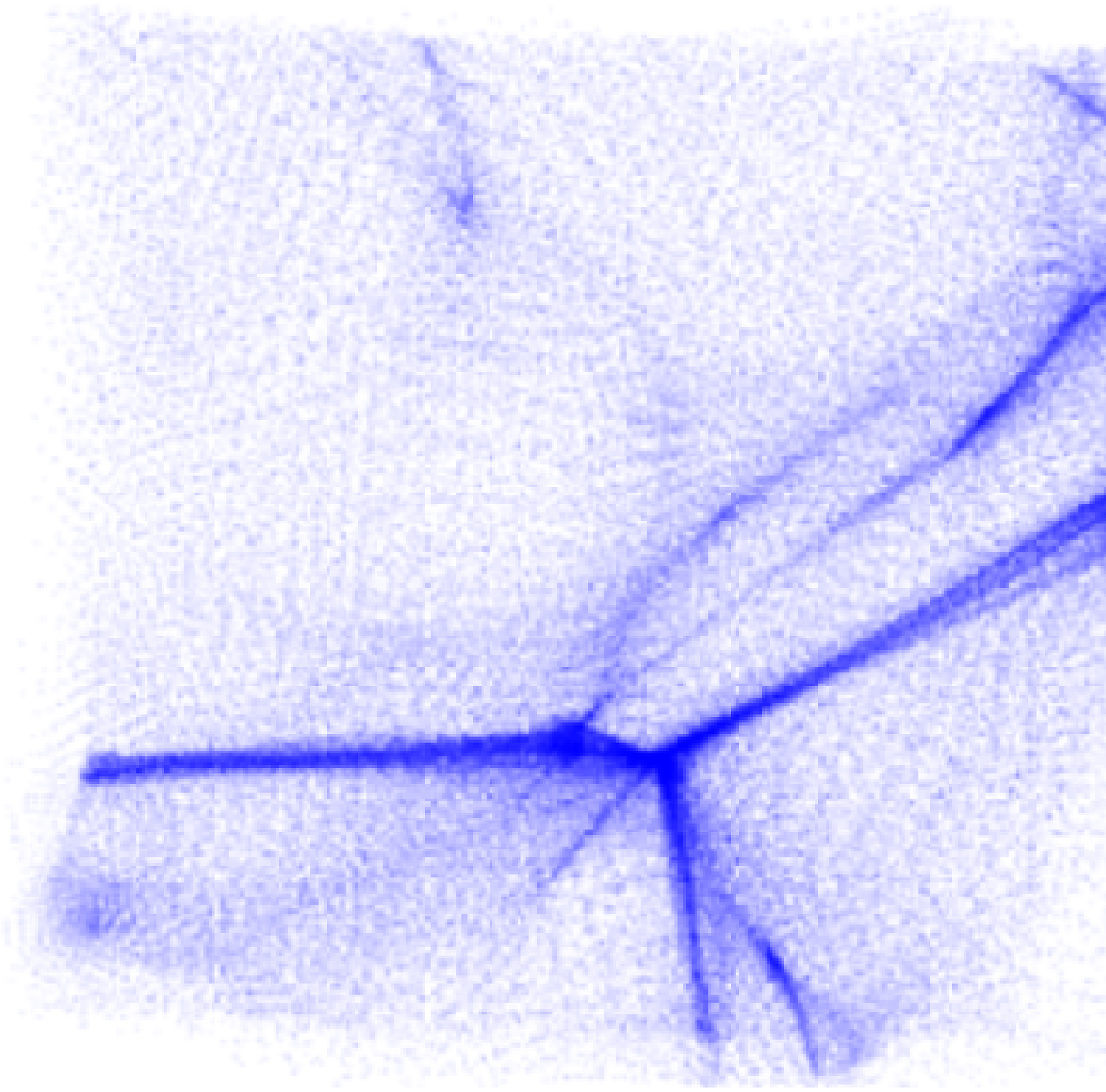}
    \includegraphics[scale=0.35]{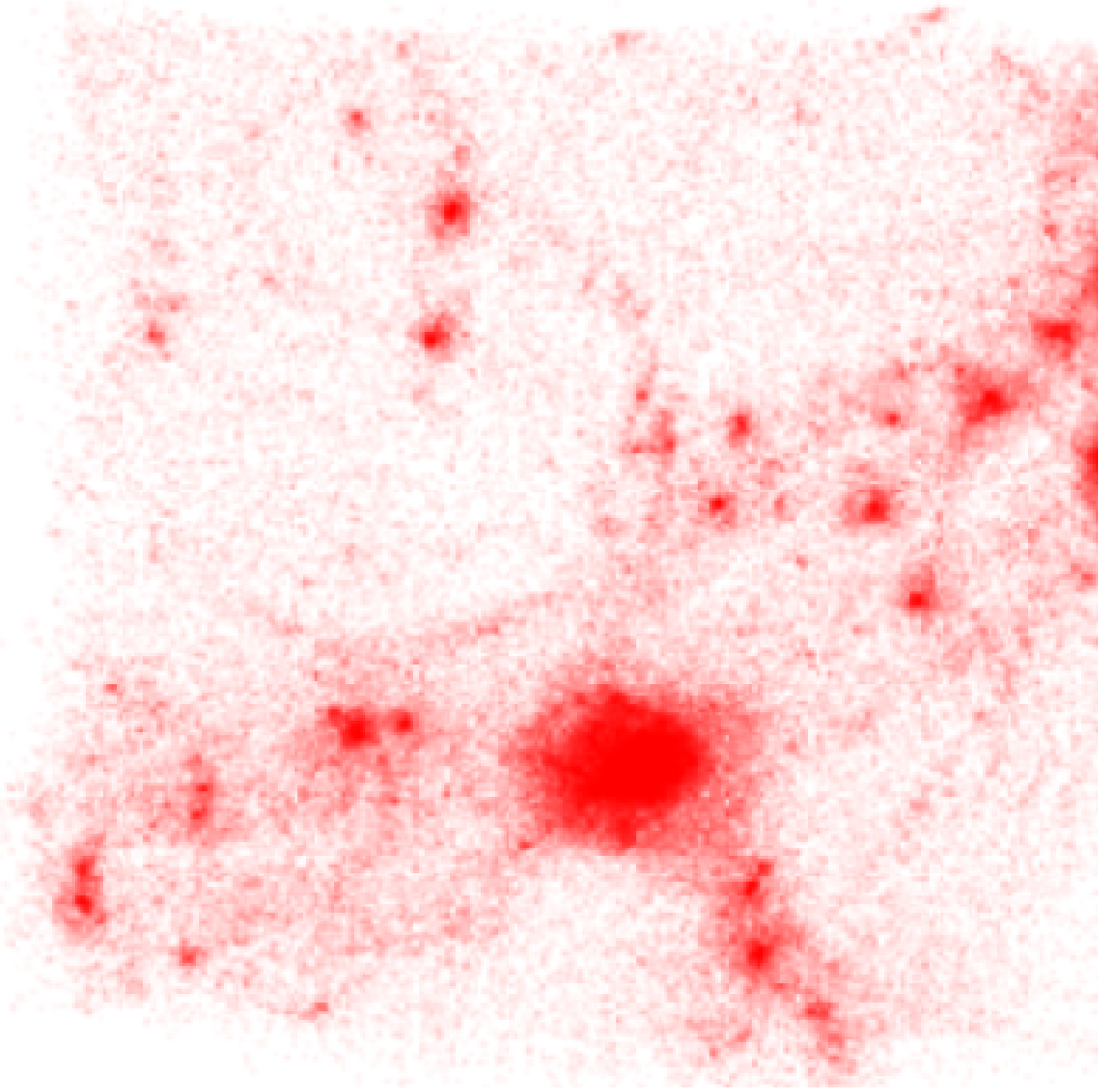}
    \end{center}
    \caption[1]{
    \label{sub_all_den}
    Upper: Gas simulation particles in a small region of the 
    filament shown in Figure~\ref{fig_1_2_2}.  Lower:  Same view
    showing just dark matter particles.  The width of the images is
    about $200 \unit{kpc}$.
    }
    \end{figure}
}

\newcommand{\cmdshell}{
    \begin{figure}
    \begin{center}
    \leavevmode
    \includegraphics[scale=0.24]{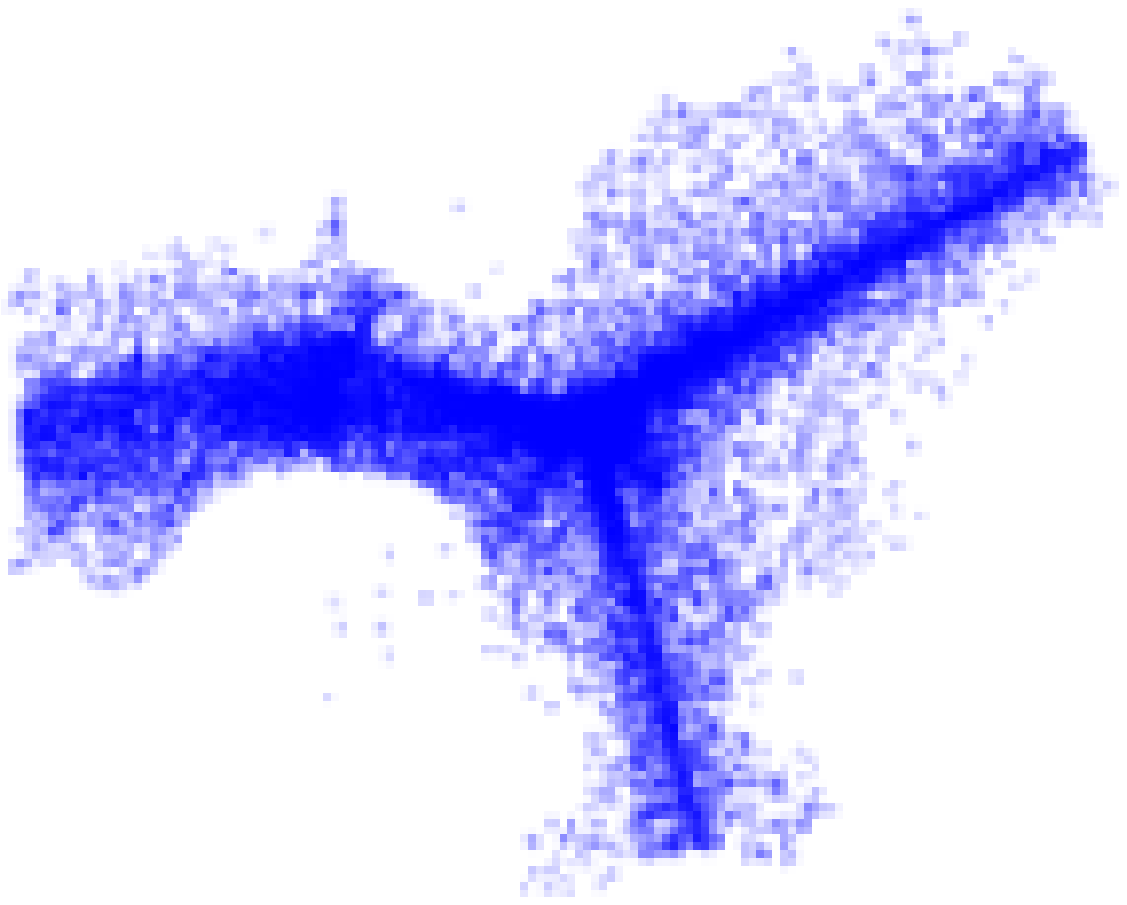}
    \includegraphics[scale=0.24]{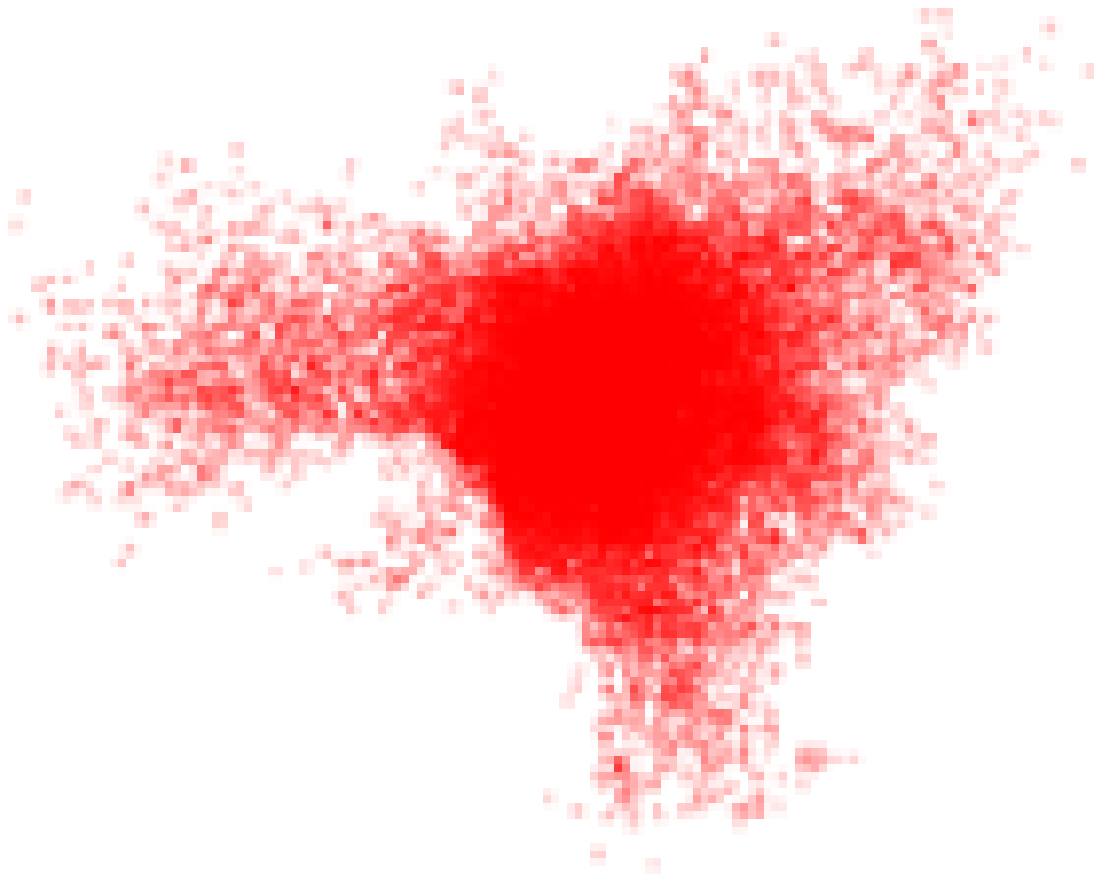}
    \includegraphics[scale=0.24]{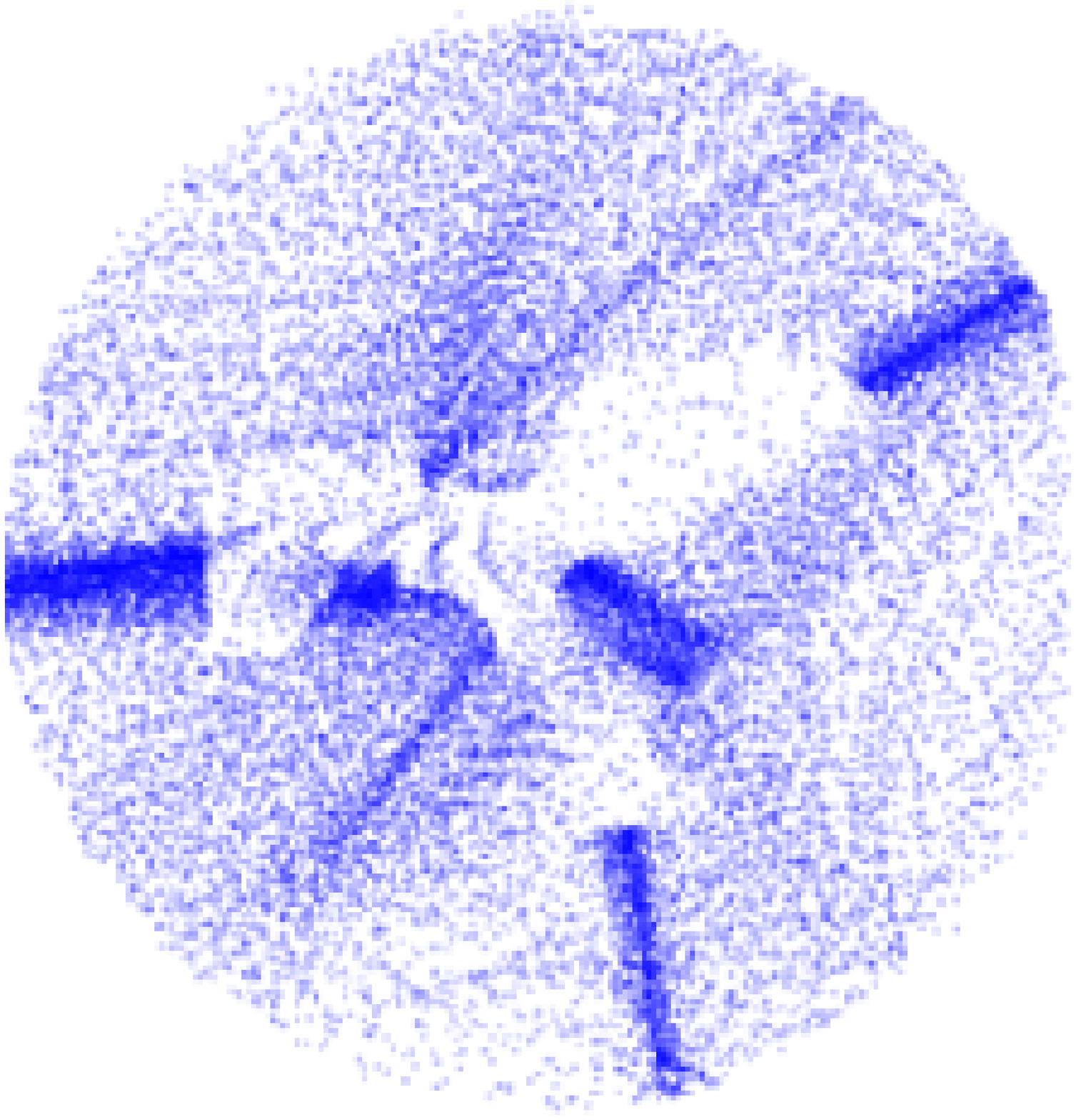}
    \includegraphics[scale=0.24]{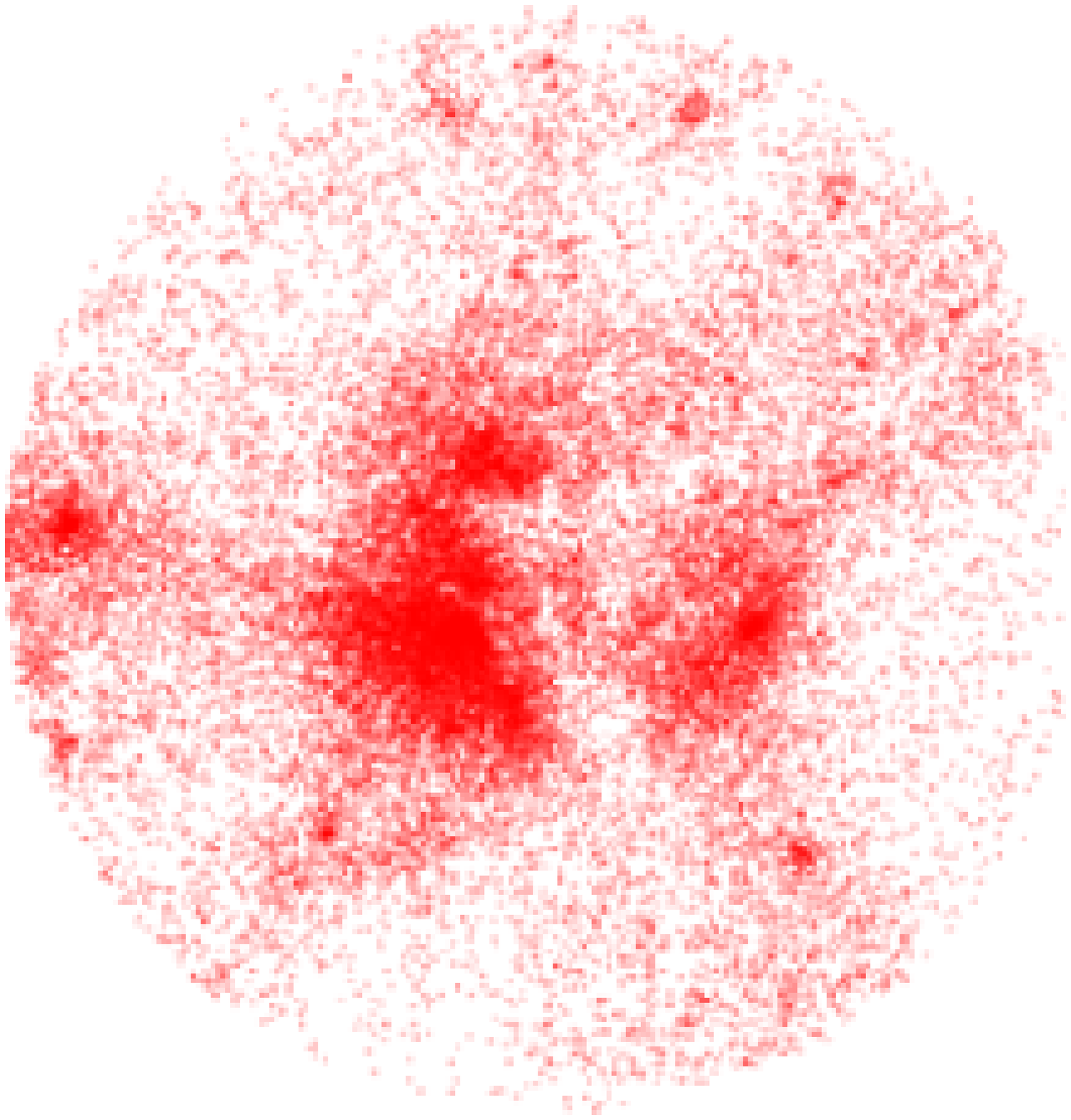}
    \end{center}
    \caption[1]{
    \label{shell}
    Bound and unbound gas (blue) and dark matter (red) particles 
    within a radius of $55.4 \unit{kpc}$ (proper) of the
    second largest galaxy at redshift of 5.1. Upper left: gas bound 
    to galaxy, upper right: dark matter bound to galaxy, lower left:
    gas not bound to galaxy, lower right: dark matter not bound to
    galaxy.
    }
    \end{figure}
}

\newcommand{\cmdtriadtilt}{
    \begin{figure}
    \begin{center}
    \leavevmode
    \includegraphics[scale=0.198]{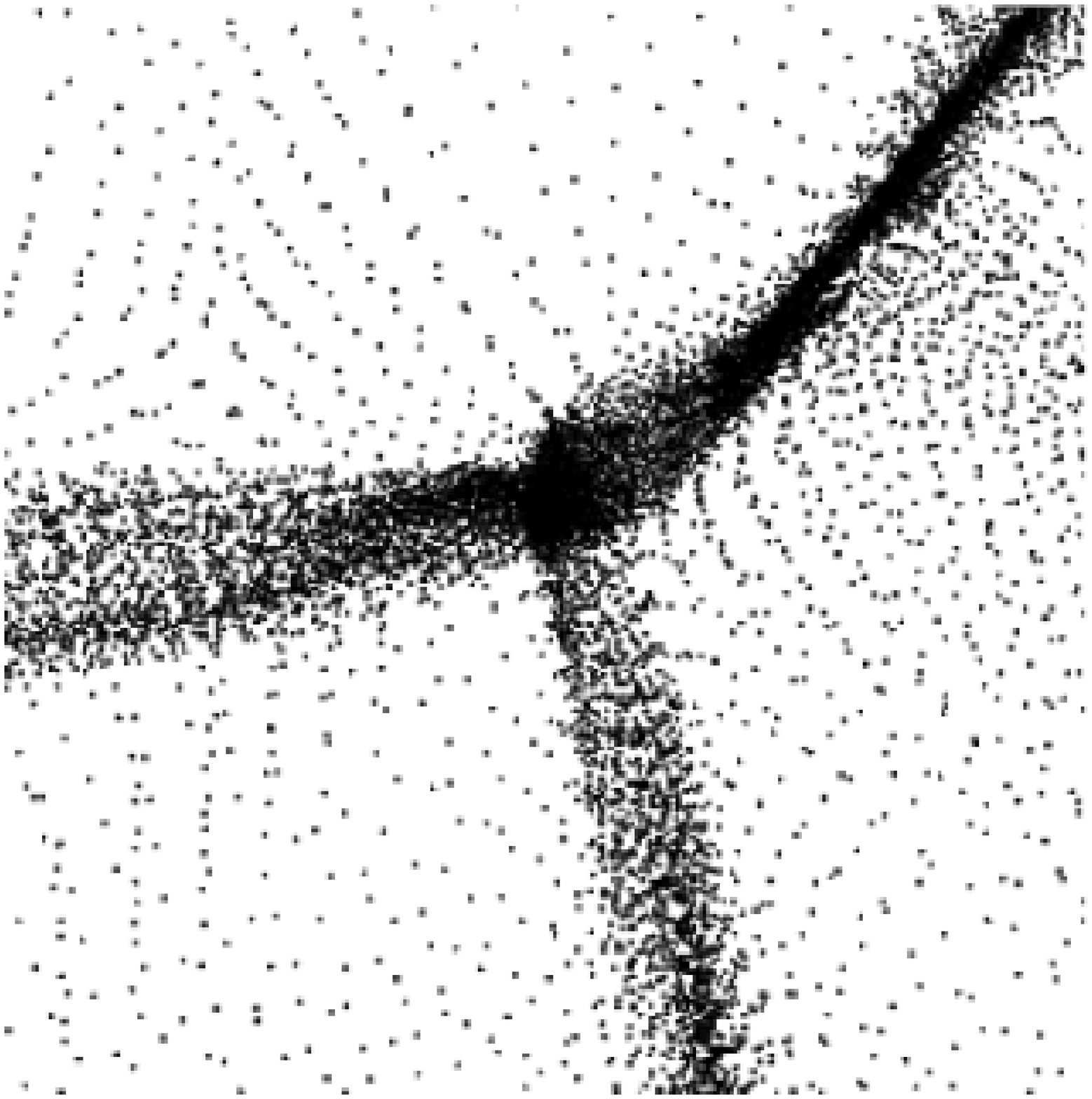}
    \includegraphics[scale=0.198]{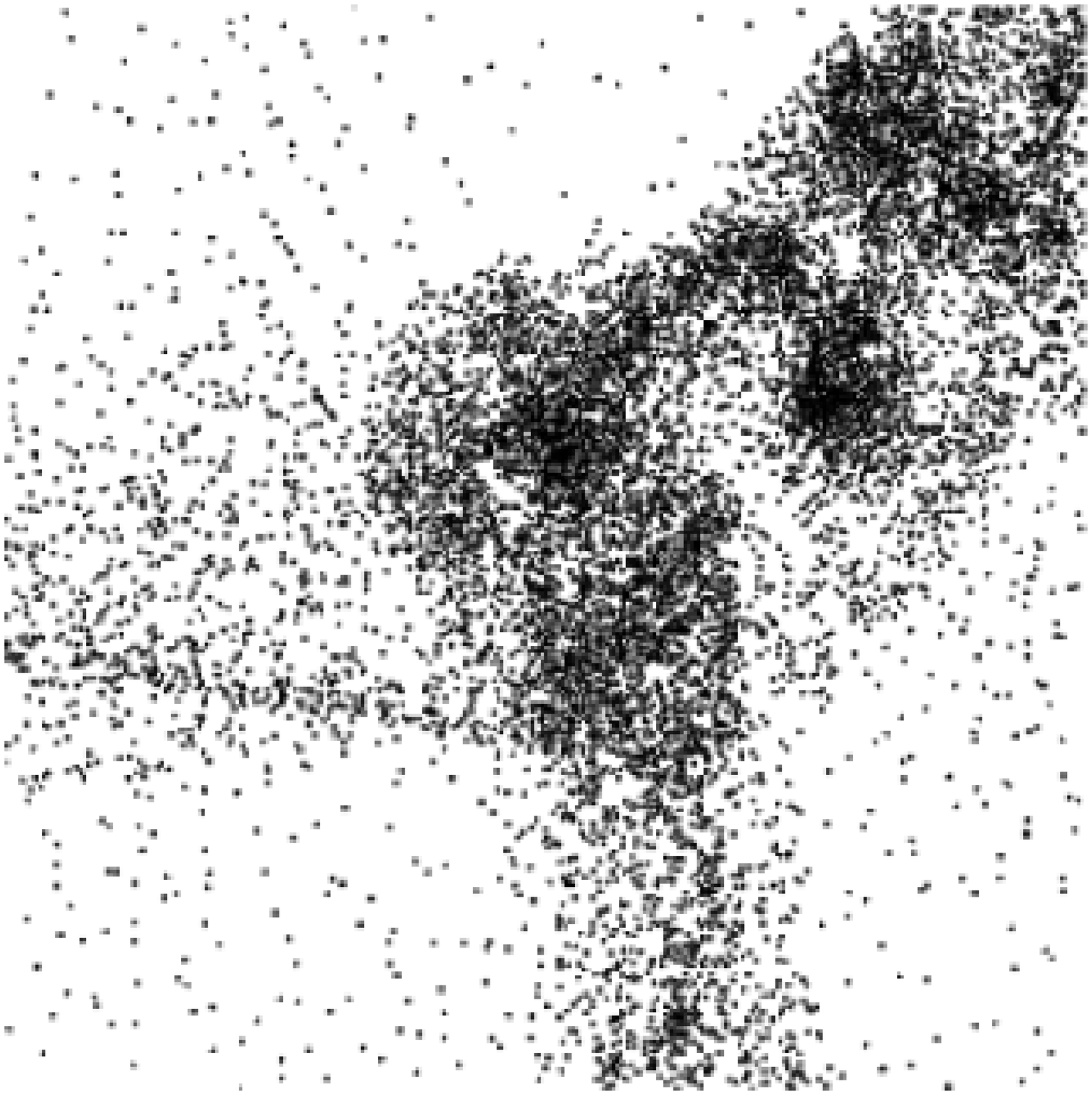}
    \includegraphics[scale=0.198]{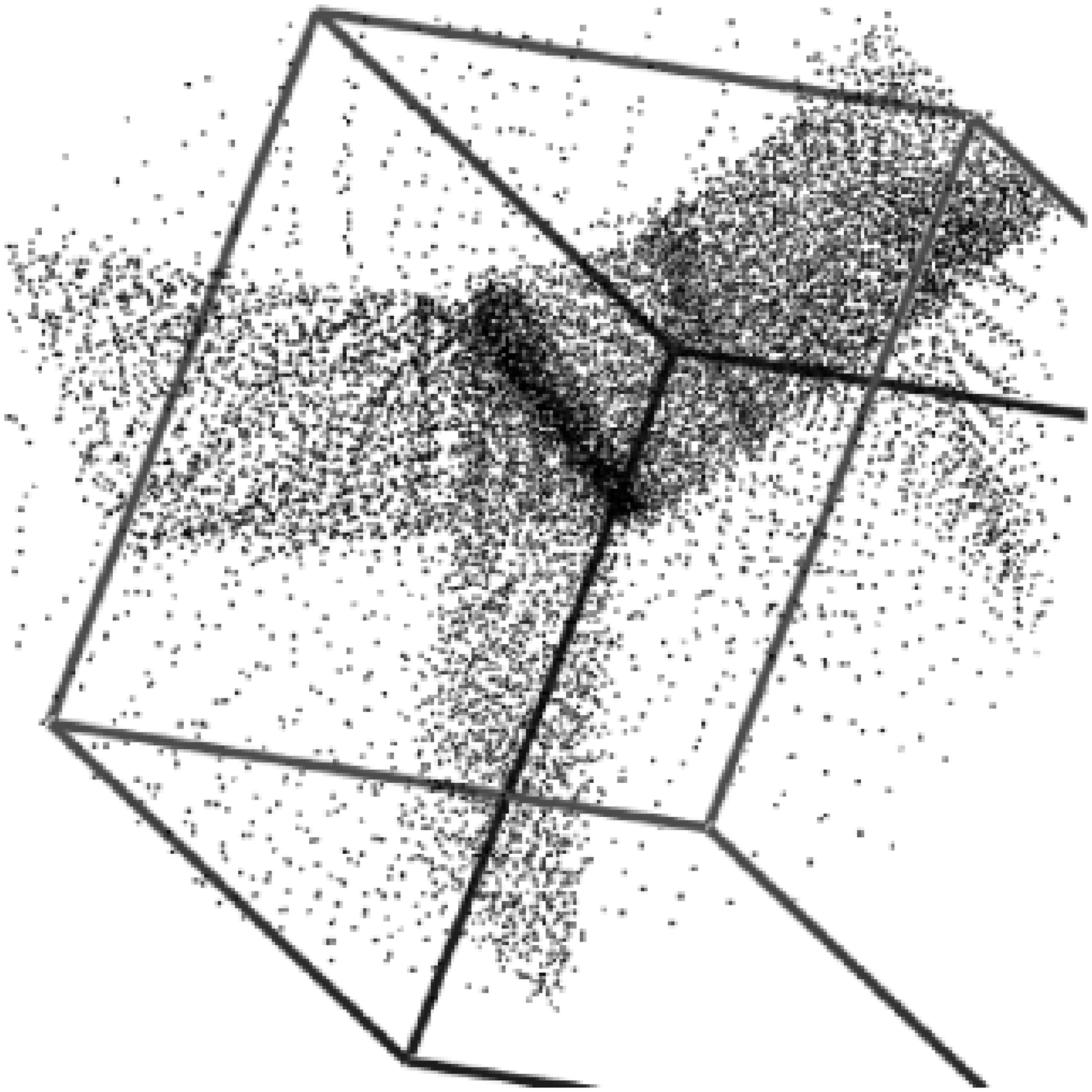}
    \includegraphics[scale=0.198]{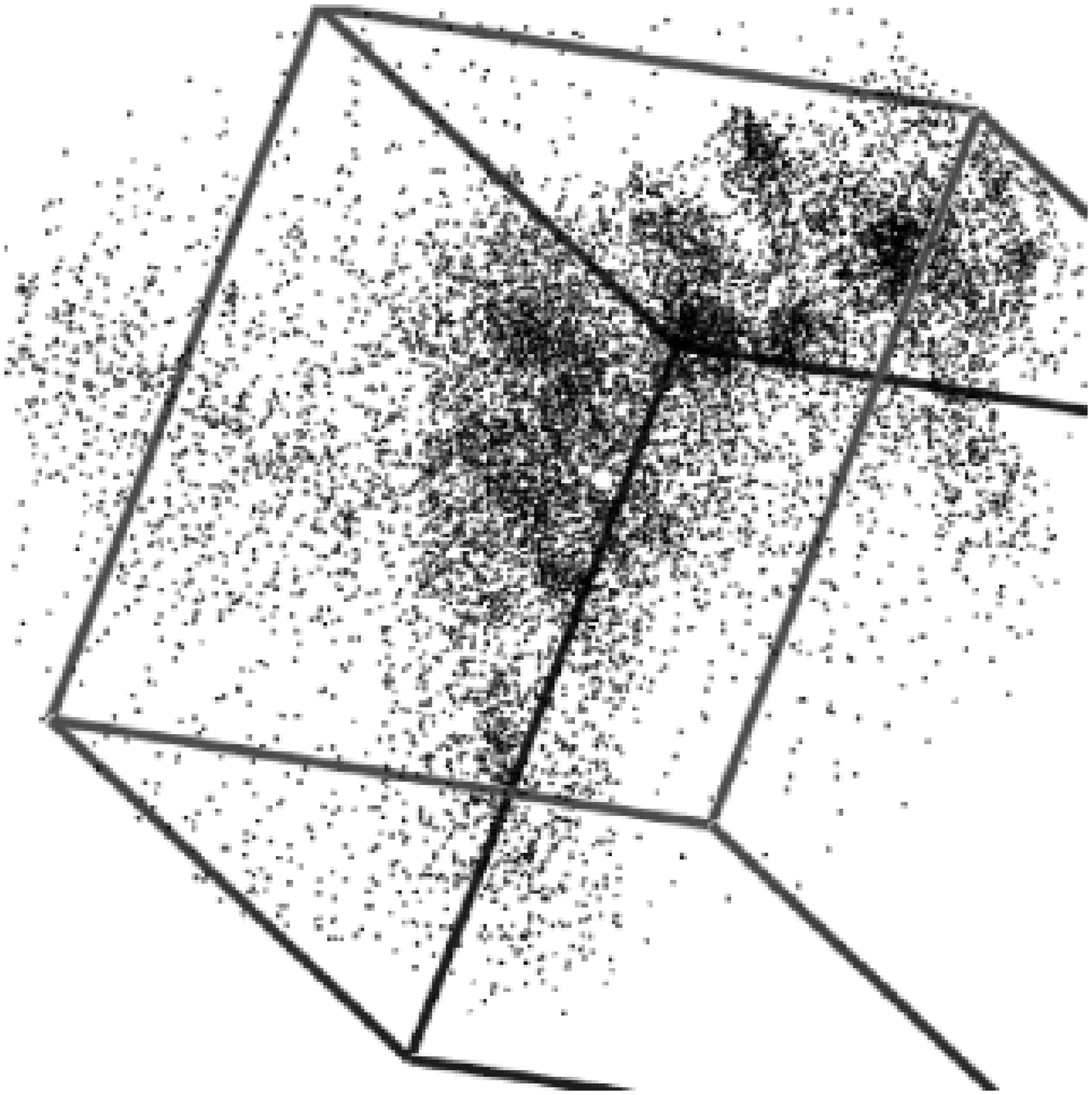}
    \end{center}
    \caption[1]{
    \label{triad_tilt}
    Sheets of gas radiating from a filament.  Shown is the
    distribution of gas and dark matter
    around the axis of a filament from the galaxy shown in 
    Figure~\ref{shell}.  Upper panel shows gas particles (left) and
    dark matter particles (right) in a projection on to a plane perpendicular
    to the filament.  The axis of the filament is at the center going 
    into the page.
    In the lower panel these structures are shown at a different angle
    to show the three sheets formed by the gas particles (left) and
    the more irregular distribution of the dark matter particles (right).
    The dimension of each square image, and of the box, is $110.8 \unit{kpc}$ proper.
    }
    \end{figure}
}

\newcommand{\cmdhpixcontour}{
    \begin{figure}
    \begin{center}
    \leavevmode
    \includegraphics[width=3.2in]{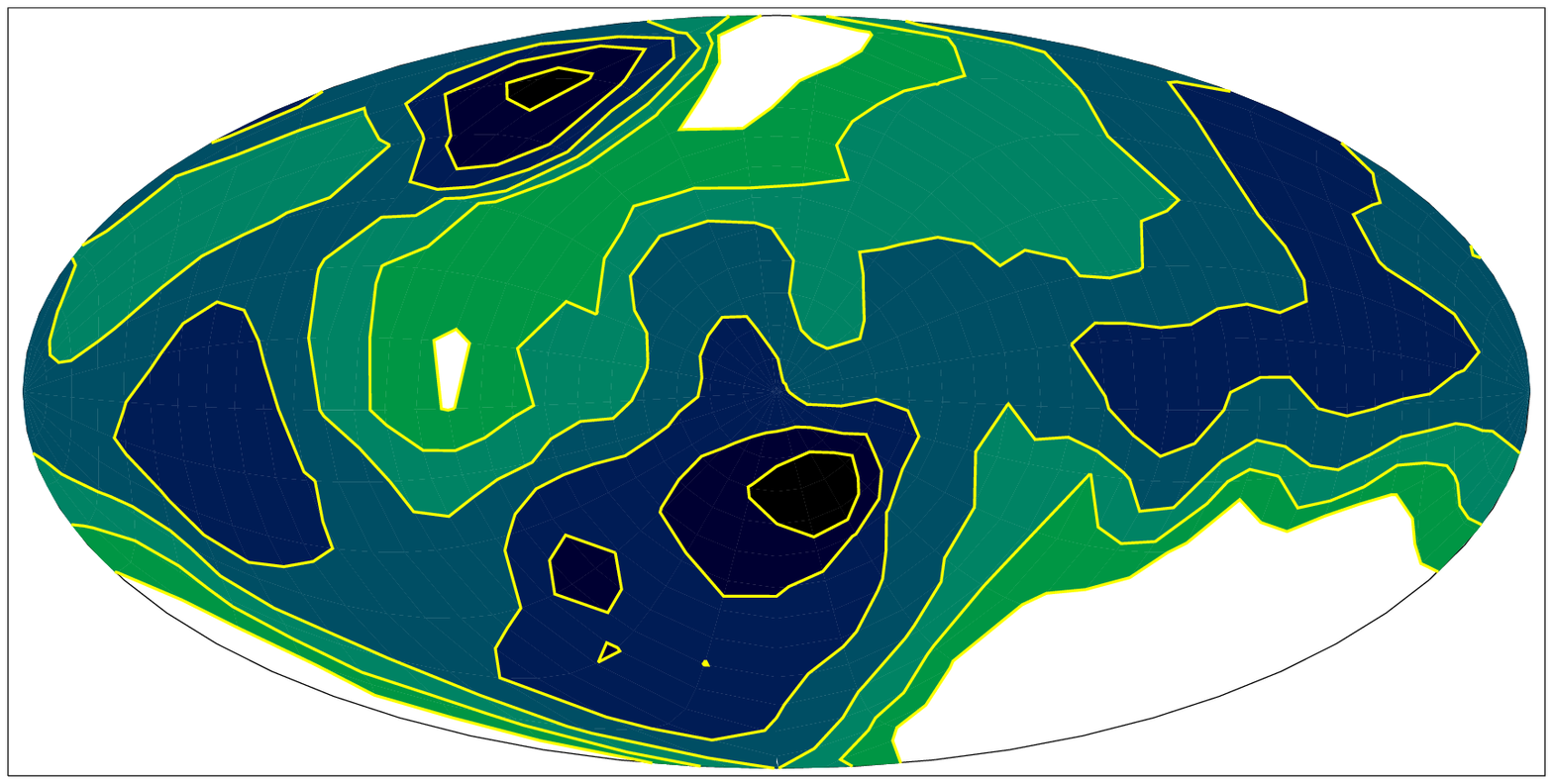}
    \includegraphics[width=3.2in]{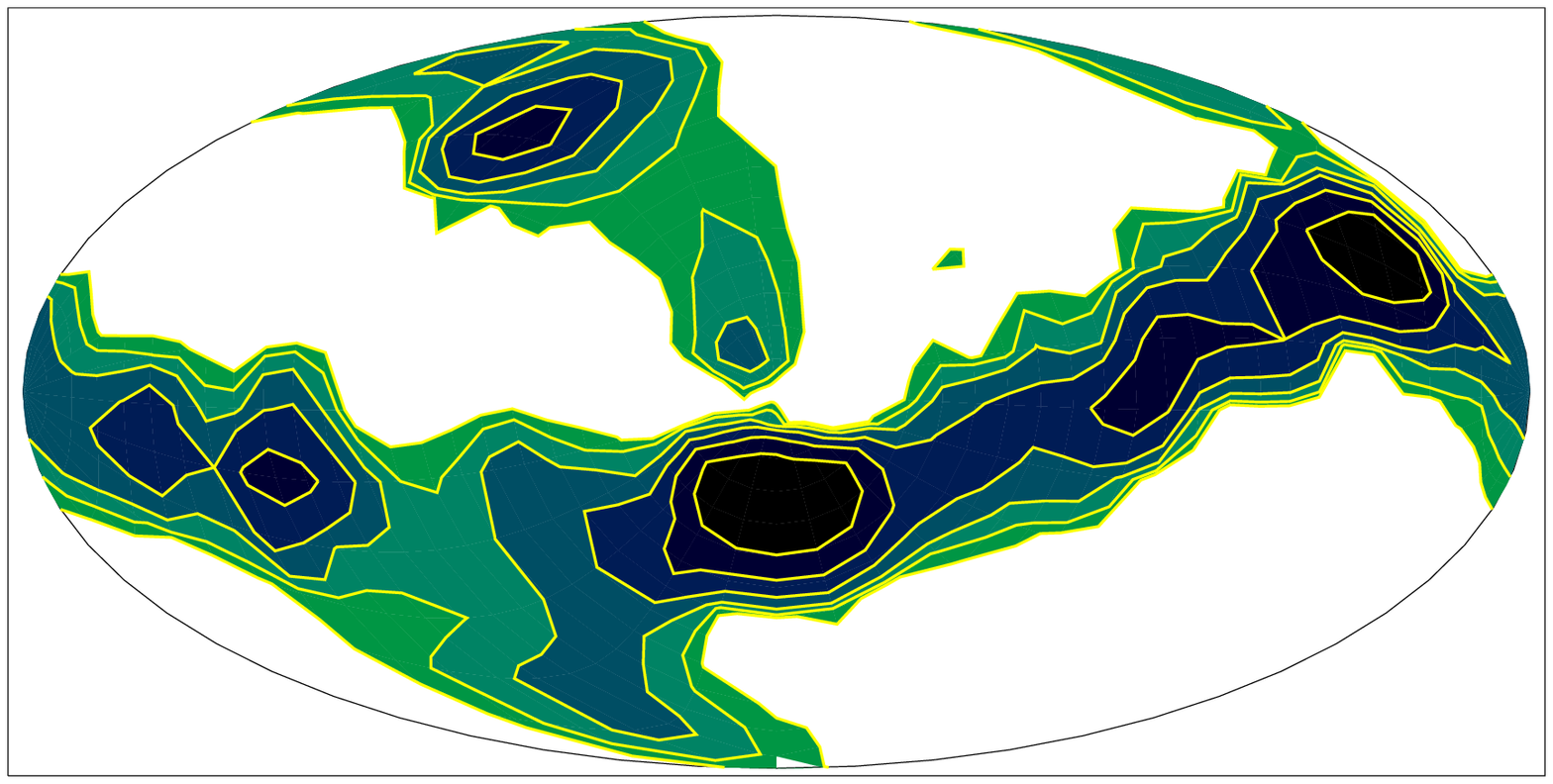}
    \includegraphics[width=3.2in]{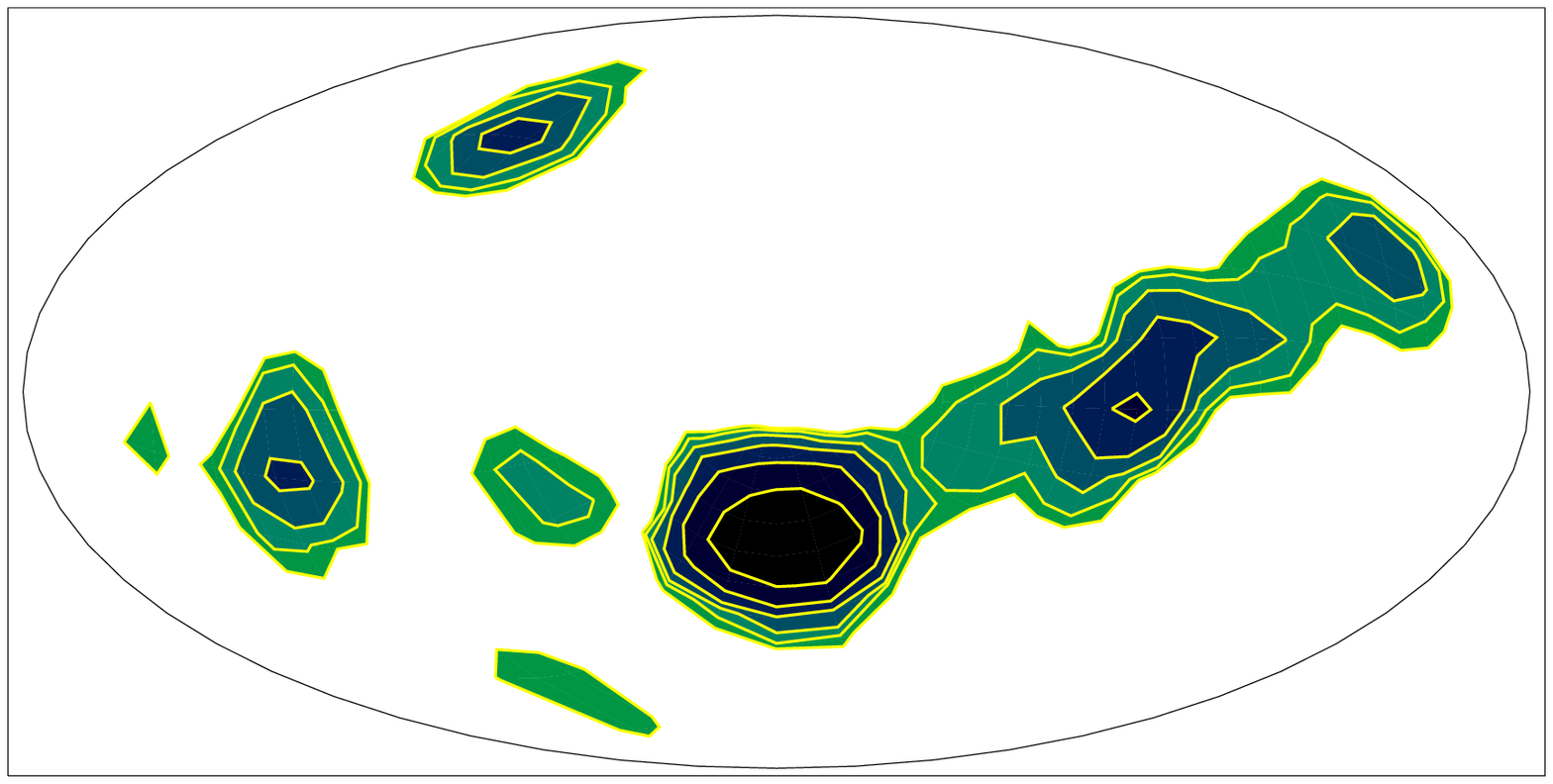}
    \includegraphics[width=3.2in]{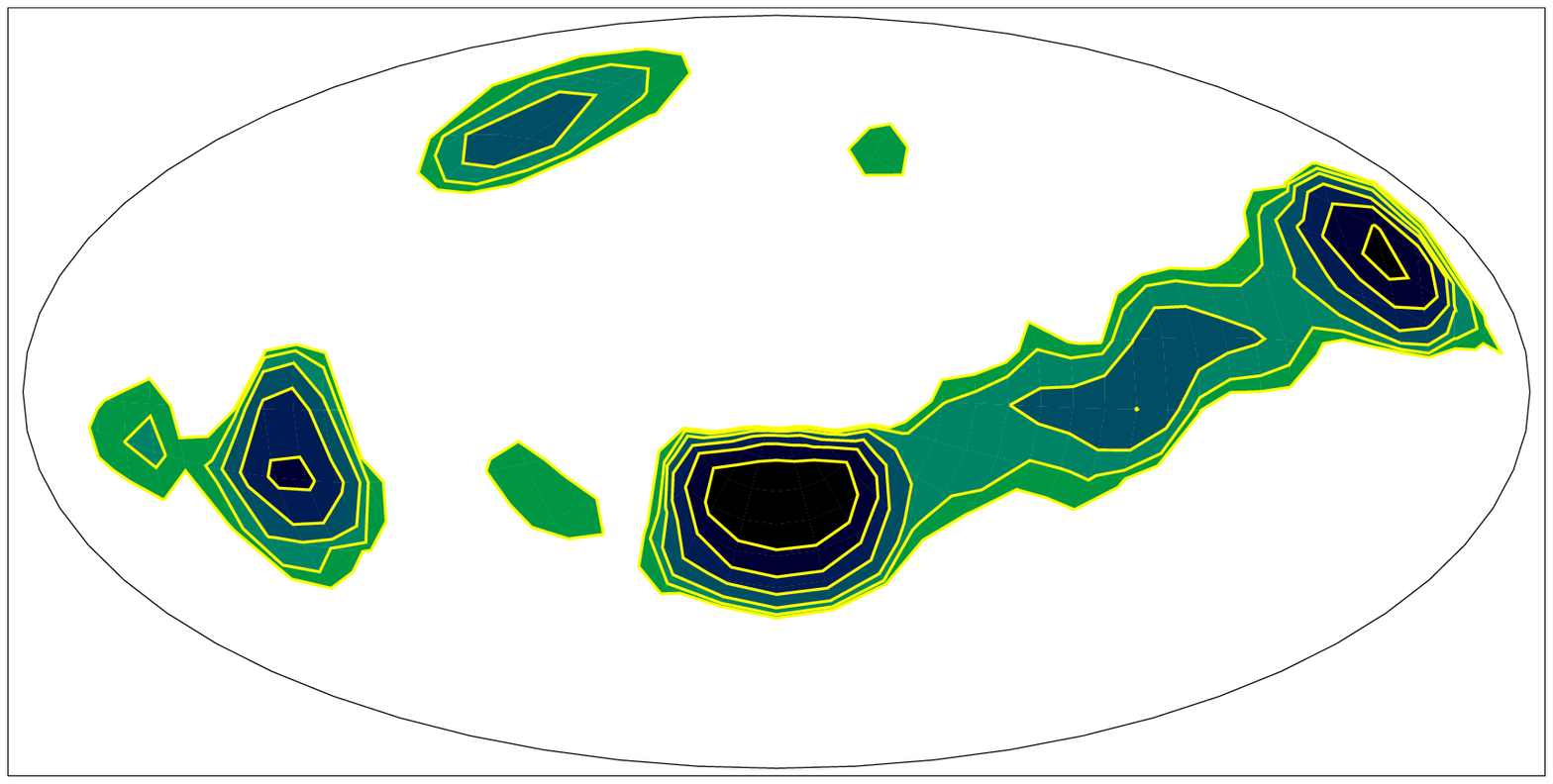}
    \includegraphics[width=3.2in]{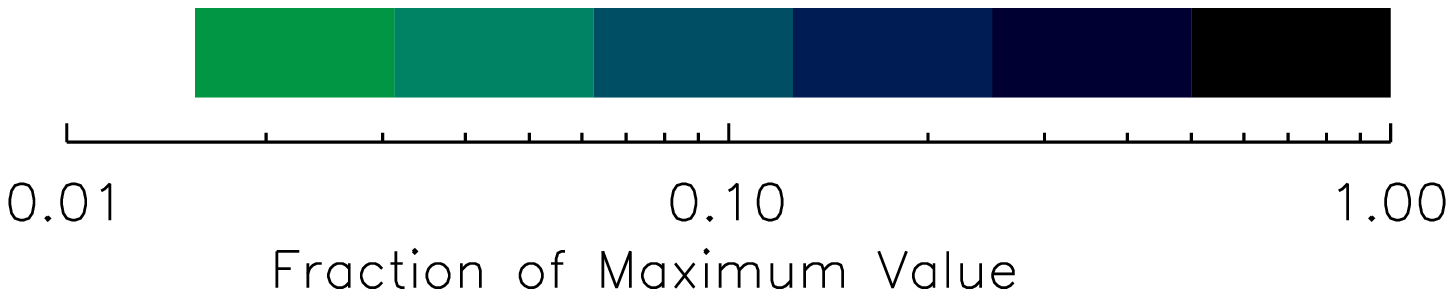}
    \end{center}
    \caption[1]{
    \label{hpix_contour}
Angular distribution of matter about the second largest galaxy, in
Hammer-Aitoff projection.  In descending order the maps show
dark matter, gas, neutral hydrogen, and stars.
Shown are contours of masses in the solid angles subtended by 192
HealPix pixels for a shell from 22.2 to $55.4 \unit{kpc}$.
For each map there are six
contour levels with a factor of two spacing.  The highest contour
for each map is one half the maximum of that map. 
    }
    \end{figure}
}

\newcommand{\cmdcyldistrib}{
    \begin{figure}
    \begin{center}
    \leavevmode
    \includegraphics[scale=.3]{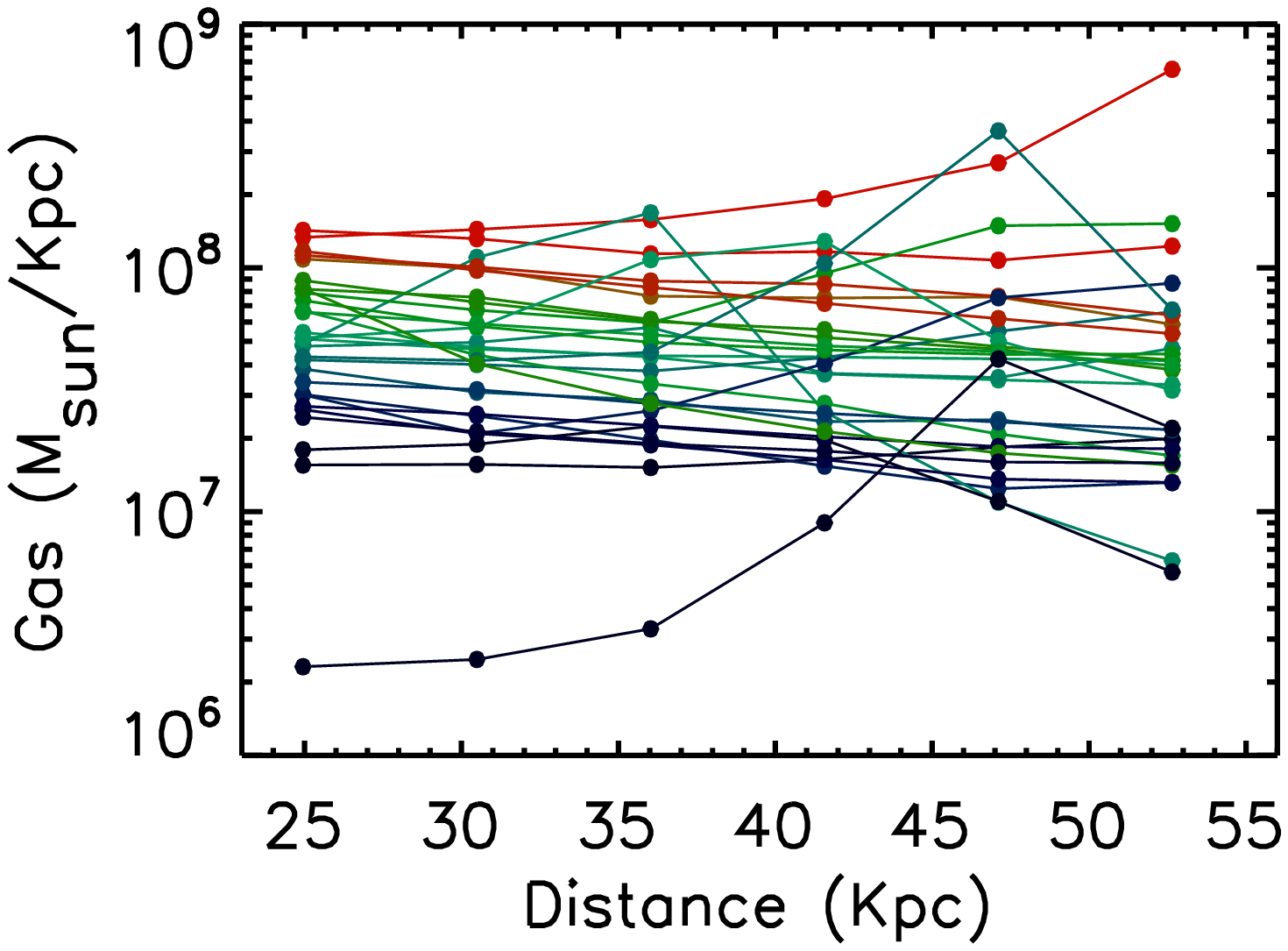}
    \includegraphics[scale=.3]{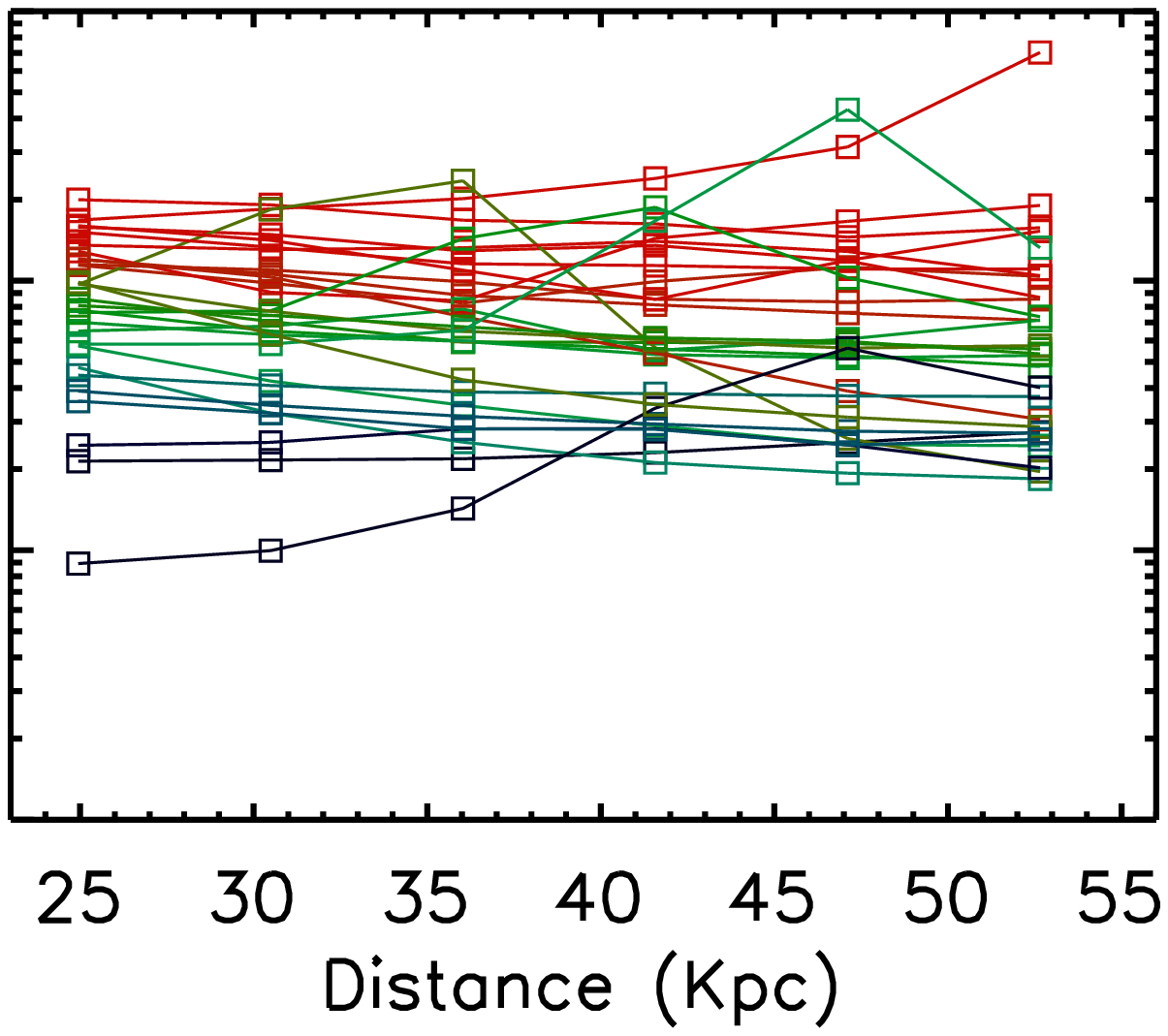}
    \end{center}
    \caption[1]{
    \label{cyl_distrib}
Linear density of gas along each of the 29 filaments around the
10 largest galaxies.  Each curve
represents one filament, and each point represents the gas contained
in a segment of a coaxial cylinder around that filament of radius
$8 \unit{kpc}$ (left) or $16 \unit{kpc}$ (right).  
See Figure~\ref{cartoon} for a drawing
of the geometry.  The abscissa is distance from the
galaxy center along the filament. The points are plotted at the centers
of the segments.
    }
    \end{figure}
}

\newcommand{\cmdfilprofile} {
    \begin{figure}
    \begin{center}
    \leavevmode
    \includegraphics[scale=0.28]{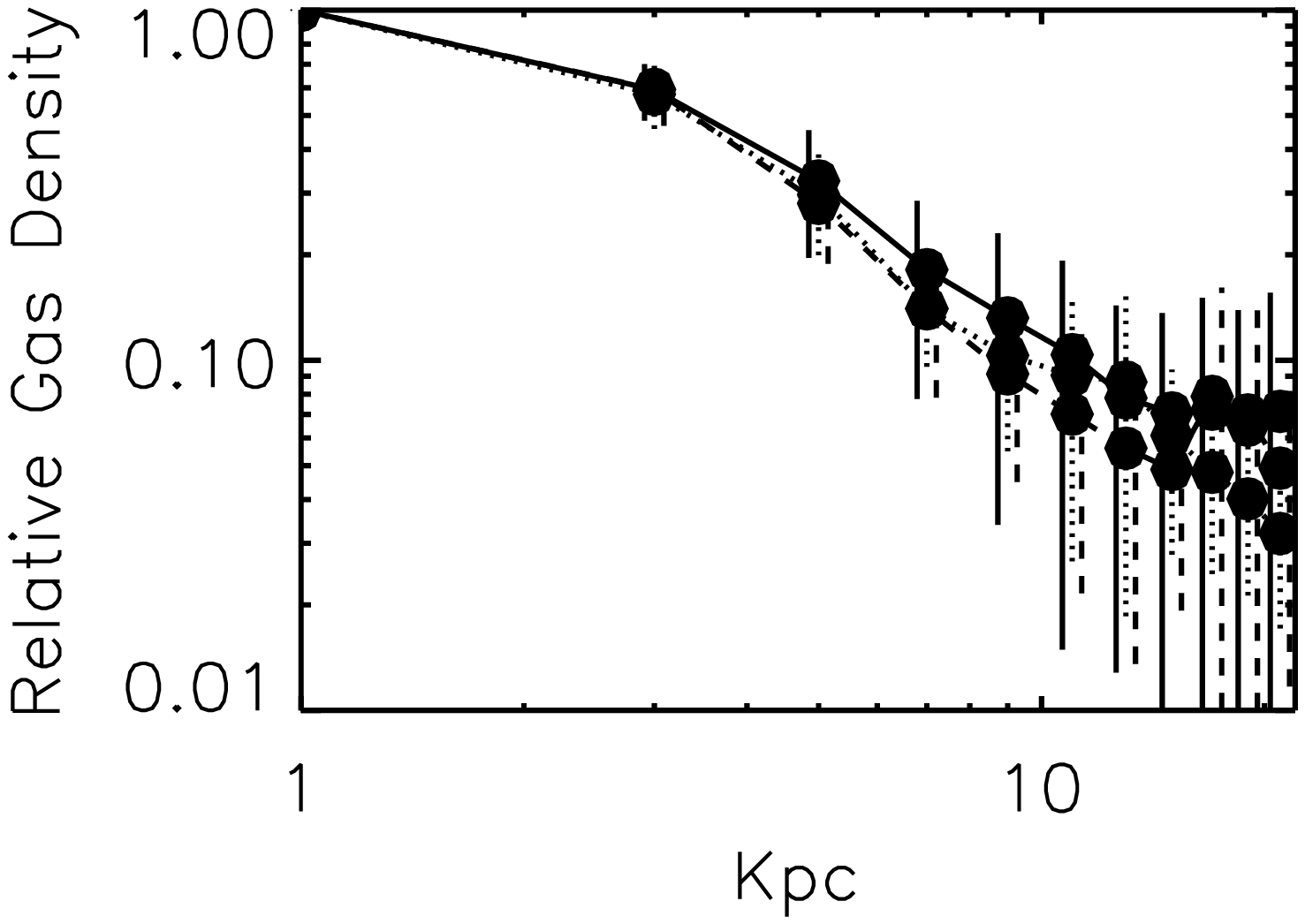}
    \includegraphics[scale=0.28]{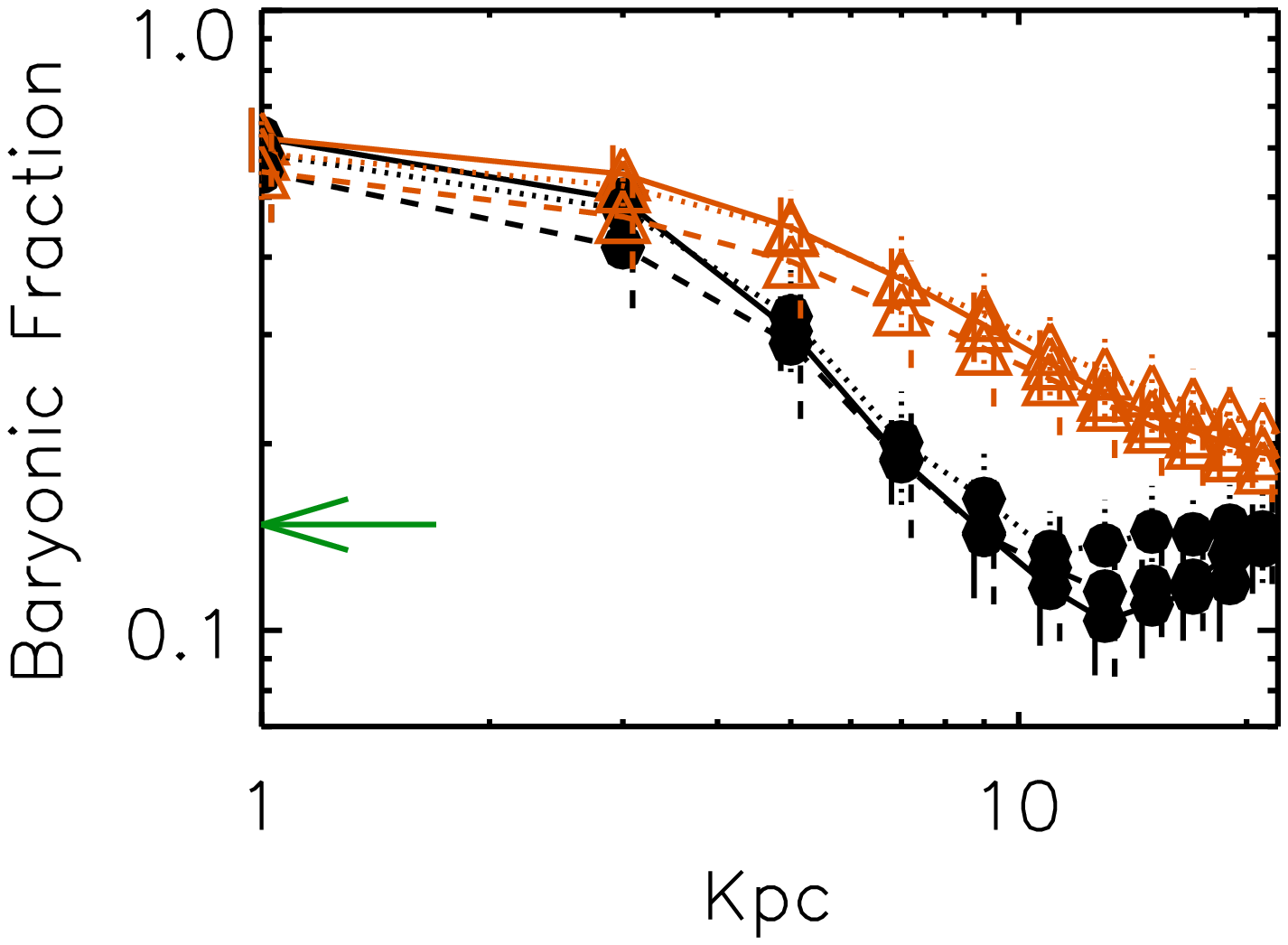}
    \end{center}
    \caption[1]{
    \label{fil_profile}
Average cross-sectional profiles of filaments
at each of three distances
from the galaxy.
The left and right panels show respectively the gas density
and baryonic fraction as a function of distance from the center
of the filament.  In the right panel the black curves with the black
filled circles show the baryonic fraction in the individual radial
bins, while the orange curves with the triangles show cumulative
baryonic fractions for regions extending outward and including the bin.
The green arrow in the right graph shows the cosmic mean baryon fraction.
For these graphs, each filament from $22.2$ to $55.4 \unit{kpc}$ from the
center of the galaxy
was divided into 3 equal segments.
Solid, dotted, and dashed lines show profiles
for the inner, middle, and outer segments
averaged over all 29 filaments.
Gas densities and baryonic fractions were computed on cylindrical
shells having a radial width of 2 kpc.
In the left panel the data for each filament segment 
were normalized to the gas density
in the lowest radial bin for that particular filament.
Vertical lines show two sigma errors.
Symbols are plotted at the centers of the radial distance bins.
    }
    \end{figure}
}

\newcommand{\cmdfilsheet}{
    \begin{figure}
    \begin{center}
    \leavevmode
    \includegraphics[scale=0.4]{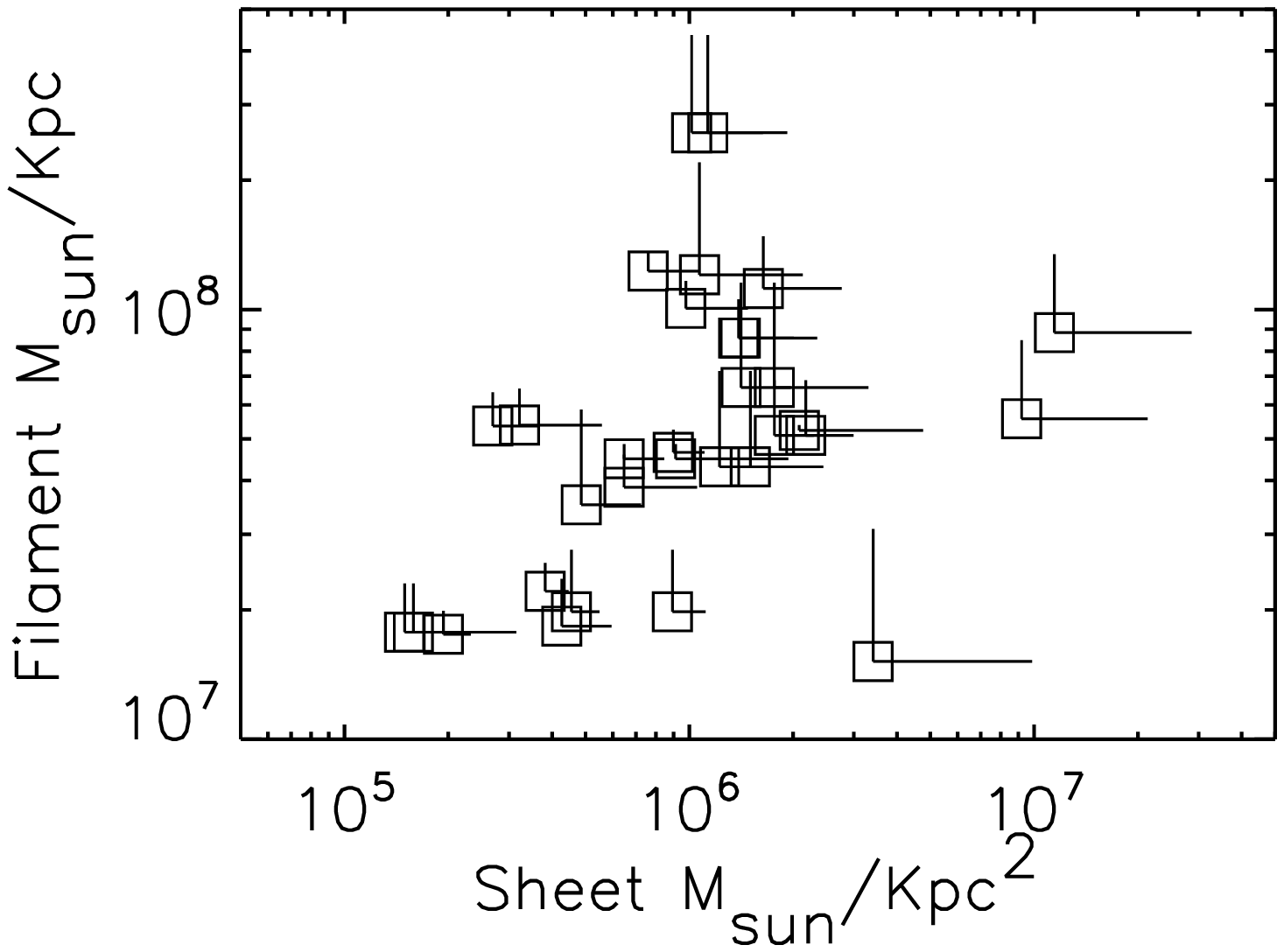}
    \end{center}
    \caption[1]{
    \label{fil_sheet}
Surface density of gas in sheets compared to the
linear density of the parent filament. Horizontal and vertical
lines show the standard deviations of the densities obtained from
averaging nine equal rectangles of the sheet.
For clarity in a log-log plot only the error bars on the positive sides
of each data point are drawn. Each rectangle measures $11.1 \unit{kpc}$ in
the direction of the filament and $15.8 \unit{kpc}$ along the sheet in a
direction perpendicular to the filament.  The innermost boundary
of the region analyzed is a cylinder of radius $8 \unit{kpc}$ 
coaxial with  the filament.  This limit is used to exclude
material from the filament itself.
    }
    \end{figure}
}

\newcommand{\cmdslabbf}{
    \begin{figure}
    \begin{center}
    \leavevmode
    \includegraphics[scale=0.4]{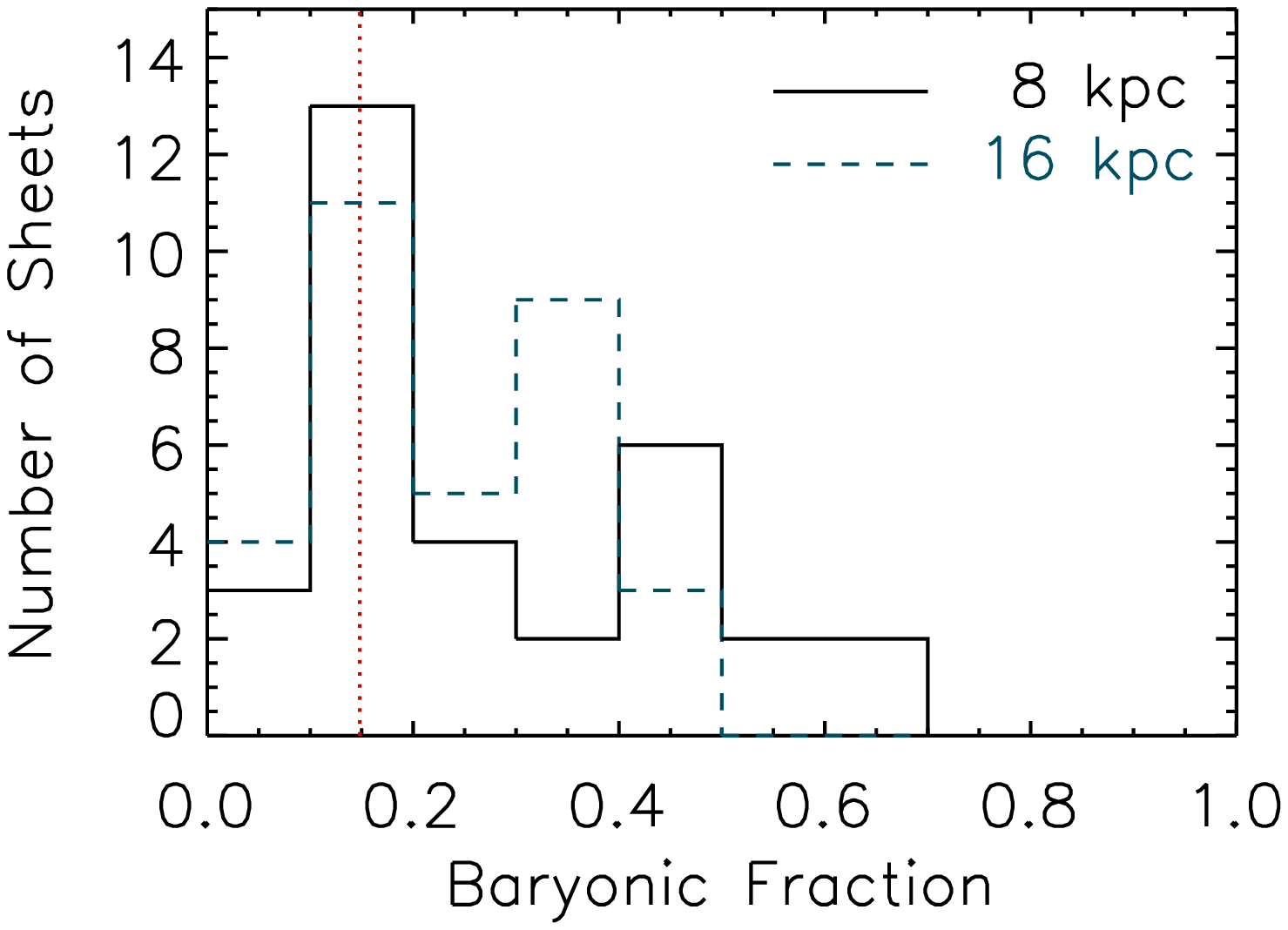}
    \end{center}
    \caption[1]{
    \label{slab_bf}
Histograms of the baryonic fractions of the 32 sheets around the 
29 filaments.  These fractions are
calculated on slabs 8 and $16 \unit{kpc}$ 
in thickness centered on the sheet (see Figure~\ref{cartoon}).
The region considered for each 
sheet extended from 8 to $55 \unit{kpc}$ from the filament 
in a perpendicular directon and between 22.2 and 55.4
kpc from the galaxy along the filament.  The vertical dotted red
line marks the cosmic mean baryon fraction.
    }
    \end{figure}
}

\newcommand{\cmdrichfractbsize} {
    \begin{figure}
    \begin{center}
    \leavevmode
    \includegraphics[scale=0.4]{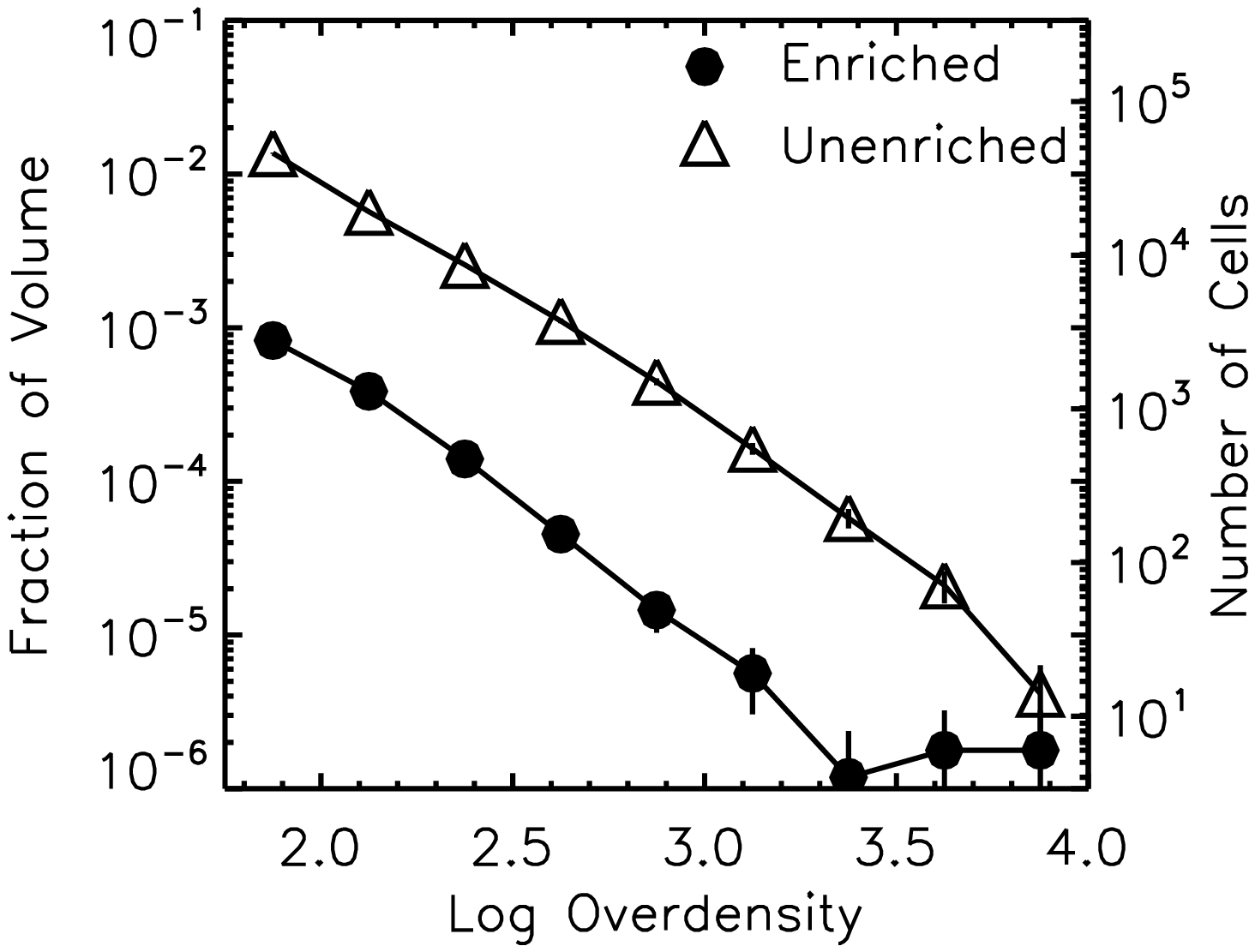}
    \includegraphics[scale=0.4]{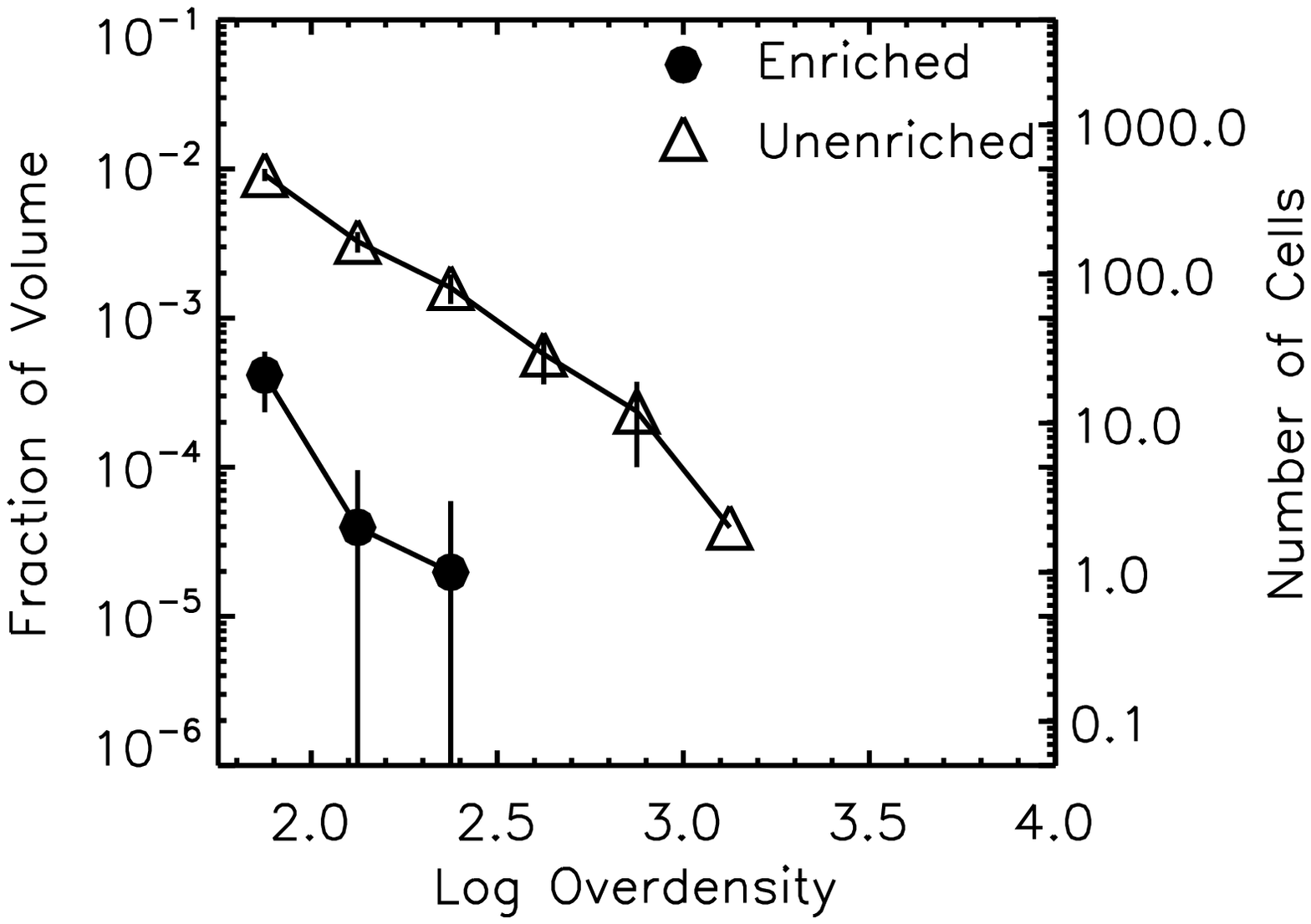}
    \end{center}
    \caption[1]{
    \label{rich_fract_bsize}
Effect on baryonic fraction of varying the grid cell size.
Same as Figure~\ref{rich_fract},
but for cell sizes of
$4.1 \unit{kpc}$ proper ($25 \unit{kpc}$ comoving)
in the upper graph,
and
$16.3 \unit{kpc}$ proper ($100 \unit{kpc}$ comoving)
in the lower graph.
    }
    \end{figure}
}

\newcommand{\cmdfighetc} {
    \begin{figure}
    \begin{center}
    \leavevmode
    \includegraphics[scale=0.5]{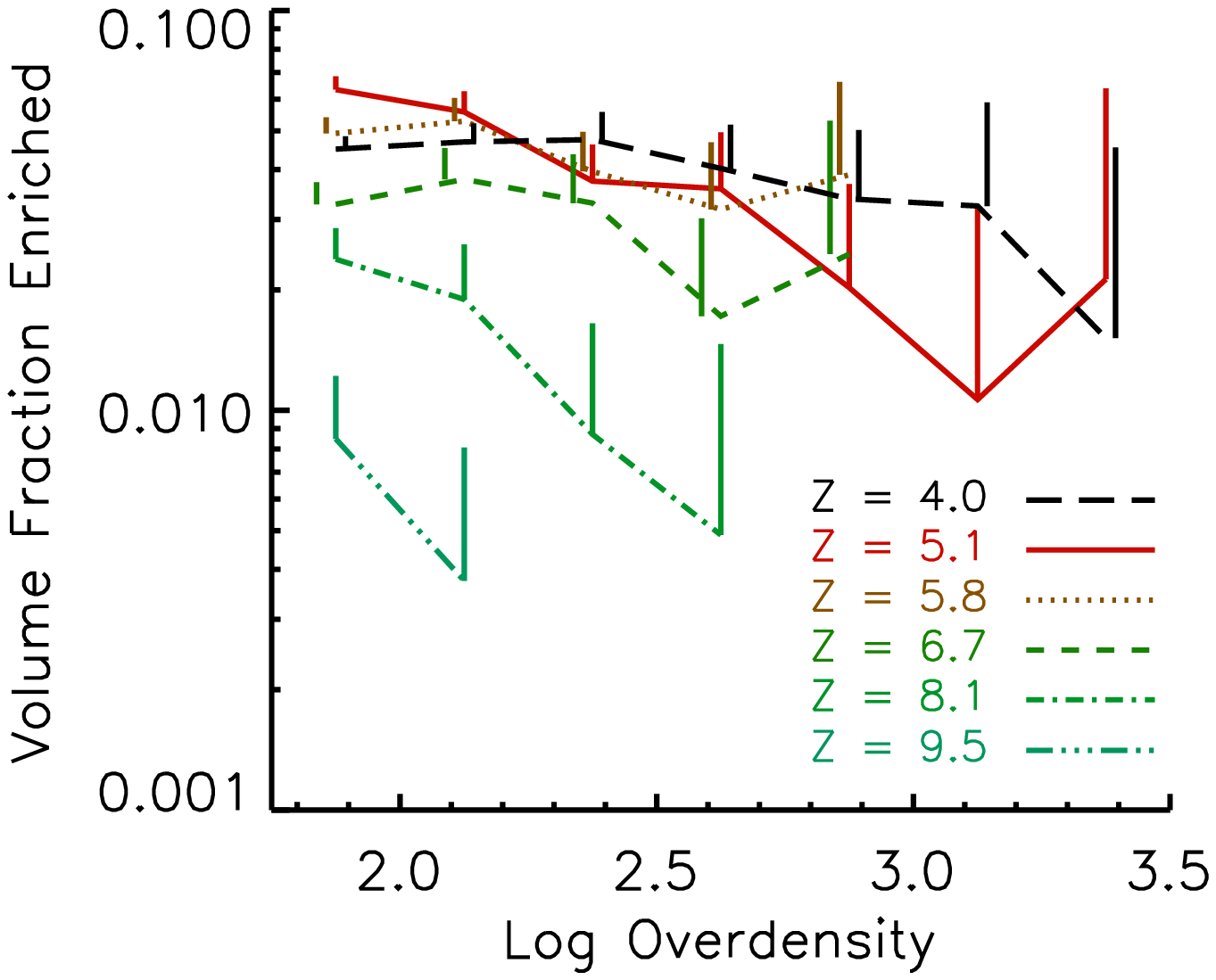}
    \end{center}
    \caption[1]{
    \label{fig_h1_xsect2_census_gas}
Volume fraction of baryon-enriched regions as a function of overdensity,
at several redshifts.
As elsewhere in this paper,
the threshold for baryon-enrichment is taken to be twice the cosmic mean.
The cell size for all redshifts is $40 \unit{kpc}$ comoving.
Vertical lines show two sigma errors (only the upper error bars  are shown).
    }
    \end{figure}
}

\begin{document}

\maketitle

\begin{abstract}
Using a high resolution cosmological simulation of reionization
we have examined the differing structures formed by gas and dark
matter at a redshift of 5.1.  Baryon-rich regions form a small number
of filaments, which connect the largest galaxies in the simulation.
More detailed 
examination of the ten largest galaxies reveals long, slender gaseous 
filaments about 5 proper kpc in width radiating from the galaxy 
centers.  Extending out from each
filament are a few smooth, thin, nearly planar gaseous sheets.  By contrast,
the dark matter concentrates into quasi-spherical bodies.  The
results
have implications for our understanding of structure 
formation in the early universe and of the Lyman alpha forest.
\end{abstract}


\section{Introduction}
\label{intro}

During the first gigayear of the universe
the first stars, galaxies, and black holes formed,
and reionizaton was completed.
Most of the matter resided in the intergalactic medium,
forming a network of sheets and filaments
(for reviews
see \citealt{meiksin_07}, \citealt{barkana_07}).
In this environment galaxies grew by accreting dark matter and 
gas to form the ``cosmic web'' of walls and filaments of 
clusters of galaxies that characterize the visible large scale structure of
the universe today.

Gravitational clustering produces sheets and filaments naturally,
because gravitational collapse of ellipsoidal maxima
occurs at different times along the three unequal axes
\citep{lin_65,
zeldovich_70,
jing_suto_02,shen06}.
Collapse occurs first along a single axis to produce a two 
dimensional sheet.
Subsequent collapse of a sheet along a second axis produces a filament.
Finally, collapse of filaments leads to quasi-spherical galactic haloes.
Because matter clusters at various scales
and with a range of overdensities,
all three kinds of structure may coexist at a single moment in time.
Cosmological simulations using a variety of cosmological models
confirm that sheets and filaments occur generically
and ubiquitously
(for discussion see \citealt{valinia_97},
\citealt{bond96},
\citealt{shandarin_95}, \citealt{schmalzing_99}, \citealt{sheth_03},
and \citealt{shen06};
for reviews of early theoretical work see \citealt{sdz_83},
\citealt{shandarin_89}).

At large scales,
baryons and dark matter
are expected to trace the same structures.  Cosmological
hydrodynamic simulations with resolutions in the tens of kiloparsecs
have borne this out, although differences in the smoothness of the gas
and dark matter have been noted (\citealt{cen94, miralda_escude_96,
hernquist_96,zhang_98}).

On smaller scales,
hydrodynamic processes may be expected
to lead to some separation of baryons and dark matter.
The purpose of the present paper is to explore this separation
using a cosmological simulation that has sub-kiloparsec resolution
and includes a detailed and careful treatment of baryonic physics
\citep{harford07}.
The simulation includes 3-dimensional radiative transfer
of radiation produced self-consistently by star formation, and it follows 
the detailed ionization and chemistry of atomic hydrogen, molecular hydrogen,
and helium. 
We focus on an epoch at high redshift following reionization.
The Jeans length changes by more than an
order of magnitude over the course of reionization 
(see \citealt{gnedin_filt03} and references therein).

The simulation
method has been validated in several respects.  First, 
galaxy luminosity functions 
compare favorably to observations (\citealt{harford07};
\citealt{harford03}).
Second, the reionization history is consistent with the spectra of high
redshift quasars \citep{gnedin_fan06}.  Finally, considerable progress has 
been made in reproducing the frequency of Lyman limit systems
\citep{kohler07}.

In examining the output of the simulation reported here,
we were struck by the marked
difference in the distribution of baryons and dark matter on scales
small enough that hydrodynamic processes are important, but large
enough that intergalactic structure is easily visible.  We noticed that
intergalactic gas tends to occur in long, nearly continuous
filaments and attached sheets, which are unusually rich in baryons
relative to the cosmic mean.  These filaments form ``backbones'' to 
which large, dark matter dominated galaxies are attached.  In contrast,
the dark matter tends to occur in many small, quasi-spherical clumps.

This paper presents evidence for these baryon-rich structures, and
characterizes them.

\section{The Simulation}
\label{simulation}
The simulation used in this paper was run with a ``Softened
Lagrangian Hydrodynamics'' (SLH-P$^{3}$M) code 
\citep{gnedin95,gnedin_bertschinger_96}.  
The simulation has a flat $\Lambda$CDM cosmology,
with values of cosmological
parameters determined by the first year WMAP data \citep{svp03}:
$\Omega_{m} = 0.27$, $\Omega_{b} = 0.04$, $\sigma_{8} = 0.91$,
and $h = 0.71$.
The mean baryonic fraction is therefore 0.148.
We focus on an epoch at redshift 5.135.
The computational expense to run this simulation to significantly later 
times is prohibitive.

Each dimension of the simulation box is $8h^{-1} \unit{Mpc}$ comoving,
and contains 256 Lagrangian simulation cells.
The gas dynamics is
followed on a quasi-Lagrangian mesh, which is gradually deformed
during the simulation to achieve better resolution in high density
regions.  In the post-processing stage, in order to compare dark matter and
baryons in a similar manner, we convert each cell of the 
quasi-Lagrangian mesh into a ``gas particle'' with the same physical
properties (mass, momentum, temperature, etc.).
The positions of an equal number of dark matter particles of constant mass
$2.73\times10^{6}\dim{M}_\odot$ are computed with the P$^{3}$M algorithm
using a softening length of 0.08 kpc proper.
The best resolution is limited to
about 2-3 times the softening length.

The reionization process is simulated by
including star formation
by the Schmidt law and
radiative transfer by the optically thin variable Eddington tensor
method (OTVET) \citep{gnedin_abel_01}.
Also included are radiative transfer effects of molecular
hydrogen, whose formation and disassociation are followed.
A two-level implicit scheme is used to compute
the effects of a hydrogen and helium plasma.

We analyze the simulation on a grid with cell
size of $40 \unit{kpc}$ comoving
($28.4 h^{-1} \unit{kpc}$ comoving),
equal to $6.52 \unit{kpc}$ proper at a redshift of 5.135.  This
cell size is about 80 times the softening length.
This choice of grid cell size,
comparable to the size of a small galaxy,
is a deliberate one.
The grid cell size is small enough to bring out
the separation of baryons and dark matter produced by hydrodynamical processes,
but large enough to map regions of overdensity as low as
$10^{1.75}$, about 56,
with good signal-to-noise.
We find this overdensity to be a convenient lower limit
to differentiate the filamentary structure.
Results for other grid cell sizes and other redshifts are
explored in an Appendix.

\cmdsimbox

The spatial resolution at any point in the simulation depends upon the 
local density of the Lagrangian simulation cells, 
which is roughly proportional to the 
local overdensity.  However, the
proportionality breaks down in regions of very
high or very low baryonic fraction.   We
have determined that all of the baryon-enriched grid cells with a minimum
overdensity of 56 at z = 5.1 have a minimum of
60 gas particles, each of which corresponds to a cell of the
simulation.  The unenriched grid cells may have fewer gas particles,
and consequently lower resolution.


Unless otherwise noted all distances in this 
paper are proper.

\cmdgalacct

\cmdrichfract

\cmdfiga
\cmdsuballden

\section{Baryon-Rich Filaments}\label{baryrichfil}

The upper panel of Figure~\ref{simbox} illustrates how
the baryon-rich regions
in the simulation form a network of filaments.
The baryon-rich regions, plotted in translucent green, are grid cells
whose total overdensity
is at least 56, and whose baryon fraction
is at least twice the cosmic mean.
Both gas and stellar matter in the simulation
are included as baryons in this paper.
The stellar mass in a region is
usually much less than the gas mass,
and generally we have not examined it separately.

The baryon-rich filaments
are associated with the largest galaxies in the simulation.  
The upper panel of Figure~\ref{simbox}
shows, as black spheres,
the 88 largest galaxies in the
simulation, those with a total mass of at least $10^{10} \unit{M}_{\sun}$
in dark matter and baryons.
Galaxies were identified
with the DENMAX algorithm \citep{bertschinger91}
as gravitationally bound density peaks
containing at least 100 simulation particles.  The most
massive galaxy in the simulation is $1.2\times10^{11}\dim{M}_\odot$.

Figure \ref{galacct} shows how many galaxies in each mass range are
within two grid cells of a baryon-rich cell (we define such cells
as ``nearby'' hereafter).  Above a mass of
$10^{10}\dim{M}_\odot$, 83\% of the galaxies are nearby, while below
this range only 11\% are.  The nearby galaxies account for 85\% of
the galactic stellar mass.  Because many of the smaller galaxies
are devoid of stars, only 38\% of the galactic dark matter is
nearby.

Although the largest galaxies in the simulation are near
the baryon-rich filaments,
the galaxies themselves generally
have baryonic fractions close to the cosmic mean. 

The baryon-rich regions are relatively rare compared to unenriched regions.
The lower panel of Figure~\ref{simbox}
shows unenriched cells
with an overdensity of 56 or greater.
The unenriched cells outnumber the enriched cells by a factor of about 20.
The unenriched regions,
though clustered,
are not as strikingly filamentary
as the enriched regions.

Figure~\ref{rich_fract}
shows that
baryon-rich cells constitute about 1/20th of the volume of
unenriched cells, almost independent of overdensity.
The Figure shows only overdensities 56 or above
because the determination of baryon fraction becomes
increasingly uncertain at lower overdensities.
The Figure suggests a slight trend
for greater enrichment
at lower overdensity,
but the statistical significance is not large.

Figure~\ref{fig_1_2_2}
shows a zoom-in
of the longest enriched filament in Figure~\ref{simbox}.
The zoom-in
shows large galaxies as clumps of unenriched grid
cells arranged along a baryon-rich backbone.
The lower panel of the Figure shows the backbone by itself,
with gaps at the positions of some of the large galaxies.
We expect galaxies to concentrate
gas at their centers, and many in the simulation do.
However, the central regions
are usually too small to show up as baryon-rich
on the adopted $6.5 \unit{kpc}$ grid scale.

Figure~\ref{sub_all_den} zooms in still further
to show individual simulation 
particles.  The upper panel shows just the gas particles concentrated
into filaments.  
The lower panel shows the same field of view with just the dark
matter particles.  The dark matter, although broadly following the
filaments, tends to assume quasi-spherical shapes.

\section{Filaments Radiate from Centers of Galaxies}.
In section~\ref{baryrichfil} we showed that the baryon-rich regions of
the simulation form a system of filaments that connect the 
largest galaxies.  In this section we explore the properties of the
filaments in more detail by focussing on regions surrounding the
10 largest galaxies and the second largest galaxy in particular.

\cmdshell

Figure~\ref{shell} shows four images of a $55.4 \unit{kpc}$ (proper) 
radius region centered on the second largest galaxy.  The two upper
images show the gas particles (left) and dark matter particles (right) 
that are bound
to the galaxy as indicated by DENMAX.  Each lower image shows the
corresponding unbound particles.  The three major spokes of bound 
gas in the uppper left image are nearly coplanar.

\cmdhpixcontour

To isolate individual filaments for quantitative study,
we examine a shell extending
from $22.2 \unit{kpc}$ to $55.4 \unit{kpc}$
around each of the 10 largest galaxies.
The inner radius encompasses the bound region of the galaxy,
while the outer radius is large enough to reveal the filamentary structure
clearly, but not so large as to encroach on neighbouring large galaxies.
On average, 10\% of the solid angle of the shell contains
65\% of the gas and 50\% of the dark matter.

Figure~\ref{hpix_contour} shows contour maps of a 192-pixel
HealPix\footnote{We gratefully acknowledge the use of the HealPix software
package obtained from http://healpix.jpl.nasa.gov.}
tiling \citep{gorski05} of the shell around the galaxy 
in Figure~\ref{shell}.  The dark matter (top panel)
shows the smoothest distribution, while gas, neutral hydrogen, and stars 
(succeeding panels) are progressively more concentrated.

For objectivity we adopted a simple formal procedure for
selecting the peak pixels that define the filaments.  First we
select those pixels that contain at least $3.6\times 10^{8} \unit{M}_{\sun}$
of gas.  This minimum mass requirement is imposed so that a radial
segmentation of a pixel into six equal parts is expected
to contain at least $6\times 10^{7} \unit{M}_{\sun}$ of gas,
which is more than one humdred times
the fiducial mass of a gas particle in the simulation.  Some smaller
filaments, apparent to the eye, are thus omitted.   Having made
this selection, we refer to the contour maps to ensure that only one 
pixel, the maximum, is chosen from each contour peak.  This procedure
yields a total of 29 filaments around the 10 galaxies.
Two of the 29 filaments overlap along their length,
but the overlap is only about 30\% of the length analyzed.

\cmdcyldistrib

Figure~\ref{cyl_distrib} shows that
the linear mass density of gas in the filaments generally varies along the 
filament by less than a factor of two or three.
The range of values among the filaments 
is about an order of magnitude, with a few outliers.
The Figure shows the mean density in cylinders of radius 
8 and $16 \unit{kpc}$
about a radial axis centered on the peak pixel.

\cmdfilprofile

Figure~\ref{fil_profile} shows further evidence
that the properties of filaments do not vary greatly along their length.
The Figure shows that the
average cross-sectional profile of gas density and baryonic fraction
of filaments is approximately independent of distance from the galaxy.
Each of the three black curves with filled circles in each graph 
show the profile, averaged over all 29 filaments,
of one of three equal segments spanning the distance from
22.2 to $55.4 \unit{kpc}$ from the galaxy.

The lower panel of
Figure~\ref{fil_profile}
shows that the average baryonic fraction (black curves with filled circles)
exceeds twice the cosmic mean out to about $5 \unit{kpc}$,
and remains above the cosmic mean out to about $9 \unit{kpc}$,
beyond which it dips slightly below the mean.
This is
consistent with the idea that there has been a separation of
baryons and dark matter, with baryons becoming more centrally concentrated
in the filaments.
The Figure also shows the cumulative
baryonic fraction
(orange curves with triangles)
within the radius.
The cumulative baryonic fraction remains elevated out to
beyond $20 \unit{kpc}$.
At least some of this enhanced baryon fraction
can be attributed to sheets, described in the next section,
at whose intersections the filaments are found.

\cmdtriadtilt

\begin{figure}
\fbox{\scalebox{3} {\parbox{0.5in}{
    \includegraphics[scale=0.025]{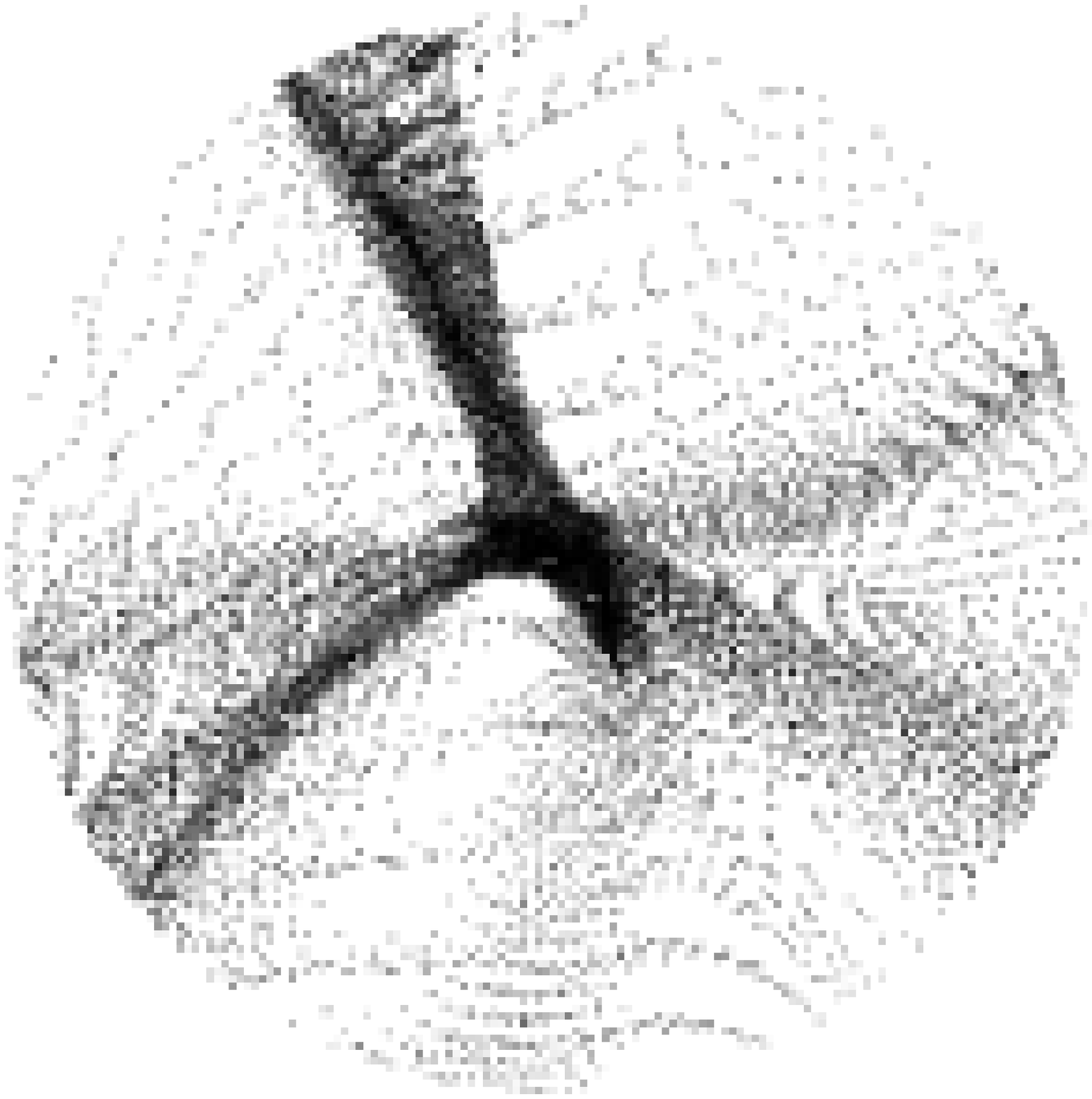}
    \includegraphics[scale=0.025]{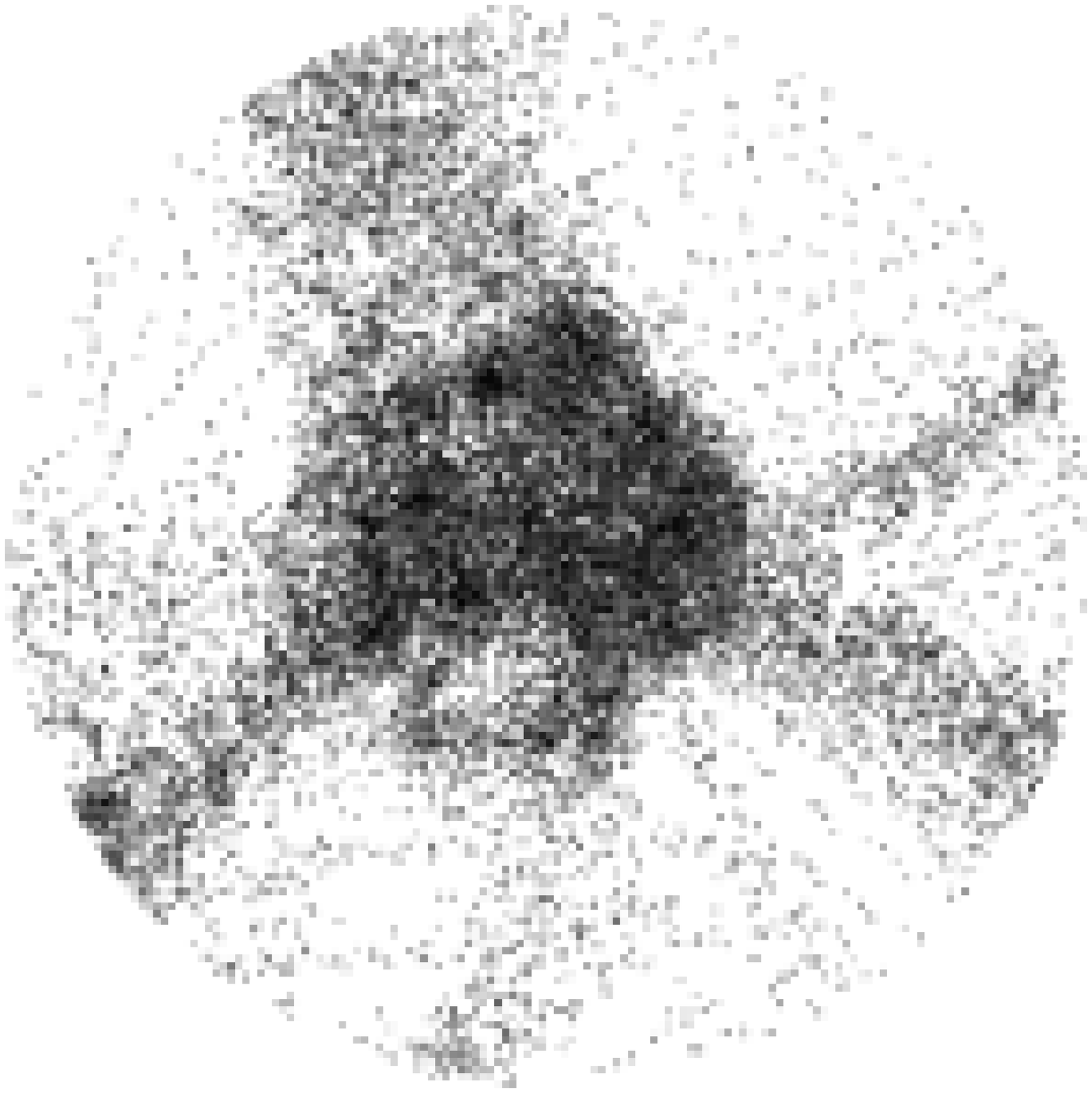}
    \includegraphics[scale=0.025]{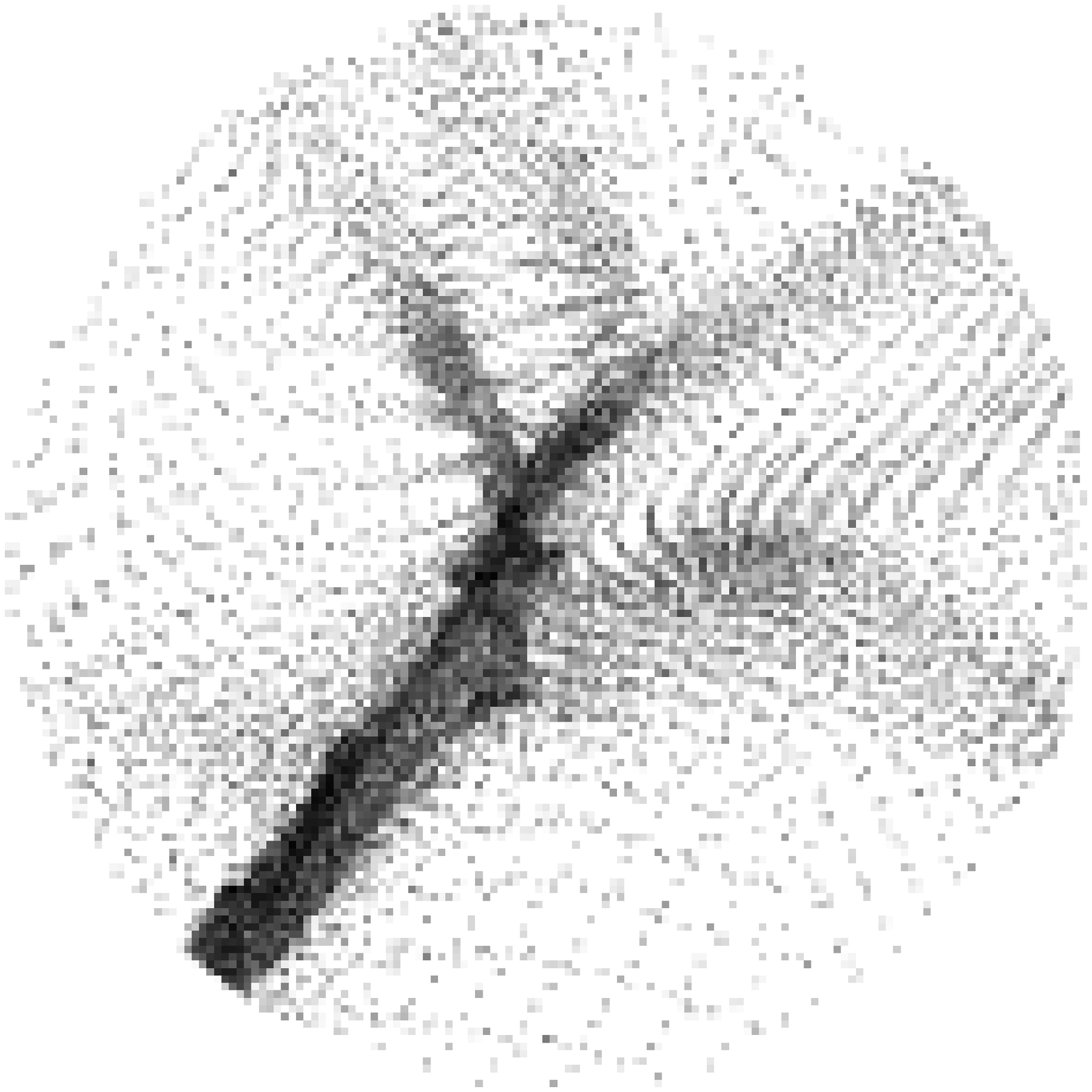}
    \includegraphics[scale=0.025]{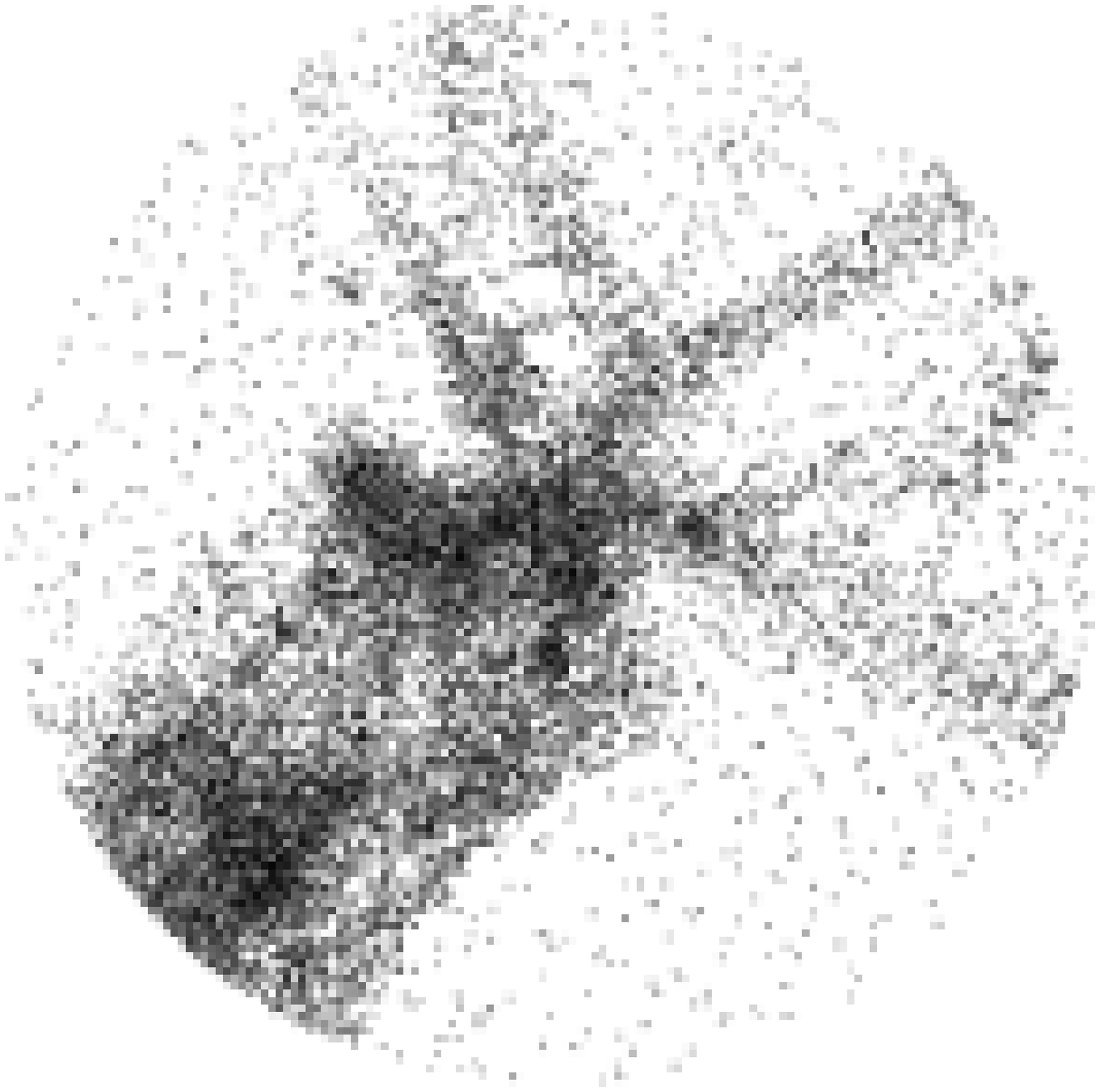}
    \includegraphics[scale=0.025]{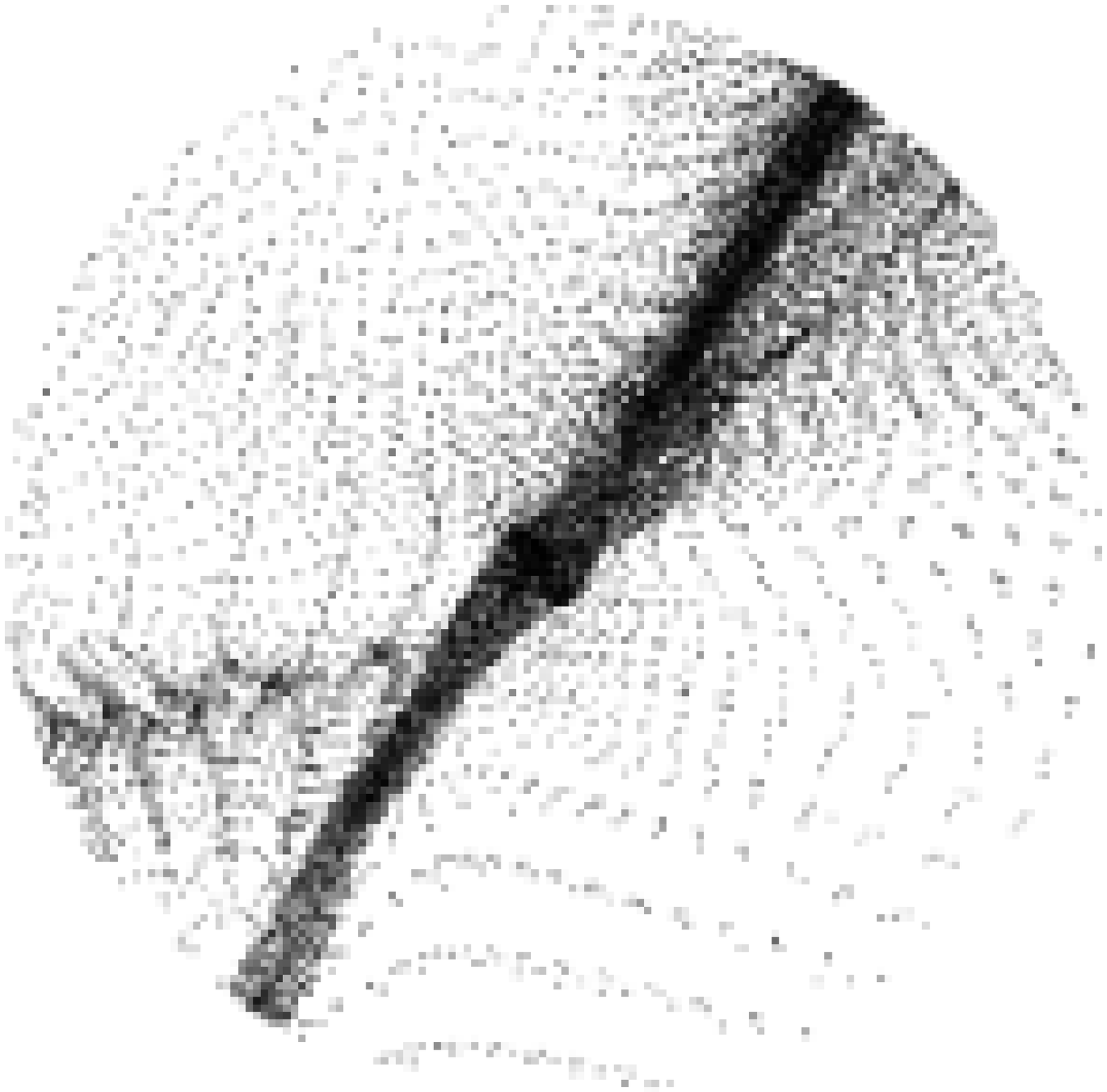}
    \includegraphics[scale=0.025]{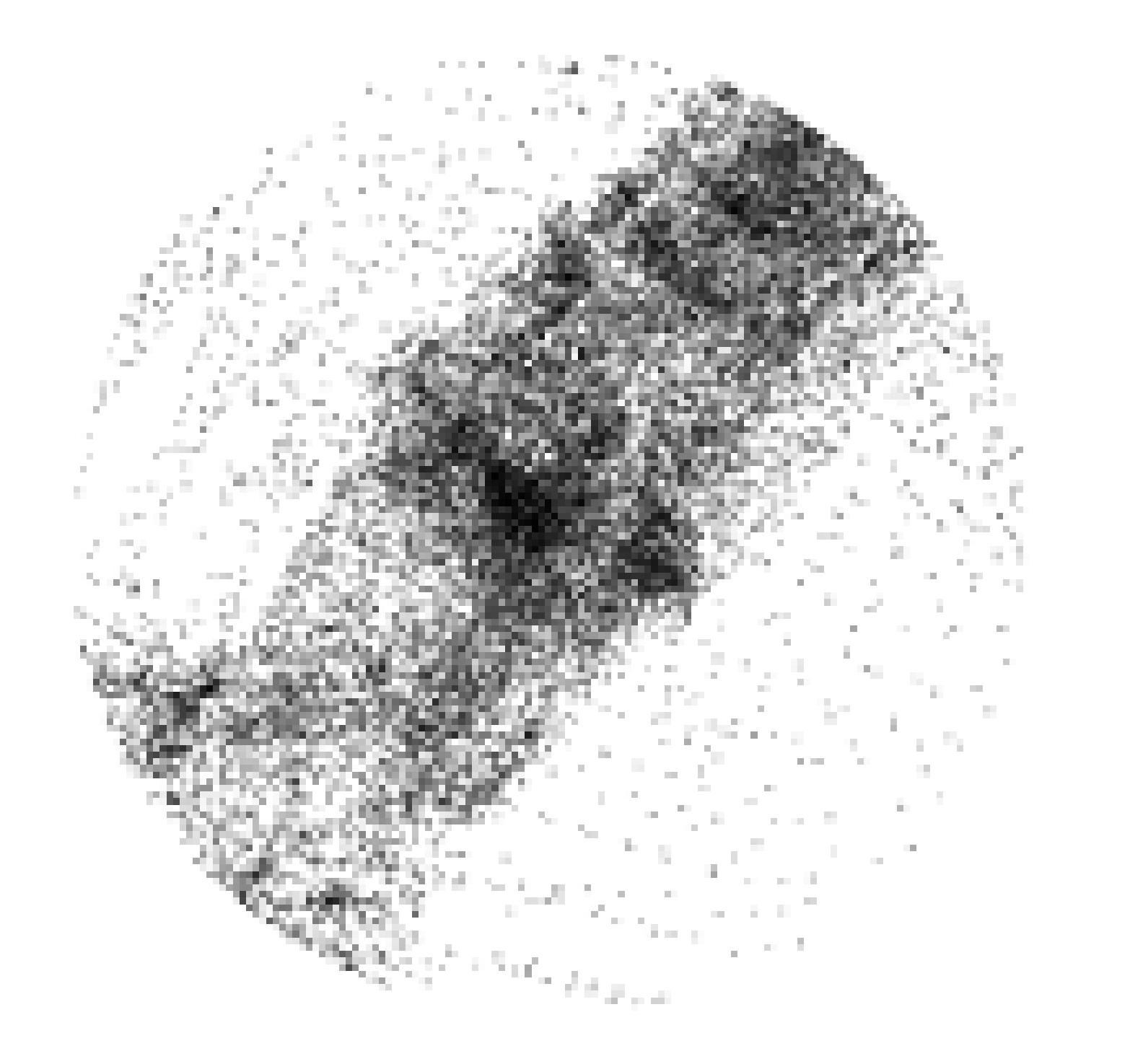}
    \includegraphics[scale=0.025]{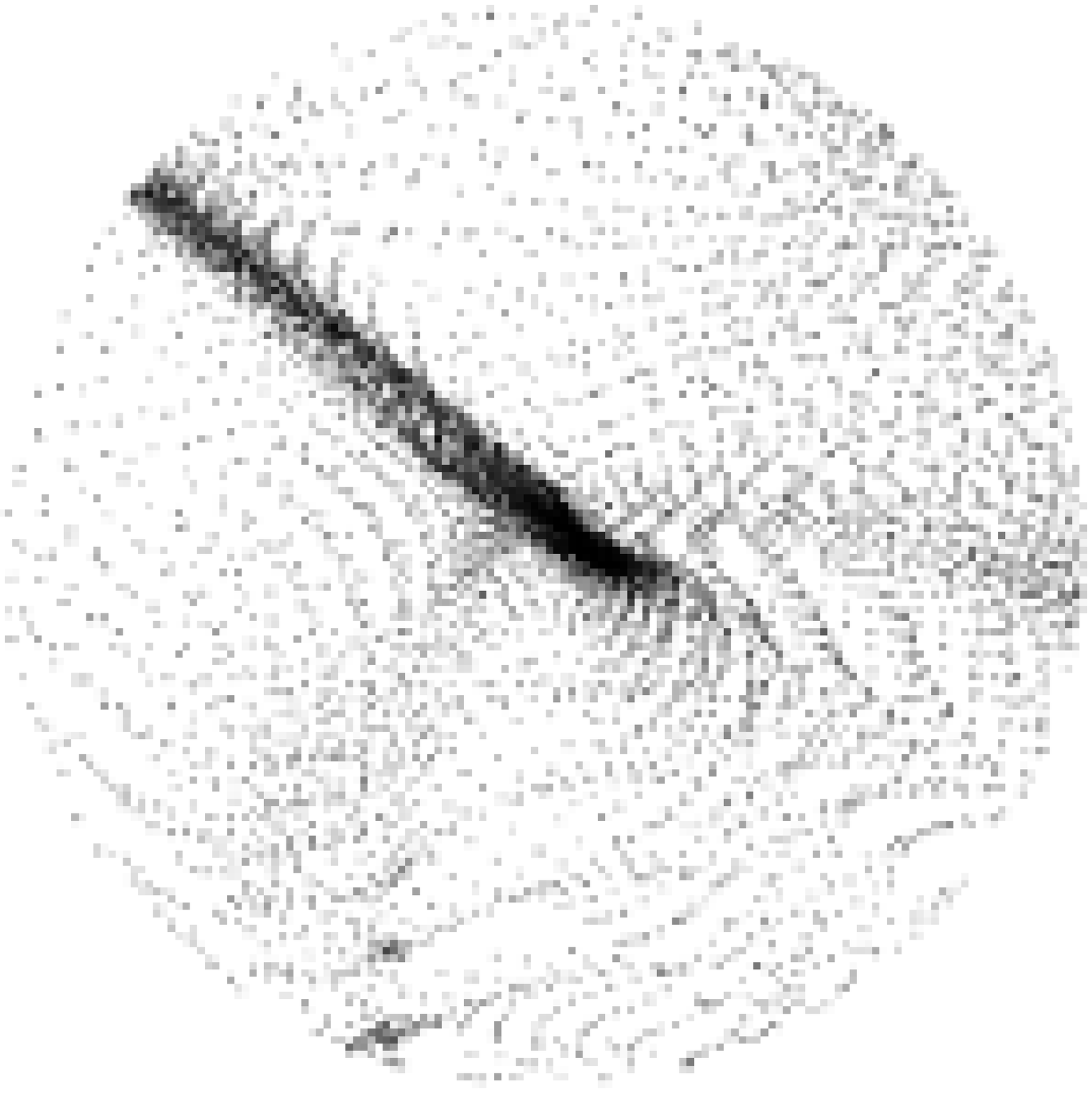}
    \includegraphics[scale=0.025]{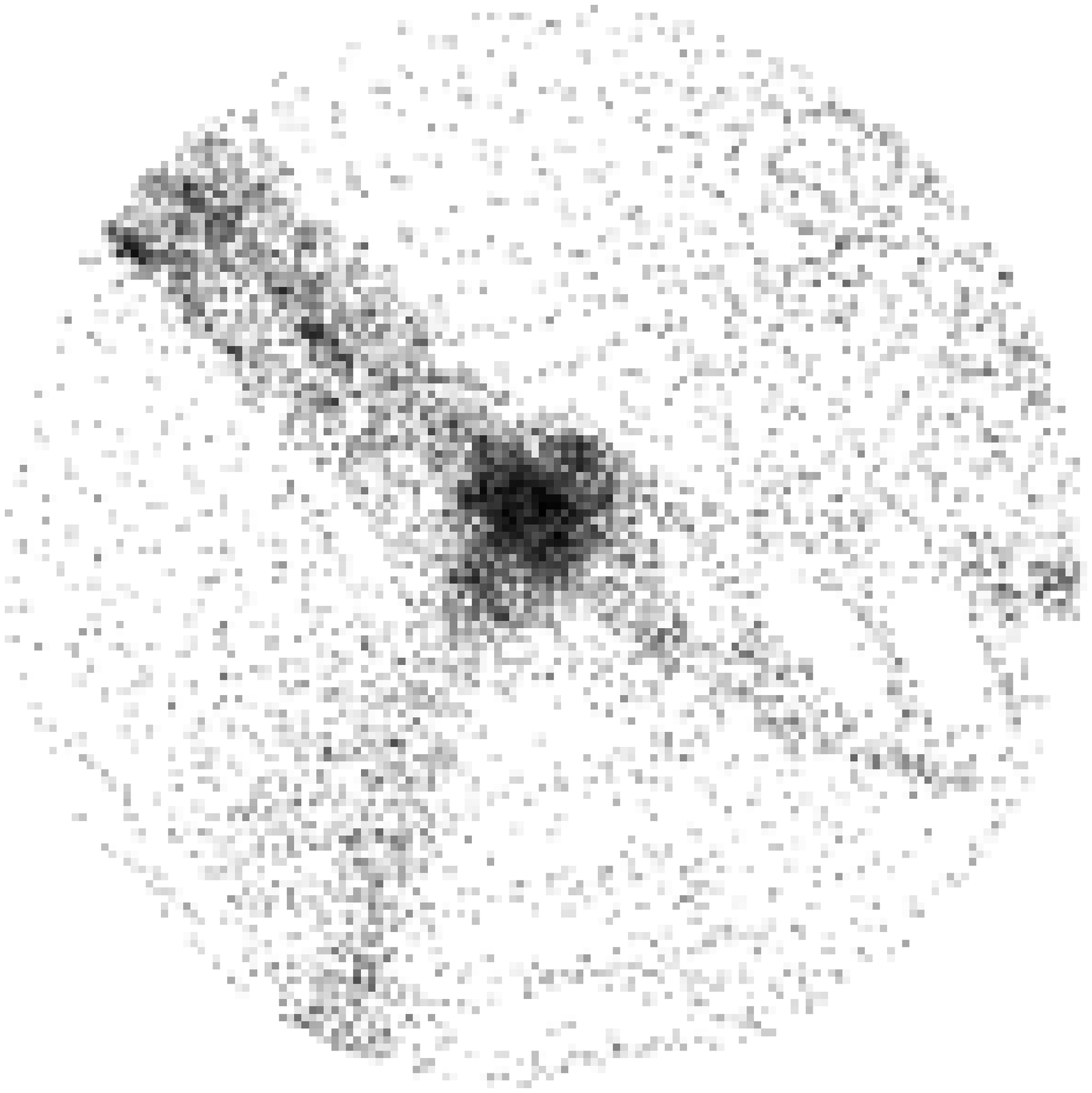}
    \includegraphics[scale=0.025]{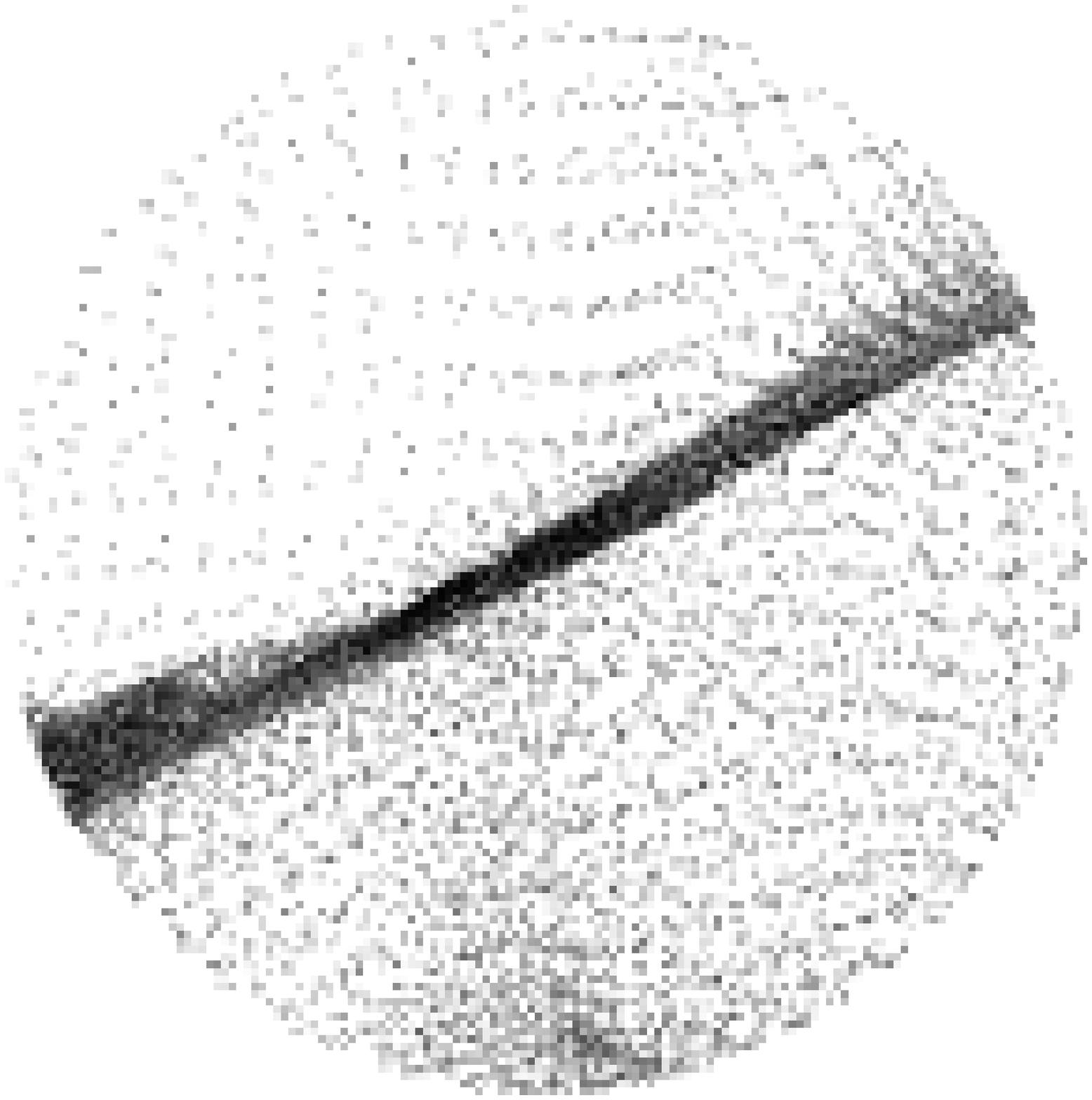}
    \includegraphics[scale=0.025]{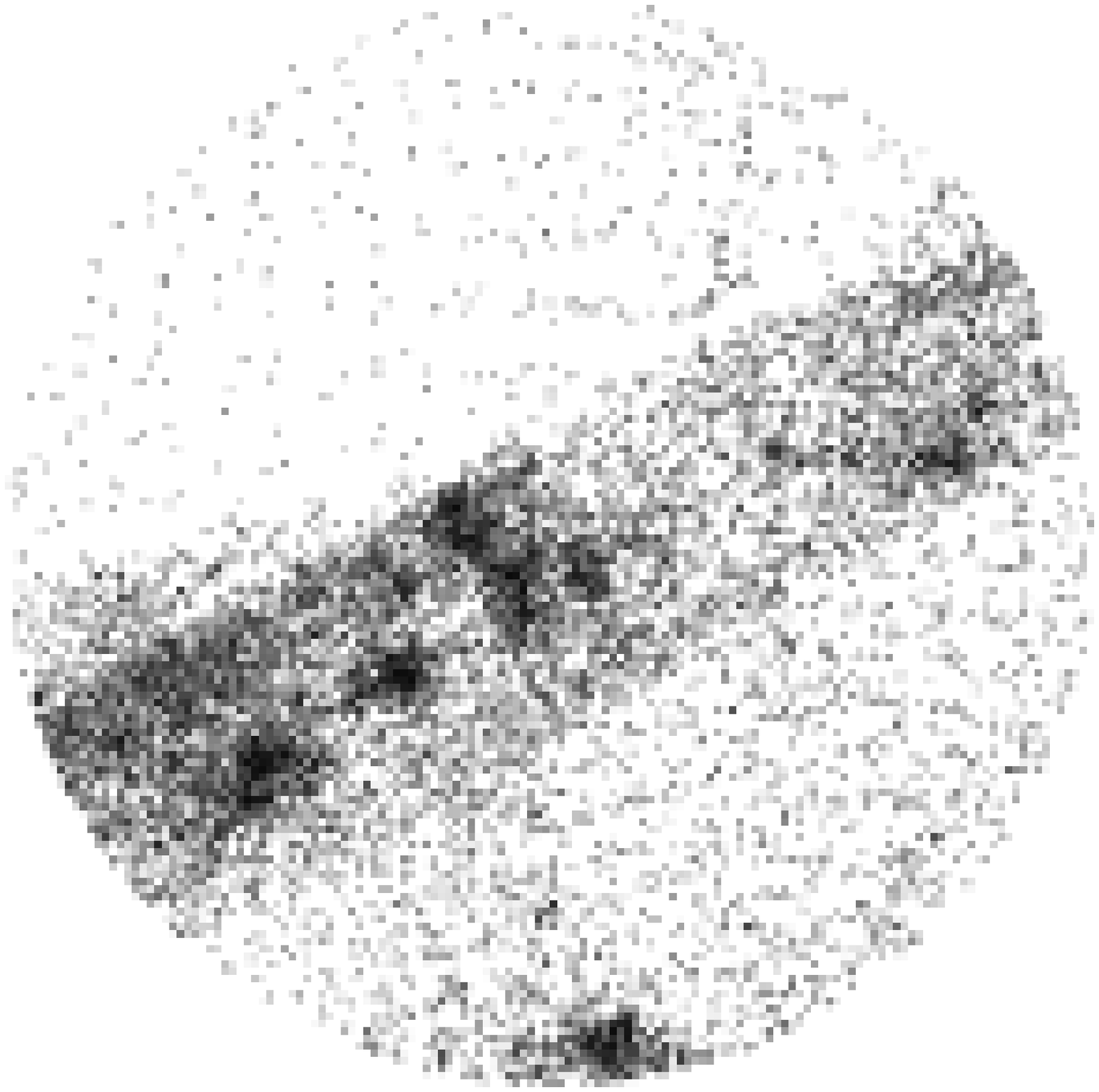}}}
}
\fbox{\scalebox{3} {\parbox{0.5in}{
    \includegraphics[scale=0.025]{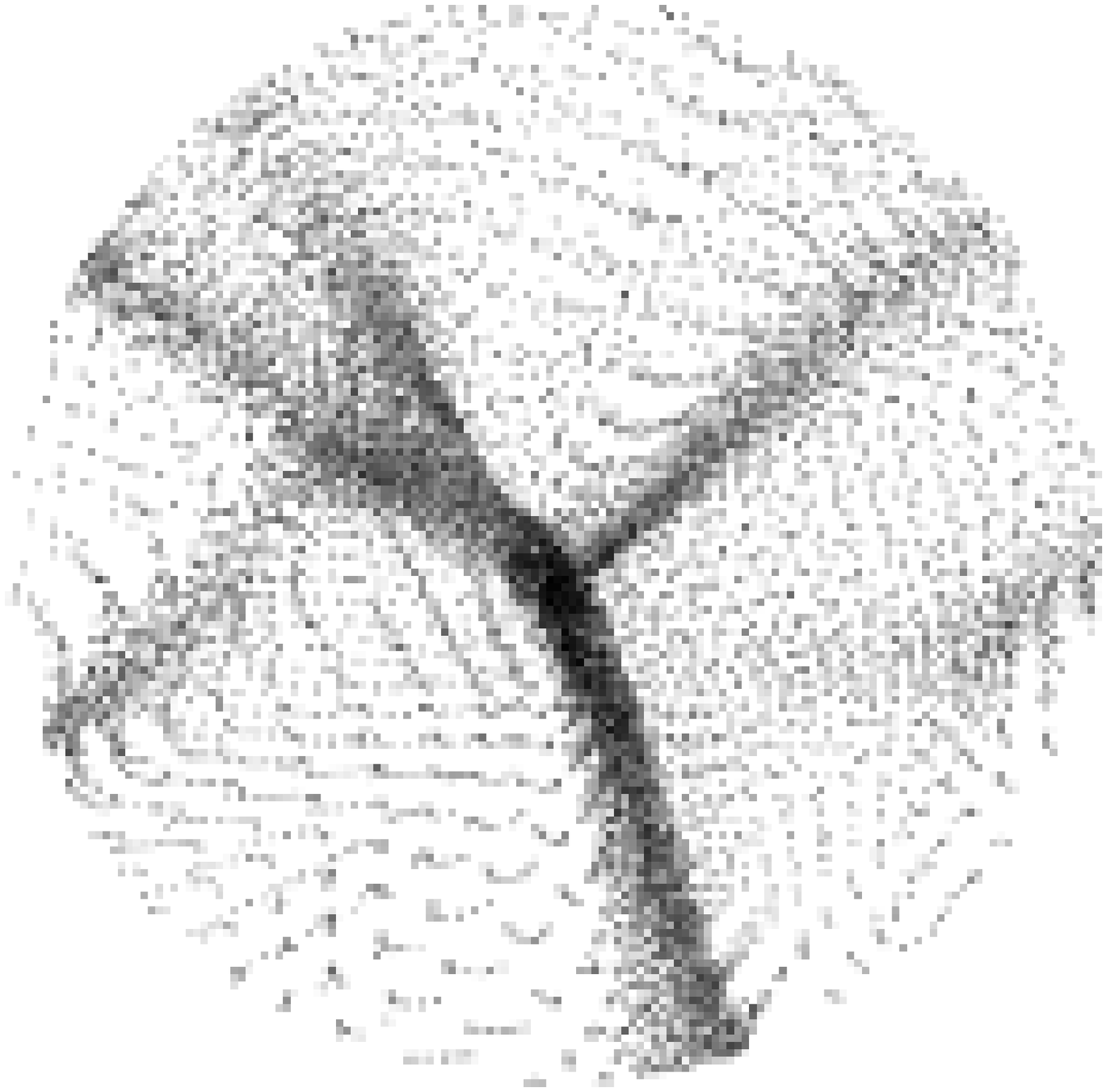}
    \includegraphics[scale=0.025]{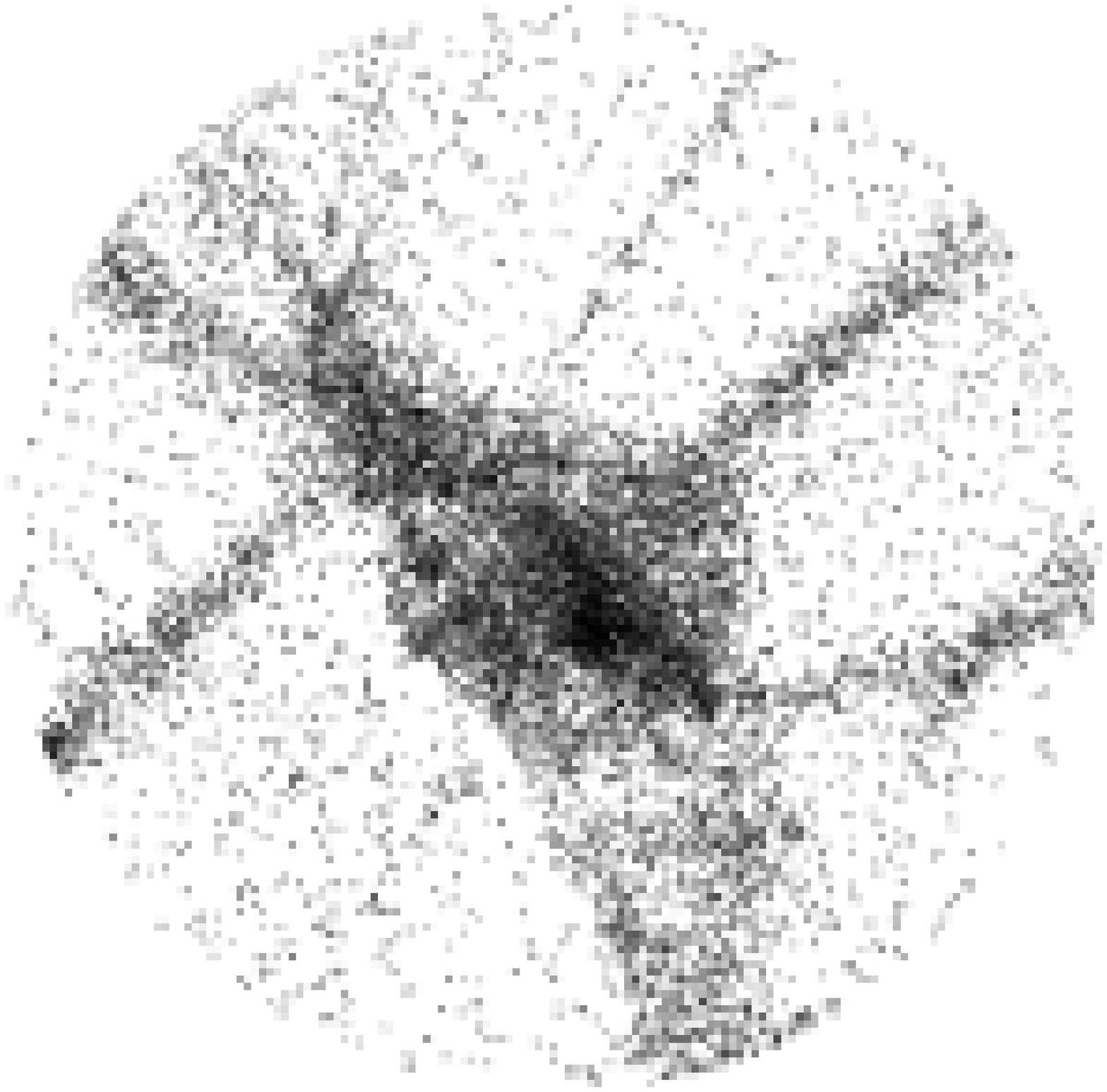}
    \includegraphics[scale=0.025]{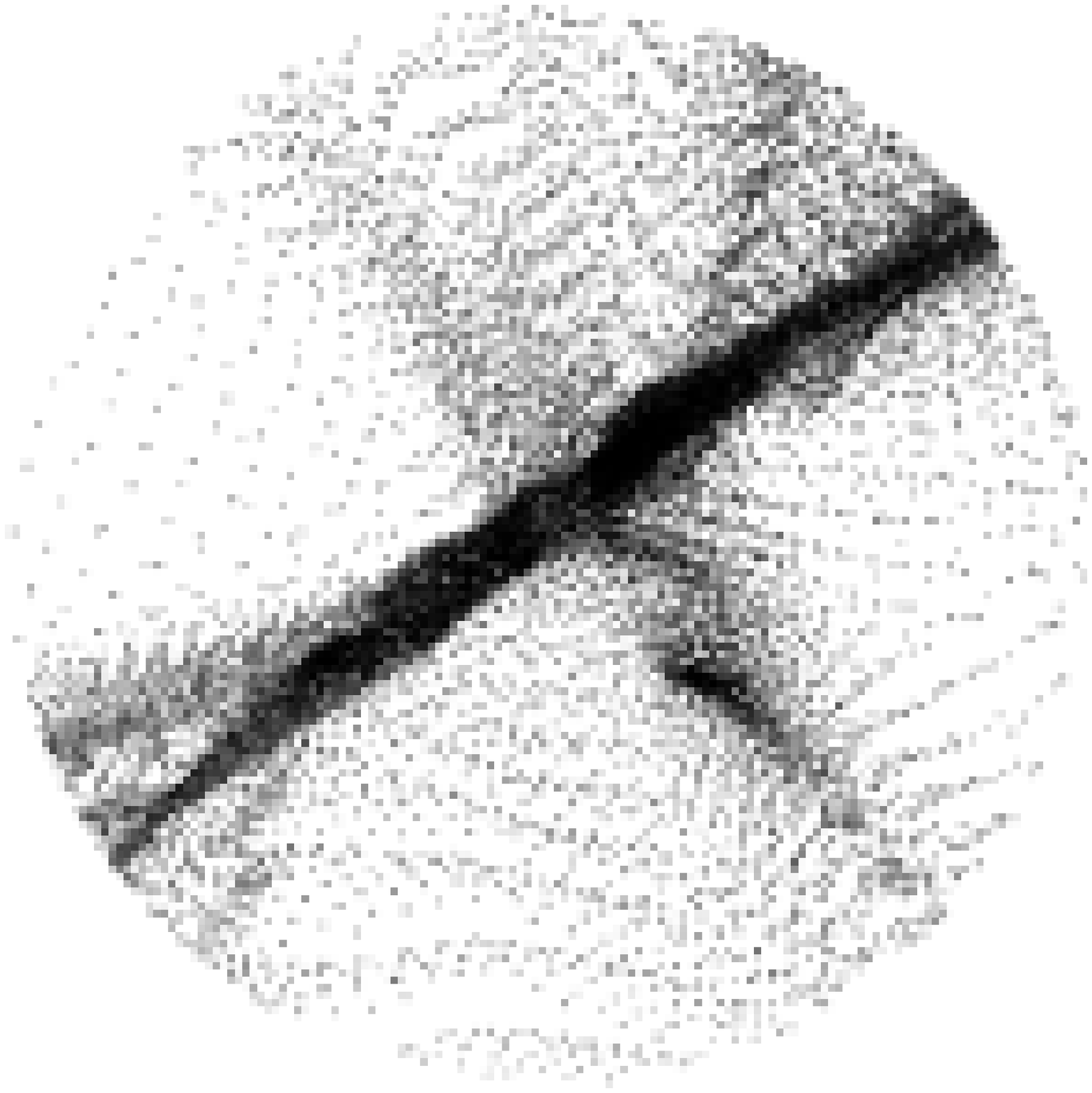}
    \includegraphics[scale=0.025]{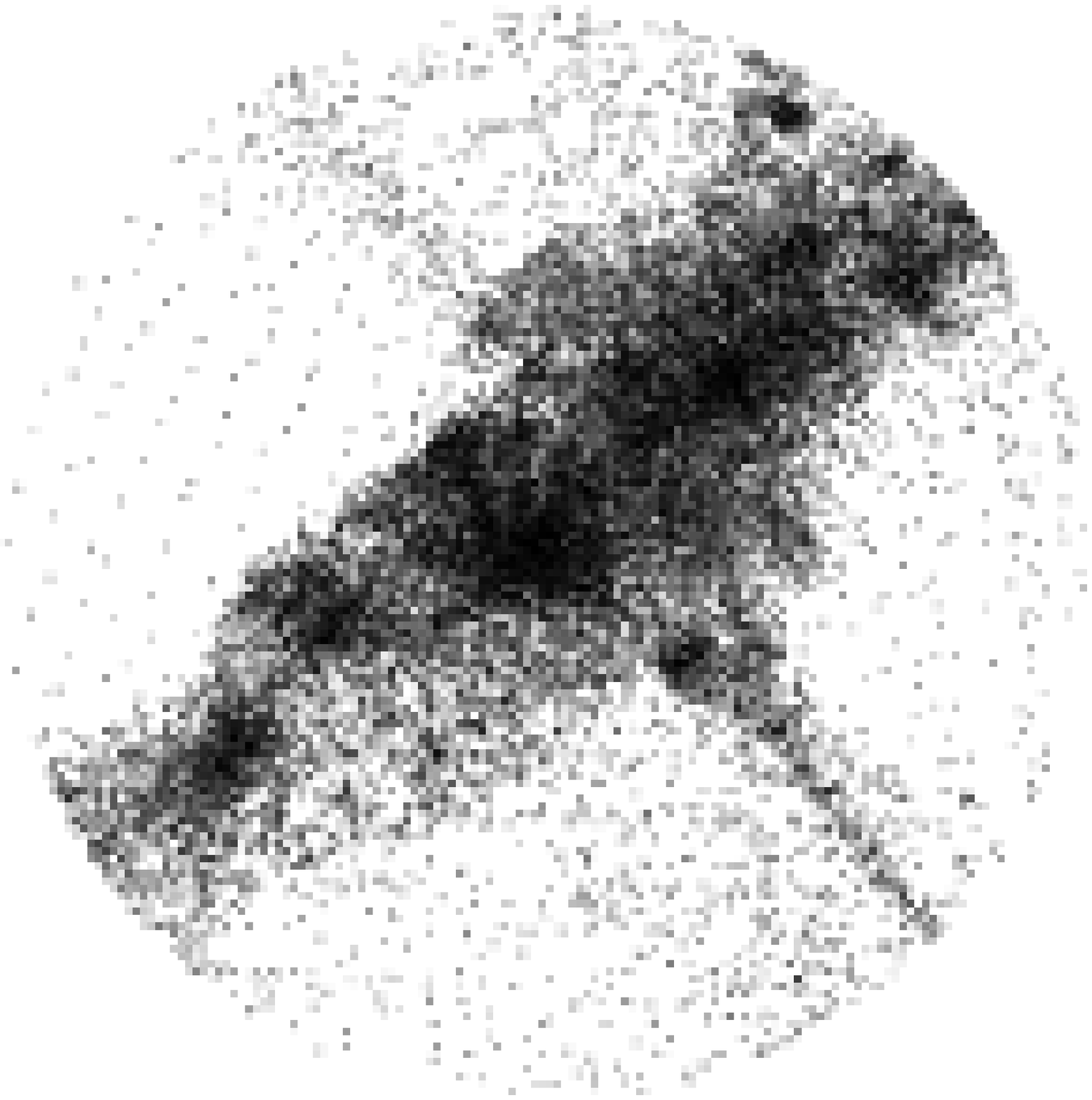}
    \includegraphics[scale=0.025]{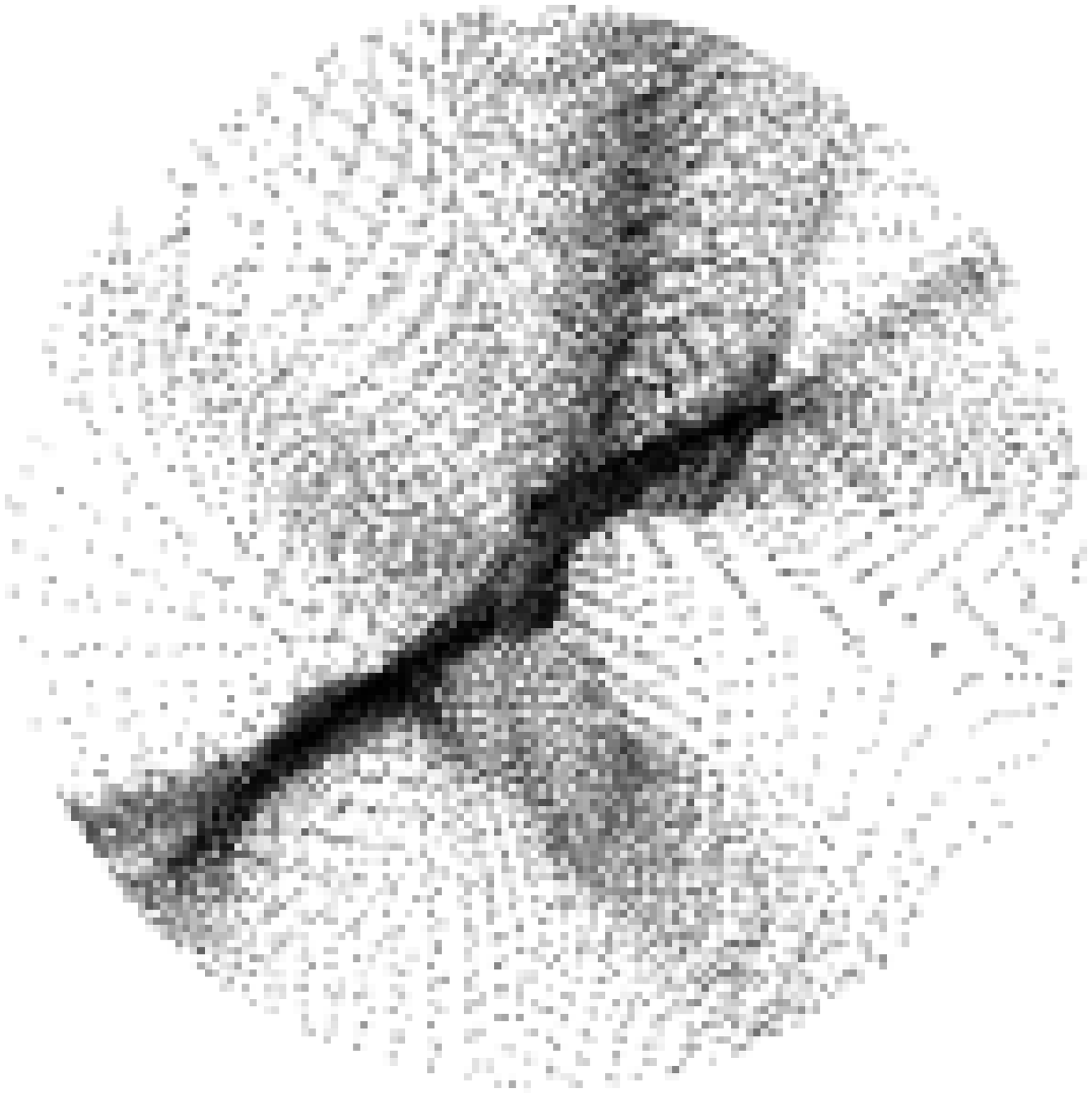}
    \includegraphics[scale=0.025]{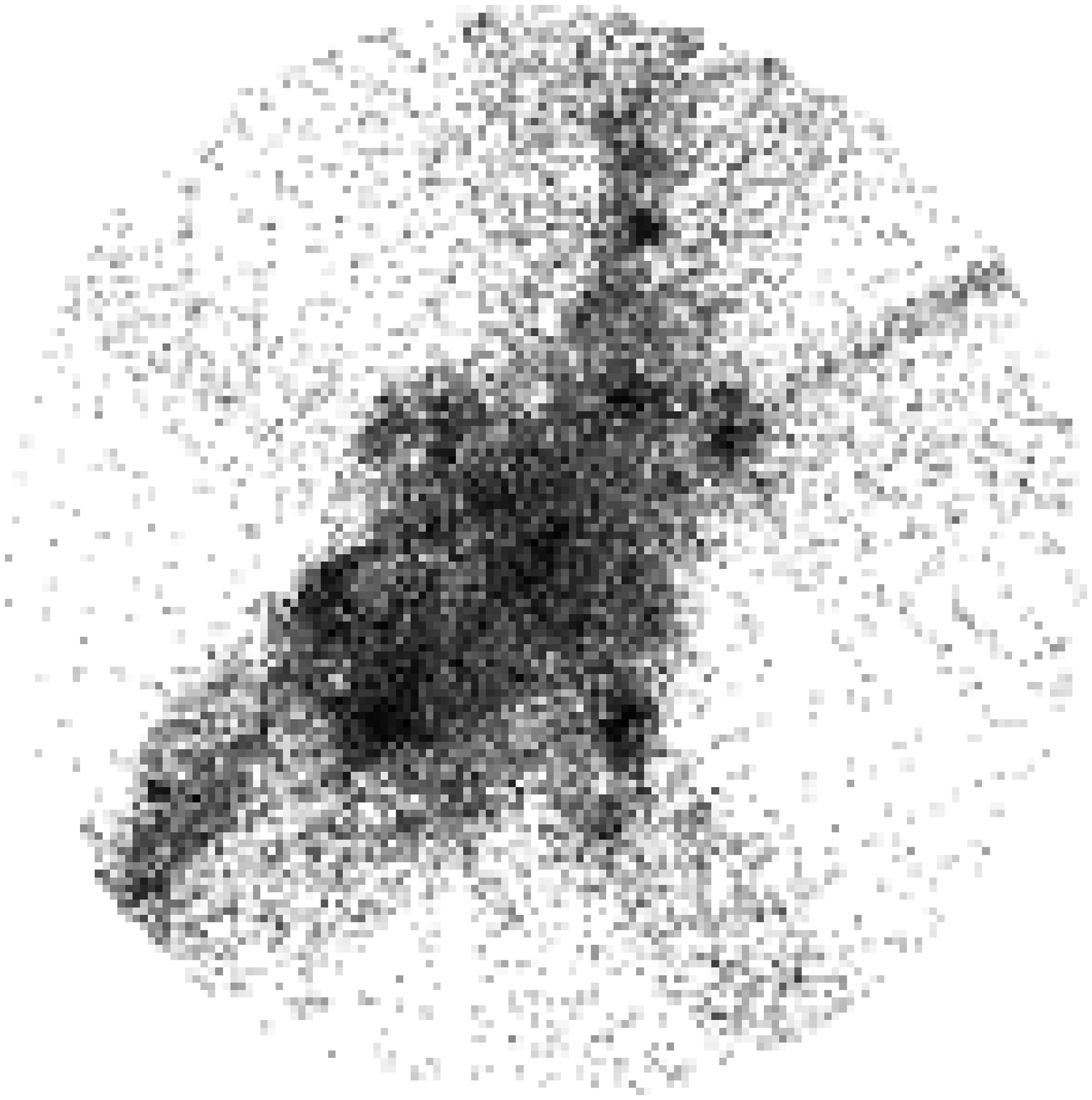}
    \includegraphics[scale=0.025]{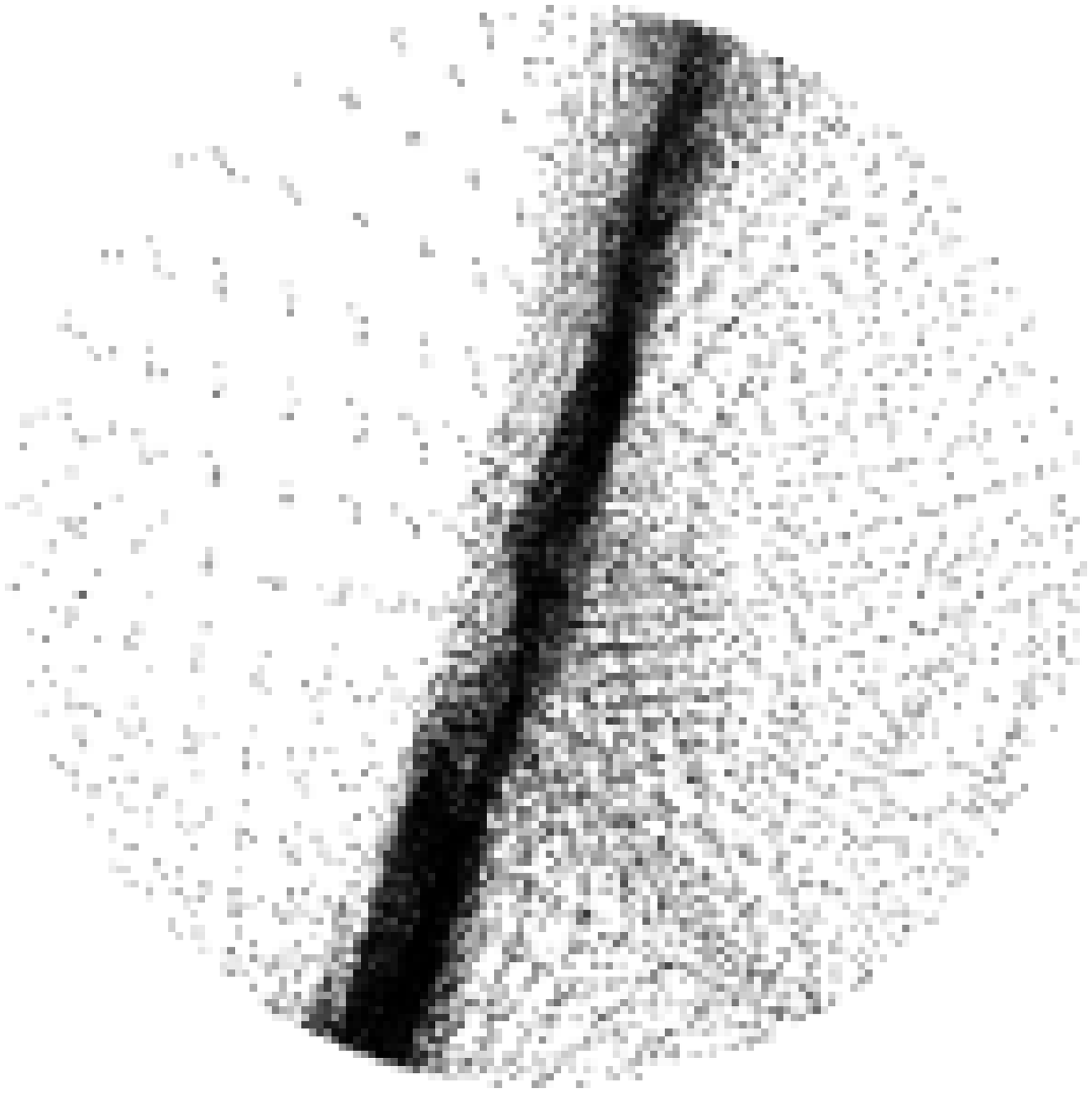}
    \includegraphics[scale=0.025]{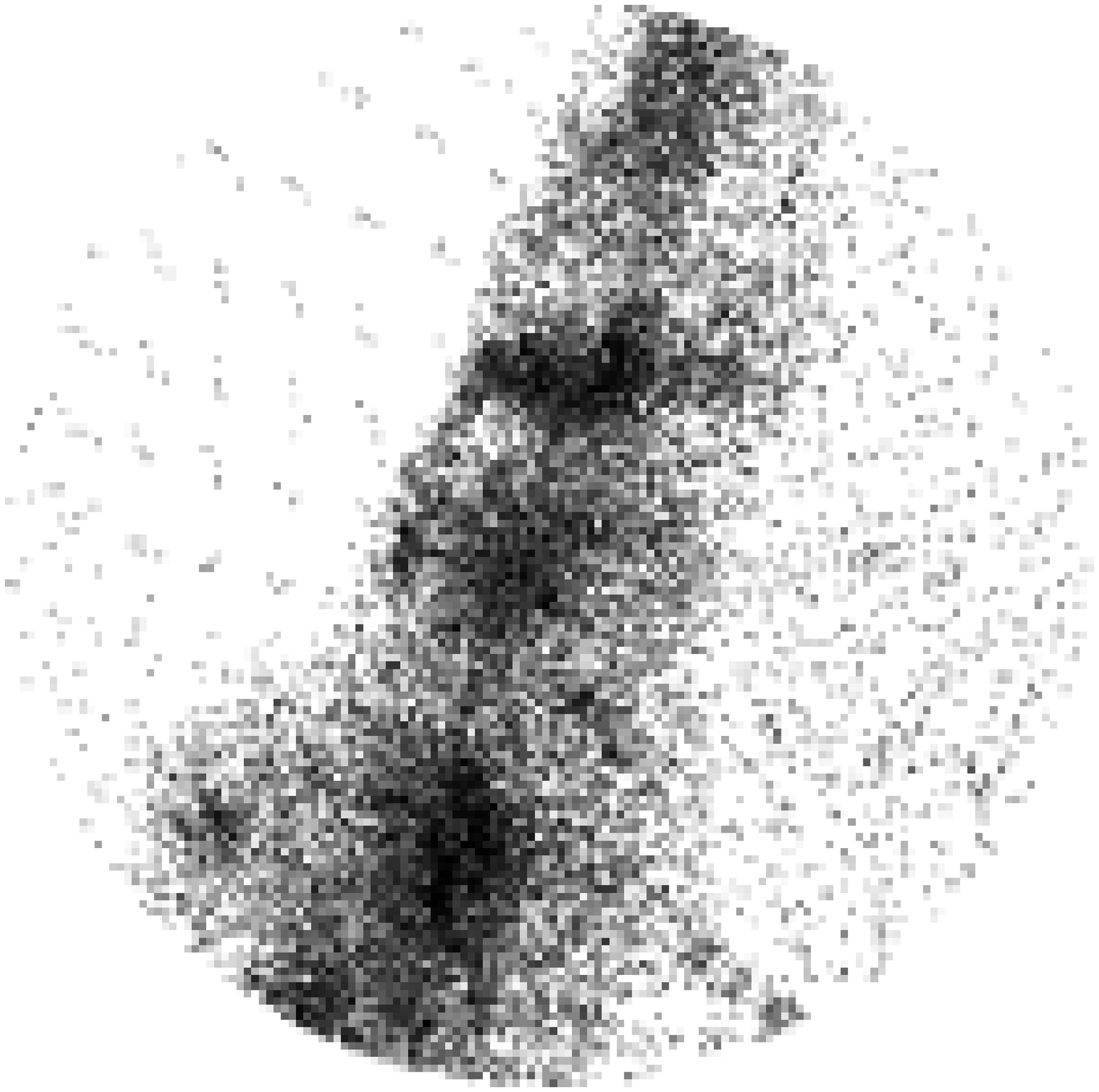}
    \includegraphics[scale=0.025]{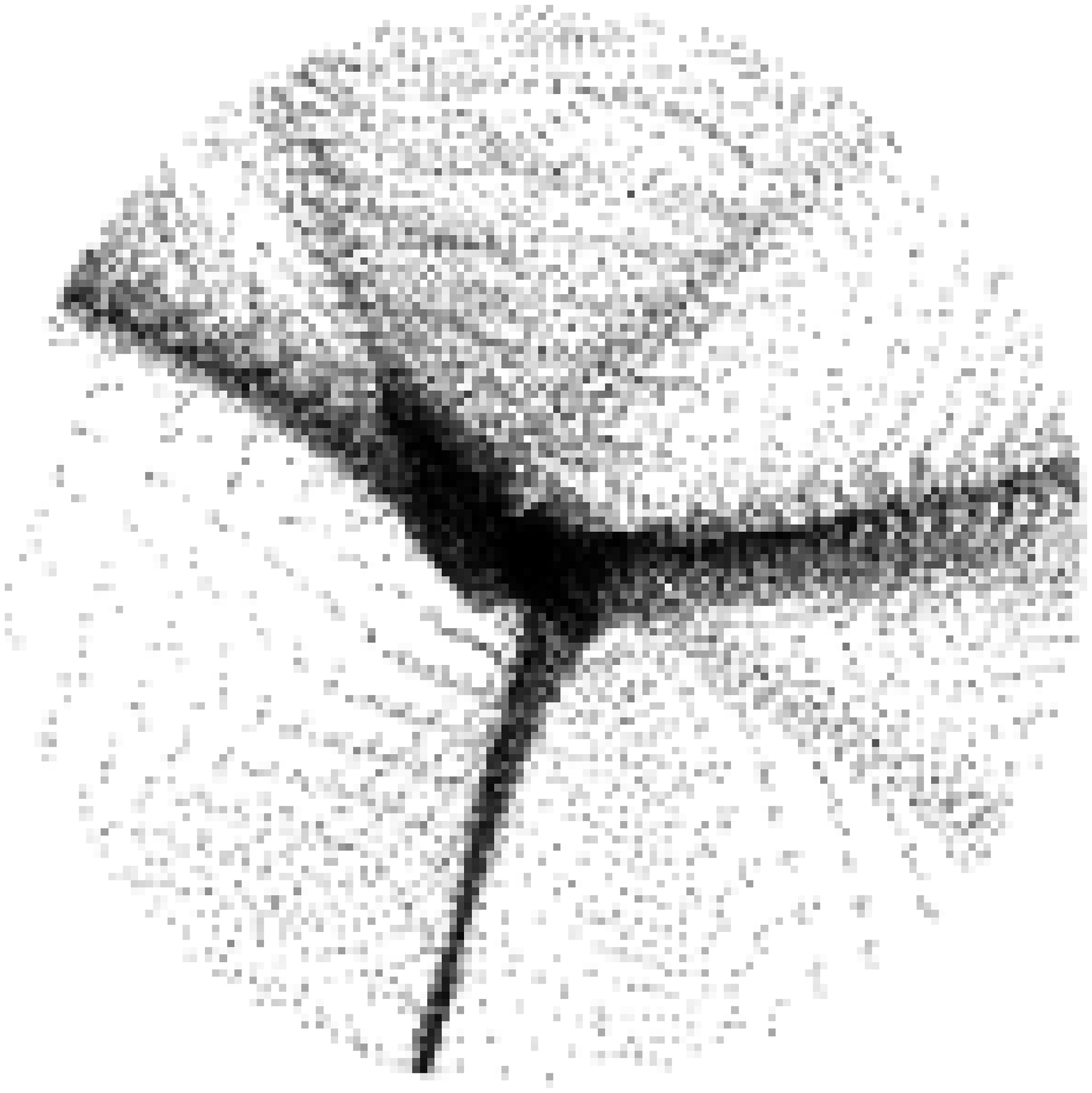}
    \includegraphics[scale=0.025]{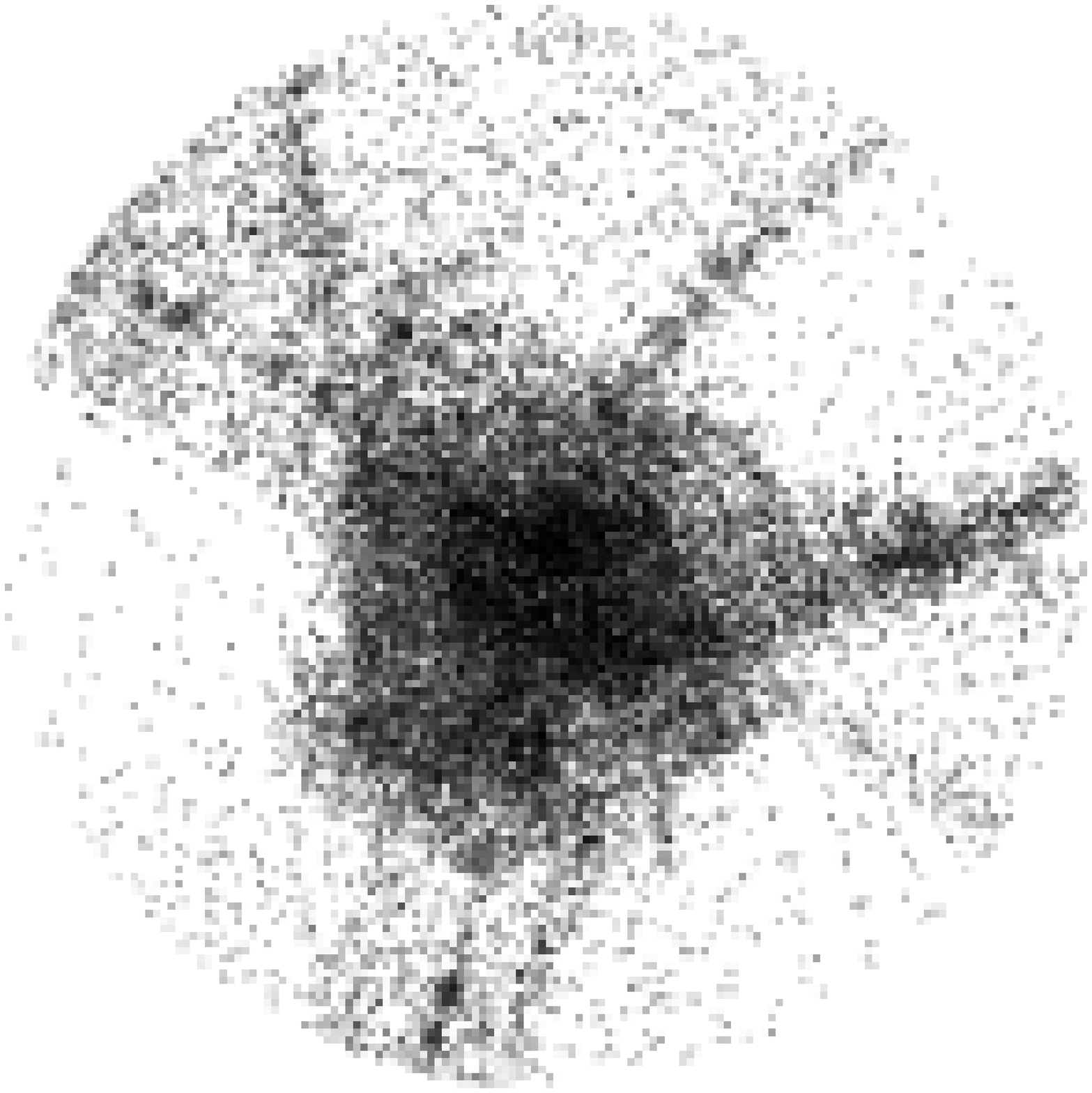}}}
}
    \caption[1]{
    \label{triad}
Sheets of gas radiating from the other filaments around the three
largest galaxies.  
Each horizontal pair of images represents the same filament with
the gas particles on the left and the dark particles on the 
right.  Each filament is viewed as a projection on a plane
perpendicular to the filament with the filament in the center
going into the page.  Each image is $110.8 \unit{kpc}$ on a side.  Only
the portion along the filament from 22.2 to $55.4 \unit{kpc}$ from the center
of the galaxy is shown.
    }
\end{figure}

\cmdcartoon

\cmdslabbf

\cmdfilsheet

\section{Filament Associated Sheets}

Radiating from the filaments are thin sheets of gas.
Figure~\ref{triad_tilt} shows three sheets extending from a single
filament of the galaxy shown in Figure~\ref{shell}.  The images
on the left show gas particles, and those on the right dark matter
particles.  The top panel
shows a projection on to a plane perpendicular to the filament, with
the filament in the center.  The top left image shows the three
sheets edge on.  The lower left image shows this complex from another
angle so that the three sheets are fully visible.  The filament 
lies at the intersection of the three sheets.  Outlines of a
surrounding box are shown to enhance the three-dimensional
effect.  The filament is perpendicular to the front face of this
box.  The image on the lower right shows the same view as on the left
but with dark matter
particles only.  There is some tendency for the dark matter to
concentrate  in the vicinity of the sheets, but it tends to form 
quasi-spherical haloes rather than the thin, extended sheets formed
by the gas.

Figure~\ref{triad}, a montage of images of the remaining 10 filaments
from the three largest galaxies (omitting the one shown in
Figure~\ref{triad_tilt}), gives an idea of the variation seen.
Each image is a view centered on a single filament running
perpendicular to the page.  The images are arranged in horizontal pairs with
the gas particles shown on the left and the dark particles on the
right.  The spokes of gas radiating from the filament are the 
sheets seen edge on.  Viewed face on, the sheets resemble those in the
image in the lower left section of Figure~\ref{triad_tilt}.
Although dark matter clusters near the sheets, it is more diffusely
distributed than the gas.

To quantify the gas density and baryonic fraction of the sheets,
we averaged over rectangular slabs
whose geometry and selection is diagrammed in Figure~\ref{cartoon}.
To select sheets,
we take a cylinder of radius $55.4 \unit{kpc}$
coaxial with each filament,
and divide it into 16 equal angular sectors.
We exclude from the volume analyzed an inner cylinder of radius
$8 \unit{kpc}$
to eliminate the central region of the filament itself.
We select as a sheet those sectors
that contain at least 20\% of the gas mass in the cylindrical shell.
Having identified a sheet in this fashion,
we measure the gas density and baryon fraction
in a rectangular slab
extending from 8 to $55 \unit{kpc}$ away from the filament,
and either 8 or $16 \unit{kpc}$ in thickness.

This selection process produced
32 sheets from the 29 filaments around the 10 largest galaxies.
Figure~\ref{slab_bf} shows the baryonic fraction of each sheet,
measured in slabs of thickness
8 or $16 \unit{kpc}$ centered on the sheets.
The Figure shows that there is a tendency for the sheets to be enriched
in baryons.

Figure~\ref{fil_sheet}
compares the surface density of gas of each of the 32 sheets
with the linear density of gas in the associated filament.
The surface densities of gas vary over a range two orders of magnitudes
and have a positive correlation with the filament densities that is
marginally significant at the $2 \sigma$ level\footnote{
Spearman's rho for rank correlation is 0.53,
which for a two-tailed distribution is significant at the $95\%$ level.
To compute this, sheets attached to the same filament were averaged,
resulting in 15 pairs for rank order analysis.
}.

\section{Discussion and conclusions}
\label{discussion}

Using a high resolution cosmological simulation of reionization,
we have found evidence for a significant population of baryon-rich filaments
at a redshift of 5.1.
The filaments are associated with the largest galaxies,
and can be found radiating away from their centers.
The baryon-rich filaments are at the intersections of sheets,
which also tend to be enriched in baryons.
The dark matter is distributed more irregularly,
tending to form quasi-spherical bodies.

The present work helps to fill a somewhat neglected gap
between the linear regime of gravitational collapse and the highly
nonlinear regime of galactic interiors.
Our investigation has highlighted
the contrasting distribution of baryons and dark matter
on $\sim 6 \unit{kpc}$ scales where hydrodynamic processes
are important, but intergalactic structure is still readily delineated.
The simulation investigated in this paper was originally carried out
in order to explore reionization,
and therefore includes detailed baryonic physics,
including 3-dimensional radiative transfer of ionizing radiation
self-consistently generated by star formation,
that is often neglected in other kinds of cosmological simulation.
The results point to a need for careful treatment of baryonic processes
in modeling moderately nonlinear scales.

At large, linear scales,
baryons and dark matter are expected to follow the same distribution.
In the opposite limit of small, highly non-linear scales,
the baryonic fraction within the virial radius of a collapsed halo
is close to the mean \citep{crain_07}.
It may therefore seem suprising that
at intermediate, mildly non-linear scales,
the difference between the distribution of baryons and dark matter
can be as prominent as found in this paper.
Linear theory predicts that initially the difference in the clustering
of baryons and dark matter is described by a filtering scale
\citep{gnedin_hui_98},
below which pressure forces tend to smooth the baryons
compared to the dark matter.
The present simulation indicates that, at least in regions where
the separation of baryons and dark matter is most marked,
the separation is more complex, and anisotropic.
Baryons tend to concentrate toward the centers of filaments and sheets,
in directions perpendicular to the filament or sheet,
but are more smoothly distributed than dark matter
in directions along the filament or sheet.
The central concentration of baryons compared to dark matter
is consistent with simple models of one-dimensional collapse into filaments
and sheets \citep{shandarin_89, anninos95}
although this consistency is perhaps surprising
given that evolution is much more complicated in three dimensions than in one.
Another unexpected finding is that the baryon-rich filaments and sheets
remain coherent
over scales much larger than the filtering scale of
$4 \unit{kpc}$ proper
($24 \unit{kpc}$ comoving)
predicted at this redshift by \citet{gnedin_filt03}.

Undoubtedly much of the difference between baryons and dark matter
in the simulation reported in this paper can be attributed
to the fact that baryons are collisional whereas dark matter is collisionless:
parcels of collapsing gas cannot pass through each other,
whereas dark matter can.
Probably another important effect is that the energy of collapse
of the baryons into filaments and sheets
is channelled into gas pressure,
mediated at least in part by co-axial shocks, as suggested by
\citet{cen94}, \cite{miralda_escude_96}, and \citet{zhang_98},
which can smooth the baryons
in directions along the filaments and sheets.

The baryon-rich filaments may have important implications for our
understanding of structure formation in the universe.  
First,
charting the course of reionization depends heavily upon uncertain
assumptions about the escape fraction of ionizing radiation from
galaxies.  The highly aspherical distribution of the surrounding gas
should be taken into account in mapping
the journey of such a photon from its origin near the
center of the galactic halo.  In this regard it is 
interesting that the filaments
appear to extend essentially unaltered into nearly the center of the galaxy.

Second, the filaments are likely to be important in the process of 
gas accretion on to galaxies.
As discussed by \citet[and references therein]{keres05},
two general modes of accretion are
distinguished, a hot mode and a cold mode.
In the hot mode, which dominates the growth of the largest galaxies,
gas, shock heated to the virial temperature of the
dark matter halo,  accretes in a spherically symmetric fashion.
In the cold mode, which dominates for smaller galaxies,
such as those seen in our simulation,
gas accretes in a directional manner, often along filaments
\citep{kawata_rauch_07}.
The filaments we see are obvious candidates
for cold accretion conduits.
Since the ability of a galaxy to accrete gas is important for 
sustained star formation, 
the evolution of these structures may in turn help us to 
understand the star formation history of the universe.

Third,
the planar distribution of Milky Way satellite galaxies may have resulted
from an early sheet \citep{libeskind05},
possibly of the kind described in this paper.

The baryon-rich structures discussed in this paper
may be potentially observable in the Lyman alpha forest
(for reviews see \citealt{rauch98}; \citealt{meiksin_07}),
although a precise assessment of their observational impact
will require further work that goes beyond the scope of this paper.
Descendents of these structures may play a role in producing
the warm-hot intergalactic medium (WHIM),
thought to be a major reservoir of baryons 
in the low redshift universe \citep{dave_01}.

\section*{Acknowledgments}
This work was supported in part by the DOE and NASA grant NAG
5-10842 at Fermilab, by NSF grants AST-0134373 and AST-0507596,
and by the National Computational Science Alliance grant AST-020018N,
and utilized IBM P690 arrays at the National Center for Supercomputing
Applications (NCSA) and the San Diego Supercomputer Center (SDSC).

\bibliographystyle{mn2e}

\bibliography{spine11}

\appendix
\section{Varying Cell Size and Redshift}

Figure~\ref{rich_fract_bsize}
shows the volume fractions enriched in baryons for grid cell sizes of
4.1 and $16.3 \unit{kpc}$ proper (25 and $100 \unit{kpc}$ comoving)
at a redshift of 5.1.
These graphs are to be compared to
Figure~\ref{rich_fract}, which shows the results for the
$6.52 \unit{kpc}$ ($40 \unit{kpc}$ comoving)
cell size used predominantly in the paper.
The results for a grid cell size of $4.1 \unit{kpc}$ are consistent
with those at $6.52 \unit{kpc}$.  We preferred the larger cell size
because estimates of the baryon fraction in each cell are more
accurate, especially at lower overdensities.
Increasing the cell size to $16.3 \unit{kpc}$
leads to a considerable loss of baryon-rich cells.
This is consistent with the $\sim 5 \unit{kpc}$ radius of filaments
illustrated in Figure~\ref{fil_profile}.

\cmdrichfractbsize
\cmdfighetc  

Figure~\ref{fig_h1_xsect2_census_gas}
compares the baryon-rich fraction
over a range of redshifts
(the data for z = 4 was obtained from a lower resolution simulation).
Baryonic enrichment appears generally to increase with time.
Computational limitations prevent us from
evolving the simulation to lower redshifts.

\end{document}